\documentclass[fleqn,usenatbib]{mnras}

\usepackage{newtxtext,newtxmath}

\usepackage[T1]{fontenc}

\DeclareRobustCommand{\VAN}[3]{#2}
\let\VANthebibliography\thebibliography
\def\thebibliography{\DeclareRobustCommand{\VAN}[3]{##3}\VANthebibliography}

\usepackage{graphicx}	
\usepackage{amsmath}	
\usepackage{float}
\usepackage{fontawesome5}
\usepackage{hyperref}

\title[Population fit Stellar Streams]{Constraining the population of dark matter halo shapes using hierarchical inference with extragalactic stellar streams}

\author[D. Chemaly et al.]{
David Chemaly$^{1}$\thanks{E-mail: dc824@cam.ac.uk},
Elisabeth Sola$^{1}$,
Vasily Belokurov$^{1}$,
Sergey E. Koposov$^{1, 2}$,
GyuChul Meyong$^{1}$,
\newauthor HanYuan Zhang$^{1}$,
Denis Erkal$^{3}$
\\
$^{1}$ Institute of Astronomy, Madingley Rd, Cambridge CB3 0HA, UK\\
$^{2}$ Institute for Astronomy, University of Edinburgh, Royal
Observatory, Blackford Hill, Edinburgh EH9 3HJ, UK\\
$^{3}$Department of Physics, University of Surrey, Guildford, GU2 7XH, Surrey, UK\\
}

\date{Accepted 2026 June 12. Received: 2026 May 28; in original form 2026 January 21}

\pubyear{\the\year{}}

\begin{document}
\label{firstpage}
\pagerange{\pageref{firstpage}--\pageref{lastpage}}
\maketitle

\begin{abstract}

Stellar streams, the debris of tidally disrupted satellites, trace their host's gravitational potential and thus probe dark matter halo structure. While six-dimensional phase-space data of Galactic streams enable precise dark matter halo modelling in the Milky Way, streams around external galaxies are typically available only as low surface brightness features without kinematics (i.e. two-dimensional photometric data), providing only weak constraints when considered individually. We present a hierarchical Bayesian framework that infers the population distribution of halo flattening using only projected stream tracks. Streams are forward-modelled in \texttt{StreaMAX}~{\renewcommand\thefootnote{\protect\mbox{\protect\faGithub}}\footnotemark}, a new \texttt{JAX}-accelerated particle-spray package that achieves orders of magnitude faster stream generation when compared to traditional methods. For each stream we fit an axisymmetric dark matter halo model and obtain a posterior on the flattening. These posteriors are then combined through hierarchical reweighting to constrain the population distribution. Using mock data, we show that individual fits recover the correct flattening with modest precision and exhibit projection-induced multi-modalities. Nevertheless, aggregating these fits yields accurate and confident constraints on the underlying population distribution of dark matter halo morphologies, clearly distinguishing between oblate, spherical, and prolate populations. The total computational cost scales linearly with sample size. Our results demonstrate that ensembles of purely photometric streams carry sufficient information to constrain dark matter halo shapes in external galaxies at the population level. With the forthcoming samples from Euclid and Rubin/LSST, this approach offers a practical path to population-level inferences of halo morphology without any kinematic measurements.

\end{abstract}


{\renewcommand\thefootnote{\protect\mbox{\protect\faGithub}}\footnotetext{\url{https://github.com/David-Chemaly/StreaMAX}}}



\section{Introduction}

Stellar streams, elongated structures formed when dwarf galaxies or globular clusters are tidally disrupted by their host's gravitational field, are powerful probes of galactic potentials and tracers of galaxy assembly \citep{Johnston1995, Helmi1999}. As satellites orbit their hosts, tidal forces strip stars from the progenitor galaxy, producing coherent, narrow features that approximately follow the progenitor’s orbit and can persist for gigayears, retaining dynamical memory of their past trajectories \citep{Johnston1995, Helmi1999, Bullock2005}. Because these trajectories depend sensitively on the underlying gravitational potential, streams provide robust constraints on mass distributions, especially dark matter (DM) haloes informing parameters such as halo mass, concentration, and shape \citep[e.g.,][]{Ibata2001b, Law2010, Koposov2010, Belokurov2014, Gibbons2014, Bonaca2018}. Streams also serve as fossil records of accretion, enabling reconstructions of merger histories and halo growth \citep[e.g.,][]{Bullock2005, Deason2024, Bonaca2025}.

Early theoretical work and simulations established the sensitivity of streams to host mass distributions over large radial ranges \citep{Johnston1995, Helmi1999, Johnston2001, Ibata2001b}. Subsequent surveys and follow-up observations \citep[e.g.,][]{Grillmair2006, Malhan2018, Shipp2018, Li2022} have leveraged stream morphology and kinematics to probe the potential of the Milky Way (MW) in detail \citep[e.g.,][]{Koposov2010, Price-Whelan2014, Price-Whelan2018, Tavangar2025}. Dynamically cold, thin streams from globular clusters respond to subtle variations in the local gravitational field, exhibiting gaps, spurs, and track curvature changes that encode small scale perturbations such as dark matter subhaloes \citep[e.g.,][]{Ibata2002, Johnston2002, Yoon2011, Carlberg2012, Erkal2015, Bovy2016, Erkal2017, Banik2019}. By contrast, broader, dynamically hotter streams from dwarf galaxy progenitors constrain halo shape and mass on larger spatial scales \citep[e.g.][]{Johnston2005, Law2010, Pearson2015, Gibbons2014, Erkal2019, Vasiliev2021}.

Within $\Lambda$CDM, haloes are approximately self-similar across mass scales: they follow universal density profiles \citep{Navarro1997, Zhao2003, Ludlow2014}, follow a mass–concentration relation \citep{Wechsler2002, Maccio2008, Dutton2014, Diemer2015}, and are intrinsically triaxial if baryons are neglected \citep{Jing2002, Allgood2006, Despali2014, Bonamigo2015}. Real galaxies, however, deviate from this idealised picture. Baryonic physics (adiabatic contraction, feedback-driven core formation) can reshape inner density profiles and axis ratios \citep{Blumenthal1986, Gnedin2004, Bryan2013, DiCintio2014, Chua2019, Prada2019}; merger histories modulate triaxiality and concentration scatter \citep{Boylan2006, Fakhouri2010}; and environmental processes (e.g. tidal stripping) alter subhalo structure \citep{Hayashi2003, Pearrubia2008, Errani2020}. Non-standard DM models (e.g. self-interacting or warm DM) predict systematically different internal structures (for example, core or cusp central density) and shapes \citep{Rocha2013, Peter2013, Lovell2014, Tulin2018, Sameie2018}. Quantifying departures from self-similarity at the population level is therefore a sensitive test of galaxy formation physics and, in particular, of dark matter self-interactions through their impact on the shapes of galaxies.

Modern cosmological simulations provide increasingly detailed predictions for the distribution of halo shapes. In dark matter only runs, haloes are predominantly prolate and become more aspherical with increasing mass: \citet{Allgood2006} found a mean minor to major axis ratio of $0.54\,(M_{\mathrm{vir}}/M_{\odot})^{-0.05}$ with a Gaussian scatter around $\sim 0.1$, a scaling confirmed out to high redshift \citep{Despali2014, Bonamigo2015}. At Milky Way masses ($M_{\mathrm{vir}} \sim 10^{12}\,M_{\odot}$), $N$-body haloes typically have a minor to major axis ratio hovering around 0.6 to 0.7 \citep{Jing2002, Allgood2006, Bett2007}. Including baryonic physics substantially alters these predictions: the condensation of gas and the formation of a central stellar disc make haloes significantly rounder and more oblate. In the Illustris simulation \citep{Vogelsberger2014}, \citet{Chua2019} reported a median minor to major axis ratio of $\sim 0.7$ that is nearly invariant with both radius and halo mass once baryons are included. Zoom-in simulations of MW analogues from the Auriga project \citep{Grand2017} confirm this trend, finding that baryons make haloes rounder at all radii, with the inner regions ($r \lesssim 30$\,kpc) showing a strong correlation between halo roundness and the presence of an extended stellar disc \citep{Prada2019}. The EAGLE \citep{Schaye2015} and cosmo-OWLS \citep{LeBrun2014} simulations similarly show that haloes become more aspherical and triaxial with increasing mass, and that the stellar distribution can be misaligned with the host halo by a median of $\sim$$45^\circ$ for disc galaxies \citep{Velliscig2015}. Despite broad consensus on these trends, the predicted scatter in axis ratios remains large, and the detailed shape distribution depends on sub-grid prescriptions for star formation and feedback \citep{Butsky2016, Chua2019}. Measuring the population distribution of halo shapes from observations would therefore provide a powerful test of these models.

Within the Milky Way, the combination of astrometric, photometric, and spectroscopic data, most notably from \texttt{Gaia} \citep{GaiaCollaboration2018, GaiaCollaboration2021}, \texttt{SDSS} \citep{York2000}, \texttt{DES} \citep{DESCollaboration2018}, \texttt{LAMOST} \citep{Cui2012} \texttt{APOGEE} \citep{Majewski2017} and dedicated surveys, for example $S^5$ \citep{Li2019}, have enabled high precision, six-dimensional analyses of streams such as Palomar~5, GD-1, Sagittarius, and Orphan–Chenab \citep{Koposov2010, Kupper2015, Bowden2015, Koposov2019, Malhan2019, Vasiliev2021, Koposov2023}. These studies indicate a nearly spherical inner halo \citep{Bovy2016, Palau2023}, while at larger radii the halo appears more flattened, triaxial, and potentially misaligned with the disk \citep[e.g.,][]{Helmi1999, Law2010, VeraCiro2013, Bonaca2014, Erkal2019, Vasiliev2021, Koposov2023, Nibauer2025a}. The identification of gaps and spurs in GD-1 and Palomar~5 has further enabled tests of the predicted DM subhalo population on kiloparsec scales \citep{Ibata2002, Carlberg2012, Erkal2016, deBoer2018, Bonaca2019, deBoer2020, Banik2021, Nibauer2025c, Valluri2025}, offering insights into the small-scale DM distribution and stream evolution in time-dependent potentials \citep[e.g.,][]{Dillamore2022, Yavetz2023, Weerasooriya2025}. Beyond the MW, M31’s Giant Southern Stream, spanning a wide range of galactocentric radii, has provided stringent constraints on its halo mass despite only semi-resolved photometric and spectroscopic stellar data \citep{Fardal2013}.

Extending stream analyses to more distant galaxies introduces observational and modelling challenges. Outside the Local Group, stars are unresolved and full six-dimensional phase-space information is generally unavailable, complicating proper motion, distance, and line-of-sight velocity measurements. Pioneering modelling of projected (two-dimensional; 2D) streams dates back to \citet{Johnston2001}; since then, numerous studies have focused on low surface brightness (LSB), projected features with limited or no kinematic data \citep[e.g.,][]{Tal2009, MartinezDelgado2010, Mouhcine2010, Atkinson2013, Crnojevic2016, Carlin2016, Hood2018, vanDokkum2019, Bilek2020, Sola2022, MartinezDelgado2023, Yoon2024, Carretero2024, Sola2025}. \citet{Amorisco2015b} fitted the projected surface brightness morphology of the NGC1097 “dog-leg” stream using the progenitor remnant’s line-of-sight velocity, enabling an estimate of the host’s halo mass. \citet{Pearson2022} analysed the Dw3 stream around Centaurus~A using 2D positions and demonstrated a degeneracy between orbital velocity and halo mass. Incorporating even a single radial velocity measurement breaks this degeneracy and yields a lower bound on the host’s halo mass.

In many extragalactic systems, however, neither radial velocities nor distance gradients along the stream are available. Recent work shows that purely geometric diagnostics, such as stream track curvature and stream loop precession, still provide meaningful constraints on halo properties. \citet{Nibauer2023} demonstrated that curvature vectors of projected streams limit the orientation of the host’s acceleration field and thus constrain halo triaxiality. Complementarily, \citet{Walder2024} showed that multi-loop streams encode precession angles that inform halo radial structure, offering constraints on scale radii and concentrations without direct kinematics. In a more recent study, \cite{Nibauer2025b} have leveraged a generative model for extragalactic stream to constrain the radial profile of the host potential. These methods highlight the untapped potential of photometric morphologies alone for probing DM haloes.

Forthcoming wide-field imaging from the Rubin Observatory Legacy Survey of Space and Time (Rubin/LSST; \citealt{Ivezic2019}) and Euclid \citep{Laureijs2011} will dramatically expand the census of streams around external galaxies \citep[e.g.,][]{LSST2009, Laureijs2011, Laine2018, Ivezic2019, Pearson2019, Pearson2024, Walder2024}. This increase in detections motivates population-level methodologies that combine information from multiple streams to deliver statistically powerful constraints on the distribution of halo properties across galaxies. Our approach is informed by the extensive stream-modelling literature in the MW \citep[e.g.,][]{Johnston1998, Binney2008streams, Eyre2011, Varghese2011, Kuepper2012, Sanders2013, Gibbons2014, Sanders2014, Bovy2014, Bowden2015, Fardal2015, Amorisco2015, Hendel2015, Erkal2019, ChenLi2025} and recent advances for external galaxies \citep[e.g.,][]{Pearson2022, Nibauer2023, Walder2024}. 

In contrast to previous studies which considered fits to individual streams, we leverage hierarchical modelling to combine multiple low informative fits and obtain strong population level constraints on halo shape distributions.

In this study, we present a population based framework for analysing extragalactic stellar streams. We validate an end-to-end pipeline on simulations by fitting streams for individual systems and then performing hierarchical Bayesian inference to constrain the distribution of DM halo shapes across a galaxy population. To do so, we have developed \href{https://github.com/David-Chemaly/StreaMAX}{\texttt{StreaMAX} \textsc{\Large{\scalebox{0.8}{\faGithub}}}}, a novel, publicly available, \texttt{Python} package to rapidly model streams leveraging \texttt{JAX} \citep{jax2018github}, a high-performance numerical computing library with automatic differentiation and just-in-time (JIT) compilation. It enables efficient computation on CPUs, GPUs, and TPUs, making it especially useful for optimization and Bayesian inference. 

This work serves as a proof of concept and therefore relies exclusively on simulated data. In a following companion paper, we will apply our method to real observations. However, while several catalogues of stellar streams already exist, their depth, processing, and feature definitions vary significantly, and most identified structures are unsuitable for our modelling framework either being too short, too straight, or too disturbed by major mergers. To overcome this, we introduced in a companion paper the \texttt{STRRINGS} sample \citep{Sola2025}, a catalogue of 35 nearby galaxies hosting long, faint, and curved streams, selected for their suitability for dynamical modelling. Using residual images from the DESI Legacy Imaging Surveys (DESI-LS, \cite{Dey2019}), which enhance faint features by modelling and subtracting all bright sources, we visually inspected thousands of galaxies and segmented streams meeting our criteria. The resulting sample provides high quality, well defined stream tracks ideal for forward modelling. In future work currently underway, we will apply the methodology developed here directly to \texttt{STRRINGS} to infer the population-level distribution of dark matter halo shapes across galaxies in the Local Universe.

The paper is organised as follows. Section~\ref{sec:Methods} describes the stream-generation and forward-modelling methodology. Section~\ref{Individual_fit} presents the fitting procedure for individual simulated streams. Section~\ref{sec:hierarchical} outlines the hierarchical Bayesian model and reports population-level constraints on halo shapes. Section~\ref{sec:discussion} summarises our discussions and conclusions.

\section{Methods}\label{sec:Methods}

This work aims to infer the morphology of dark matter haloes from extragalactic stellar streams with no kinematic information. To achieve this, we first construct a three-dimensional model of a tidal stream evolving in a given gravitational potential, which is then projected onto the plane of the sky. From this projection, we extract the stream’s track defined as a one-dimensional ridgeline, an idealised, widthless trajectory tracing its central path. The parameters governing this forward-modelling pipeline will be described at the end of this section, along with the physically motivated priors adopted to limit degeneracies, constrain modalities, and ensure observational realism.

\subsection{Stream Generation}

A variety of methods exist to generate stellar streams, trading realism for computational cost. The most straightforward approximation, treating stream tracks as orbits \citep[e.g.][]{Koposov2010, Malhan2019}, is known to be biased \citep{Sanders2013}. Early work therefore used orbit fits with corrective terms \citep{Varghese2011}, followed by test-particle, streakline, or particle-spray techniques that more faithfully capture leading and trailing tails \citep{Kuepper2012, Gibbons2014}. $N$-body simulations best model disruption and self-gravity but are orders of magnitude more computationally expensive, as each particle interacts with every other one \citep{Eyre2011, Sanders2014}. Here we adopt the particle-spray method \citep[e.g.][]{Johnston1998,Kuepper2012,Gibbons2014,Fardal2015}, which has been shown to accurately and efficiently reproduce stream tracks. 

In our implementation, the progenitor is modelled as a simple analytic potential, a Plummer sphere, specified by a characteristic mass and scale radius. This analytic form provides a smooth, time-independent description of the progenitor’s self-gravity, which is required both to compute its tidal boundary and to determine the characteristic scale of the escaping debris. Rather than follow the evolution of all stars bound to the progenitor, we assume that only particles escaping through the tidal Lagrange points contribute materially to the observable stream. This avoids integrating large numbers of particles that would remain bound and never enter the debris.

At each integration step we therefore compute the instantaneous inner (L1) and outer (L2) Lagrange points. These are the saddle points of the effective potential along the line joining the Galactic centre to the progenitor, determined by the balance between the host’s gravitational field, the progenitor’s self-gravity, and the centrifugal term in the co-rotating frame \citep{King1962}. They define the two primary escape channels through which material becomes unbound during tidal stripping.

New debris particles are \textit{sprayed} from small spatial and velocity offsets around L1 and L2. These offsets serve as a proxy for the progenitor’s internal velocity dispersion and naturally generate both leading and trailing arms. In this work we adopt the dispersion prescription of \citet{Fardal2015}, which samples position and velocity perturbations consistent with the Plummer model and produces cold, kinematically coherent streams. Because we inject particles only at the tidal boundary, every spawned tracer is, by construction, likely to escape the progenitor rather than remain bound.

Once released, particles are treated as massless tracers and evolved forward in time under the combined gravitational field of the host potential and the progenitor’s analytic potential. The progenitor therefore continues to influence nearby debris until the particles move several scale radii away, at which point its contribution becomes negligible. Crucially, the debris particles do not exert forces on one another, which eliminates the $\mathcal{O}(N^{2})$ scaling of full $N$-body simulations and enables extremely fast generation of realistic streams. Despite this simplification, the particle-spray method reproduces key morphological features of tidal streams such as bifurcations, arm asymmetries, and on-sky track curvature with excellent fidelity and at orders-of-magnitude lower computational cost compared to full $N$-body simulations \citep[e.g.][]{Sanders2016, Chen2025}. Even so, a single spray realisation typically requires a few seconds, making large fitting campaigns (hundreds of thousands of streams) computationally heavy.

To address this, we built \href{https://github.com/David-Chemaly/StreaMAX}{\texttt{StreaMAX} \textsc{\Large{\scalebox{0.8}{\faGithub}}}}\footnote{\textbf{\url{https://github.com/David-Chemaly/StreaMAX}}}, a fully \texttt{JAX} \citep{jax2018github}–accelerated package that exactly reproduces conventional spray generators, such as the ones in \texttt{GALA} \citep{Price2017} and \texttt{AGAMA} \citep{Vasiliev2019}, while gaining orders of magnitude of speed-up through compilation and efficient parallelism. Crucially, spray particles evolve independently, meaning that every particle can be integrated in parallel without requiring any information on any other particle. Therefore, we can accelerate the time integration using vectorised mapping and Just-in-Time (JIT) compilation. The whole package is written in \texttt{JAX} meaning that gradients of any function can easily be obtained.

\begin{figure}
    \centering
    \includegraphics[width=1.0\linewidth]{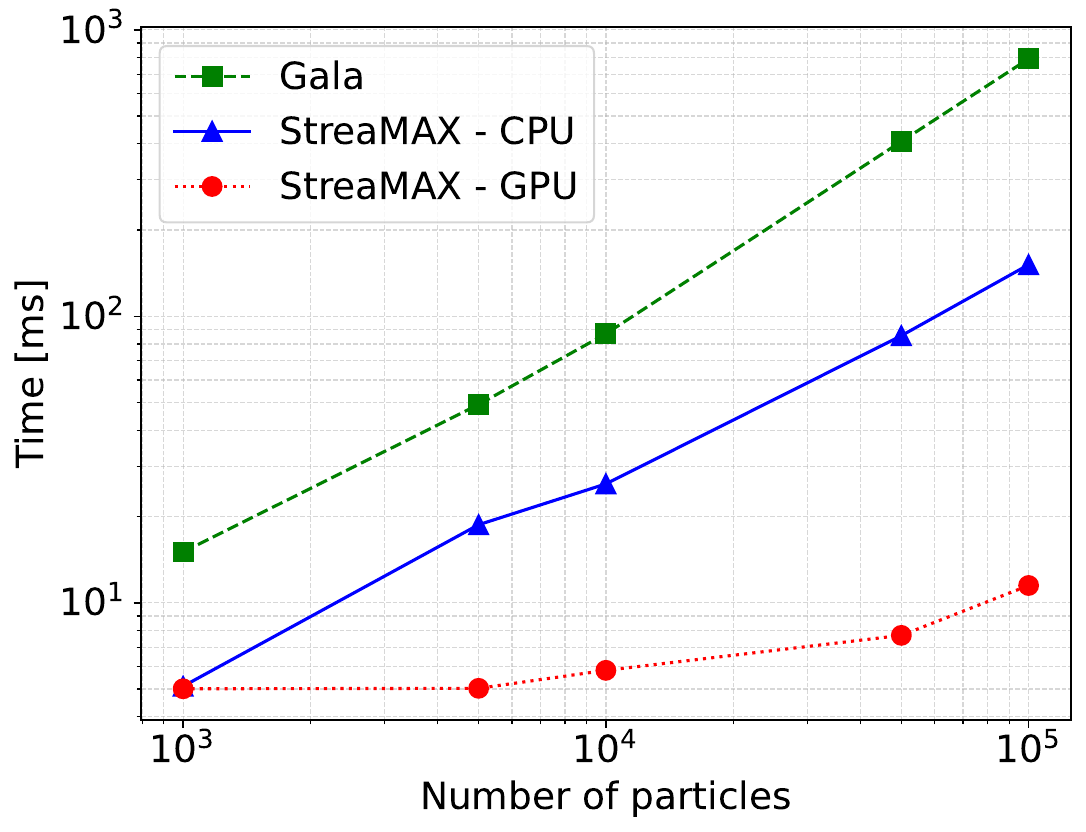}
    \caption{Time required to model a tidal stream as a function of the number of particles, using identical integration settings across \texttt{Gala} and \texttt{StreaMAX} on both CPU and GPU. CPU tests were performed on a 44-core node, and GPU tests use an NVIDIA L4. \texttt{StreaMAX} outperforms the conventional package, with the GPU version delivering the fastest and flattest scaling with particle number. For reference, a full nested sampling run for a single stream ($10^4$ particles, 2000 initial live points) takes $\sim$10~hours on the CPU.}
    \label{fig:JAX_speed}
\end{figure}

Figure~\ref{fig:JAX_speed} shows the time required to generate a stream as a function of the number of particles. All pipelines are tested using identical stream parameters and orbital integration settings. Specifically, we use \texttt{Gala} v1.11.0 and set it to use the leapfrog integrator, matching the integrator used in \texttt{StreaMAX}, with the same step size and number of particles to ensure a fair comparison. Note that \texttt{Agama} was excluded from the comparison because it uses a different integration algorithm which can not be changed. CPU benchmarks were performed on a 44-core node, and GPU results use an NVIDIA L4. On the CPU, \texttt{Gala} and \texttt{StreaMAX} exhibit similar scaling with particle number, with \texttt{StreaMAX} being the fastest across all tested sizes. When run on a GPU, \texttt{StreaMAX} achieves a substantial speedup and displays a flat scaling curve for low number of particles and a slight trend upwards at higher values. This suggests that fixed overheads dominate for low sizes where adding particles barely affects runtime. Although GPUs provide the best performance, our main analyses use CPUs due to wider availability on our computing resources. Throughout this work, we generate streams with $10^4$ particles.

The time integrations are performed using a standard leapfrog scheme \citep{Hockney1988} with fixed time steps. In this work, since the assumed potential model is smooth and well behaved (see Subsection \ref{model_params}), the number of steps is set to 100. This choice was empirically tested and shown to conserve energy and yield negligible differences compared to integrations with substantially smaller time steps. For more efficient computing, we want to have the same amount of total steps taken for every particle but, in the spray method, particles are born at different times. A particle released later has less total time to evolve, so its timestep must be smaller if it is still to take the same number of steps as earlier particles. In practice, this means early-born particles have larger timesteps and late-born particles have smaller ones. For efficient vectorisation and JIT compilation, \texttt{JAX} requires fixed array dimensions, which motivates adapting the temporal resolution as a function of integration time for each particle.

\subsection{Track Extraction}\label{sec:track}

As mentioned earlier, streams are generated in 3D, then projected onto the sky. Because the potential’s orientation is already a sampled parameter (see Section \ref{model_params}), varying the viewing direction (i.e., the line of sight from which the stream is projected) is unnecessary; a fixed projection suffices. Here, for simplicity, we choose the $XY$-plane. Note that the progenitor’s final phase-space coordinates are free parameters in the inference, so the stream’s orbital plane is not fixed and can vary relative to the line of sight even under a fixed projection choice. In this paper we fit only the stream track, intentionally discarding width and density information.

To extract the track from a modelled stream, we need to obtain the ordering of the particles along the stream. To do so, we unwrap each particle’s orbital angle individually. We begin by integrating the full 3D orbit of the progenitor. Because this orbit is continuous, the ordering of points along it is already known, and its angular coordinate can be unwrapped in the usual sense (i.e., allowing the angle to increase beyond $2\pi$ rather than resetting after each revolution). We then use the progenitor’s unwrapped angle to assign a birth angle to each stream particle. Since particles are released at different times, their true initial angle is not restricted to the interval $[0,2\pi]$. Instead, the initial angle of a particle is defined as the progenitor’s unwrapped angle at the particle’s birth time, plus the particle’s own phase offset within $[0,2\pi]$. Finally, we re-centre all particle angles by subtracting the progenitor’s final unwrapped angle. With this convention, trailing tail particles have negative angles, the progenitor sits at $0^\circ$, and leading-tail particles have positive angles. This procedure assumes monotonically increasing angles along each orbit and works robustly for our setups.

We extract the stream track by binning azimutally the unwrapped angular coordinate and averaging the particle radii within each bin. This procedure follows the method of \citet{Sola2025}, where we used it to automatically extract the tracks of segmented real external streams (see their Figure B1). The key improvement here is that our ordering and binning are fully automated, requiring no human intervention and thus enabling population-scale inference. The same bins can also provide estimates of the stream density (particle counts) and width (radial standard deviation), but these quantities are not used in the present work because our current particle-spray prescription does not produce realistic density and width profiles. This is because (i) the progenitor mass is held constant throughout the disruption, and (ii) particles are released from the Lagrange points at uniform time intervals, independent of their separation from the progenitor. Therefore, we choose to work using only the stream track which retains geometric information about the global potential shape but loses dynamical information about the progenitor mass, disruption history, and local potential fluctuations \citep{Bonaca2018}.

In this work, the angular bin size is fixed at $10^\circ$. This choice, adopted from \citet{Sola2025}, represents a balance between maximising angular resolution and maintaining sufficient information per bin to limit radial uncertainties. In STRRINGS, this binning led to an average radial distance uncertainty of approximately 2$\%$ which we include as Gaussian radial noise in our mock data. We note that in such a pipeline, we expect finer bins to yield tighter constraints on $q$, as predicted by \citet{Nibauer2023}, since there is no signal to noise penalty for going finer. To properly test how sensitive to bin size the posterior on flattening is would require a muc more realistic setup that captures the trade-off between angular resolution and signal to noise: smaller bins contain fewer photons and can yield very large radial uncertainties or even bias the extracted track in regions where the stream barely rises above the background. Lacking such a setup here, we adopt the $10^\circ$ binning from STRRINGS together with its corresponding 2\% average radial noise, so that our mocks reflect the regime in which the pipeline will be applied.

\subsection{Model Parameters}\label{model_params}

Before introducing our model parameters, we note that in this work the potential used to generate the mock streams is the same as the potential assumed during modelling. This choice is not required in general, but it allows us to test the intrinsic constraining power of our pipeline without introducing additional bias from mismatches between model and data.

\begin{figure}
    \centering
    \includegraphics[width=1.0\linewidth]{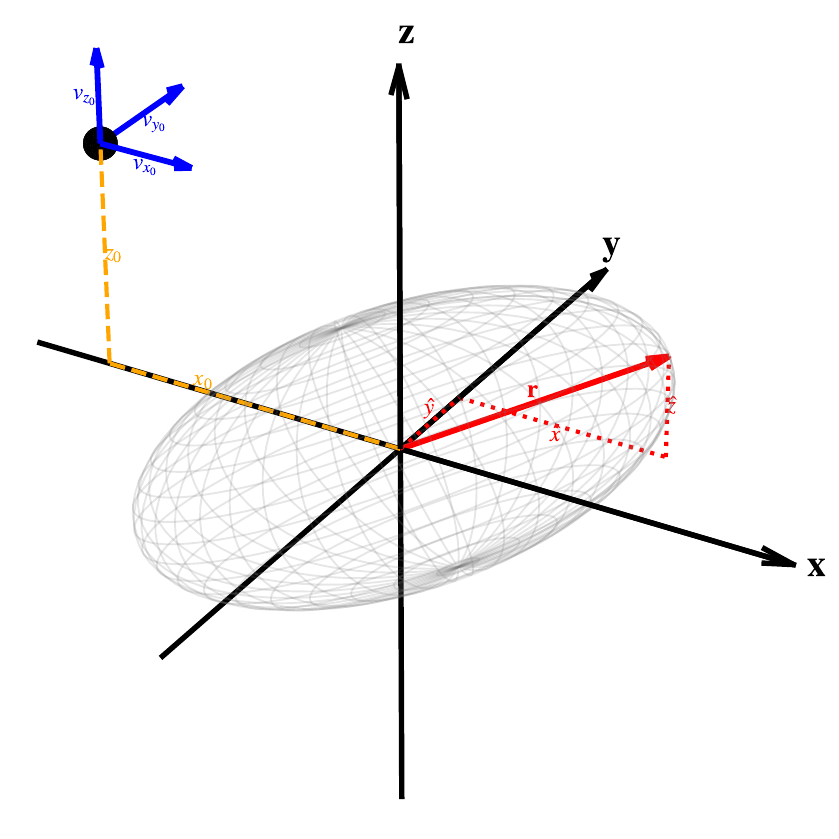}
    \caption{Schematic representation showing the coordinate system of the axisymmetric NFW potential (grey grid), the halo flattened axis $\boldsymbol{r}$, and the progenitor’s present-day phase-space location (black circle). The origin is fixed and taken as the centre of the DM halo.}
    \label{fig:schema}
\end{figure}

We model streams in an axisymmetric Navarro-Frenk-White (NFW) potential. The halo is characterised by its mass $M_{\mathrm{halo}}$, scale radius $R_s$, an axis ratio (flattening in the density) $q$, and the orientation of the flattened axis, represented by a vector $\boldsymbol{r}=(\hat{x},\hat{y},\hat{z})$. The stream progenitor is a Plummer sphere with mass $m_{\mathrm{prog}}$ and scale radius $r_s$. To backward integrate, we need its final (present-day) phase-space coordinates: $(x_0,y_0,z_0,v_{x_0},v_{y_0},v_{z_0})$. We also specify the total integration time $t_{\mathrm{int}}$. The origin is fixed at the halo centre and is not allowed to vary. This gives a total of fifteen parameters needed to model a stream. A schematic is shown in Figure~\ref{fig:schema}.

Although an axisymmetric potential can be specified with two orientation angles, sampling angles directly is prone to multimodality: periodic boundaries imply that a true solution near $0$ (or $2\pi$) induces duplicate modes at both ends of the parameter space. Instead, we parametrise the orientation of an axisymmetric halo by a nonzero vector

\begin{equation}
  \boldsymbol{r} = (\hat{x},\hat{y},\hat{z}) \in \mathbb{R}^{3}.
\end{equation}

\noindent In the prior, we draw

\begin{equation}
  \hat{x},\hat{y},\hat{z} \sim \mathcal{N}(0,1),
\end{equation}

\noindent independently, and define the auxiliary radius

\begin{equation}
  r = \|\boldsymbol{r}\| = \sqrt{\hat{x}^{2} + \hat{y}^{2} + \hat{z}^{2}} > 0.
\end{equation}

\noindent The direction of the halo symmetry axis is then given by the unit vector

\begin{equation}
  \hat{\boldsymbol{n}} = \frac{\boldsymbol{r}}{r}.
\end{equation}

\noindent To remove the trivial reflection degeneracy of an axisymmetric potential we impose

\begin{equation}
  \hat{n}_{z} \ge 0,
\end{equation}

\noindent which restricts the prior support to one hemisphere. Because the prior is spherically symmetric, this induces an isotropic prior over orientations while avoiding angular periodicity. Mathematically, the only remaining degeneracy in this parametrisation occurs on the boundary $n_z = 0$, where opposite directions of $\hat{\boldsymbol{n}}$ (e.g.\ $(1,0,0)$ and $(-1,0,0)$) correspond to the same physical model but appear as distinct points in parameter space. 

Since we only use the orientation of $\boldsymbol{r}$, its scale is redundant and introduces a new degeneracy where every combination of $(\hat{x},\hat{y},\hat{z})$ with the same orientation but different magnitude generate the same stream. To fix that, we use the magnitude $r$ (length of the vector $\boldsymbol{r}$) to define the halo flattening $q$. The goal is to transform $\boldsymbol{r}$ into a uniform distribution on [$q_{\text{min}}$, $q_{\text{max}}$].

Since $\boldsymbol{r}$ is multivariate normal with zero mean and unit covariance, the radius $r$ follows the Maxwell distribution with unit scale. Its probability density function (pdf), $f_{R}$, and cumulative distribution function (cdf), $F_{R}$, are

\begin{align}
  f_{R}(r) &= \sqrt{\frac{2}{\pi}}\,r^{2} \exp\!\left(-\frac{r^{2}}{2}\right),
  \qquad r \ge 0, \\
  F_{R}(r) &= \int_{0}^{r} f_{R}(s)\,\mathrm{d}s
          = \mathrm{erf}\!\left(\frac{r}{\sqrt{2}}\right)
          - \sqrt{\frac{2}{\pi}}\,r\,\exp\!\left(-\frac{r^{2}}{2}\right),
  \label{eq:maxwell_cdf}
\end{align}

\noindent where $\mathrm{erf}$ is the error function and $F_{R}$ is strictly increasing from $0$ to $1$ on $[0,\infty]$. We introduce an intermediate variable

\begin{equation}
  u = F_{R}(r),
\end{equation}

\noindent which by the probability integral transform satisfies

\begin{equation}
  u \sim \mathcal{U}(0,1),
\end{equation}

\noindent independently of the orientation $\hat{\boldsymbol{n}}$.

We then define the physical halo flattening $q$ by an affine mapping

\begin{equation}
  q = q_{\min} + (q_{\max} - q_{\min})\,u,
  \label{eq:q_affine}
\end{equation}

\noindent with fixed bounds $q_{\min} < q_{\max}$. Since $u$ is uniform on $[0,1]$, $q$ is exactly uniform on $[q_{\min},q_{\max}]$. In this work, we choose $q_{\min}=0.5$ and $q_{\max}=1.5$ to obtain a uniform prior $q\in[0.5,1.5]$, centred on spherical ($q=1$) with equal prior weight for oblate ($q<1$) and prolate ($q>1$).

In terms of the radius $r$, the flattening is therefore

\begin{equation}
  q(r) = q_{\min} + (q_{\max} - q_{\min})
         \left[
           \mathrm{erf}\!\left(\frac{r}{\sqrt{2}}\right)
           - \sqrt{\frac{2}{\pi}}\,r\,\exp\!\left(-\frac{r^{2}}{2}\right)
         \right].
  \label{eq:flattening_from_norm}
\end{equation}

\noindent This parametrisation yields:

\begin{enumerate}
  \item an isotropic prior over halo orientations $\hat{\boldsymbol{n}}$ (restricted to $\hat{n}_{z} \ge 0$), and
  \item an exactly uniform prior over the flattening $q$ on $[q_{\min},q_{\max}]$,
\end{enumerate}

\noindent while avoiding angular periodicity and any arbitrary normalisation of $\boldsymbol{r}$. We have verified this numerically: sampling from the prior alone and mapping to $q$ via Equation~\ref{eq:flattening_from_norm} recovers a uniform distribution, and the same result holds when running the nested sampler with a constant likelihood. From Equation~\ref{eq:flattening_from_norm}, the $q_{\min}$ boundary corresponds to the single point $r=0$, which may appear problematic, however, this has no practical consequence because the intermediate variable $u$ is explicitly rescaled to be uniform on $[0,1]$.

We, also, impose constraints on the progenitor’s final conditions. The observer is always fixed at +z-axis, so that the plane of the sky corresponds to the XY plane. First, we fix $y_0=0$, which forces the progenitor’s final position to lie in the XZ plane, regardless of any other parameters. This assumes we know the progenitor’s position in one dimension, an assumption that does not always hold. However, in our mock data, the stream contains no width or surface-density information, so the progenitor’s on-track location is far less constrained and more easily shift freely along the stream. To avoid this degeneracy, we fix one dimension of the progenitor’s final position rather than treat it as a parameter to be inferred. This choice simplifies the modelling problem but removes our ability to test constraints that depend on the progenitor’s present-day location, such as its intrinsic parameters, orbital precession, or aspects of the host’s radial structure \citep[e.g.][]{Belokurov2014, Erkal2016, Bonaca2018}. Although the progenitor’s position can sometimes be constrained from the track alone, we found that, without width or density information, doing so offers little additional benefit for our mock setup and therefore chose not to model it as a free parameter. In addition, we require $x_0>0$, $z_0>0$, and $v_{y_0}>0$. The constraint on $x_0$ simply fixes the in-plane orientation of the track; $z_0>0$ breaks the front/back degeneracy of projection; and $v_{y_0}>0$ assumes leading and trailing tails are interchangeable which is reasonable at low resolution where unresolved stars do not reveal the local “S”-shape near the progenitor \citep{Choi2007}. From our tests, these assumptions do not affect the recovered posteriors of the halo parameters but greatly speed up the inference by cutting away modalities. In particular, the front/back degeneracy ($z_0 \leftrightarrow -z_0$) is exact: the two solutions produce identical projected tracks and therefore identical posteriors on all halo parameters, including the flattening. We have verified this by running fits without the $z_0>0$ constraint, which yields two symmetric modes whose marginal posteriors on halo parameters are indistinguishable, at the cost of slower convergence. Because projected data cannot break this degeneracy, the same constraint can be applied when fitting real observations without loss of generality. In a companion paper currently in preparation, we extend the inference to include line-of-sight velocity measurements, which break this front/back degeneracy and remove the need for the $z_0>0$ constraint.

After applying the above symmetries, we infer a total of thirteen parameters summarised in Table~\ref{table:prior_orbit}. The priors are physically motivated but deliberately chosen to be wide and weakly informative. Halo and progenitor masses are given log-uniform priors; scale radii are uniform on broad ranges. We sample progenitor masses in the dwarf galaxy regime because we focus on external streams, where globular cluster streams are unlikely to be detectable. Although globular clusters and dwarf galaxies overlap in stellar mass, where globular clusters typically span $10^{5}$--$10^{7}\,M_{\odot}$ \citep[e.g.][]{Harris1996, Harris2010}, while classical dwarf galaxies range from $10^{6}$--$10^{9}\,M_{\odot}$ \citep{McConnachie2012}, their tidal debris differ significantly in surface brightness and width. For this reason, we restrict our progenitor mass sampling to the massive dwarf galaxy range. As mentioned above, the halo-axis components $\hat{x}, \hat{y}, \hat{z}$ have independent Gaussian priors (with $\hat{z}\ge 0$), while the flattening $q$ is deterministically mapped to a uniform prior $[0.5,1.5]$. The present-day positions and velocities of the the progenitor are also sampled from Gaussians, in some cases, the prior is truncated based on the restriction mentioned earlier. This ensures isotropy and limits degeneracies. The integration time is uniform. The lower bound of 1 Gyr, though arbitrary, allows sufficient time for streams to form, while the upper bound of 4 Gyr represents the timescale beyond which streams are expected to have phase-mixed enough to no longer be observable \citep[e.g.][]{Mancillas2019}. Throughout the modelling, we generate each stream using 10,000 particles. In this work, our main target is $q$; we also constrain in the orientation of the halo but all the remaining parameters are considered nuisance variables required to generate the track.

\begin{table}
\centering
\caption{Priors for the thirteen inferred parameters. $\mathcal{U}(a,b)$ denotes a uniform prior on $[a,b]$; $\mathcal{N}(\mu,\sigma)$ denotes a Gaussian prior with mean $\mu$ and standard deviation $\sigma$. Inequality constraints implement the symmetry-breaking choices described in the text.}
\label{table:prior_orbit}
\resizebox{\linewidth}{!}{%
\begin{tabular}{llll}
\hline
Parameter & Symbol & Prior & Units \\
\hline
Halo mass & $\log_{10}(M)$ & $\mathcal{U}(11,14)$ & $\log_{10}(M_\odot)$ \\
Halo scale radius & $R_s$ & $\mathcal{U}(10,25)$ & kpc \\
Halo orientation and shape & $\hat{x}, \hat{y}, \hat{z}$ & $\mathcal{N}(0,1)$ with $\hat{z}\ge 0$ & -- \\
Progenitor mass & $\log_{10}(m)$ & $\mathcal{U}(7,9)$ & $\log_{10}(M_\odot)$ \\
Progenitor scale radius & $r_s$ & $\mathcal{U}(1,5)$ & kpc \\
Final positions & $x_0, z_0$ & $\mathcal{N}(0,150)$ with $x_0>0$, $z_0>0$ & kpc \\
Final velocities & $v_{x_0}, v_{y_0}, v_{z_0}$ & $\mathcal{N}(0,250)$ with $v_{y_0}>0$ & km\,s$^{-1}$ \\
Integration time & $time$ & $\mathcal{U}(1,4)$ & Gyr \\
\hline
\end{tabular}
}
\end{table}

Lastly, to ensure that the mock data generated is both observationally realistic and sufficiently informative to warrant modelling, we generate mock observations by selecting projected tracks that meet three criteria: (i) angular coverage $>\,\pi/2$, (ii) arc length $>\,100$\,kpc and (iii) radial extent within $[10,500]$\,kpc. These conditions roughly line up to the STRRINGS sample \citep{Sola2025}. The first two ensure broad angle and spatial coverage; the third keeps the stream far enough from the bright host to be observed but close enough to be affected by the halo. For robustness, an angle bin is included in the mock data only if it contains $\ge 100$ particles, whereas during fitting we allow a lower threshold (half as many) to avoid discarding acceptable realisations due to stochastic particle losses. This minimum threshold on the amount of particles per bin also reflects the fact that only dense enough regions of the stream will be detected, and thus modeled.

\section{Individual fit}\label{Individual_fit}

Our goal is to simultaneously obtain, for each mock track, the posterior of the halo flattening $q$ and orientation for an axisymmetric NFW potential by leveraging Bayesian inference. Recall that $q$ is a deterministic function of the length of the sampled orientation vector $(\hat{x},\hat{y},\hat{z})$ (Equation~\ref{eq:flattening_from_norm}). All remaining parameters are treated as nuisance variables required to generate the stream. In Section~\ref{sec:hierarchical} we combine the marginal posteriors of $q$ from many streams to infer the DM haloes parameters at a population level.

\subsection{Likelihood}

\begin{figure*}
    \centering
    \includegraphics[width=1.0\linewidth]{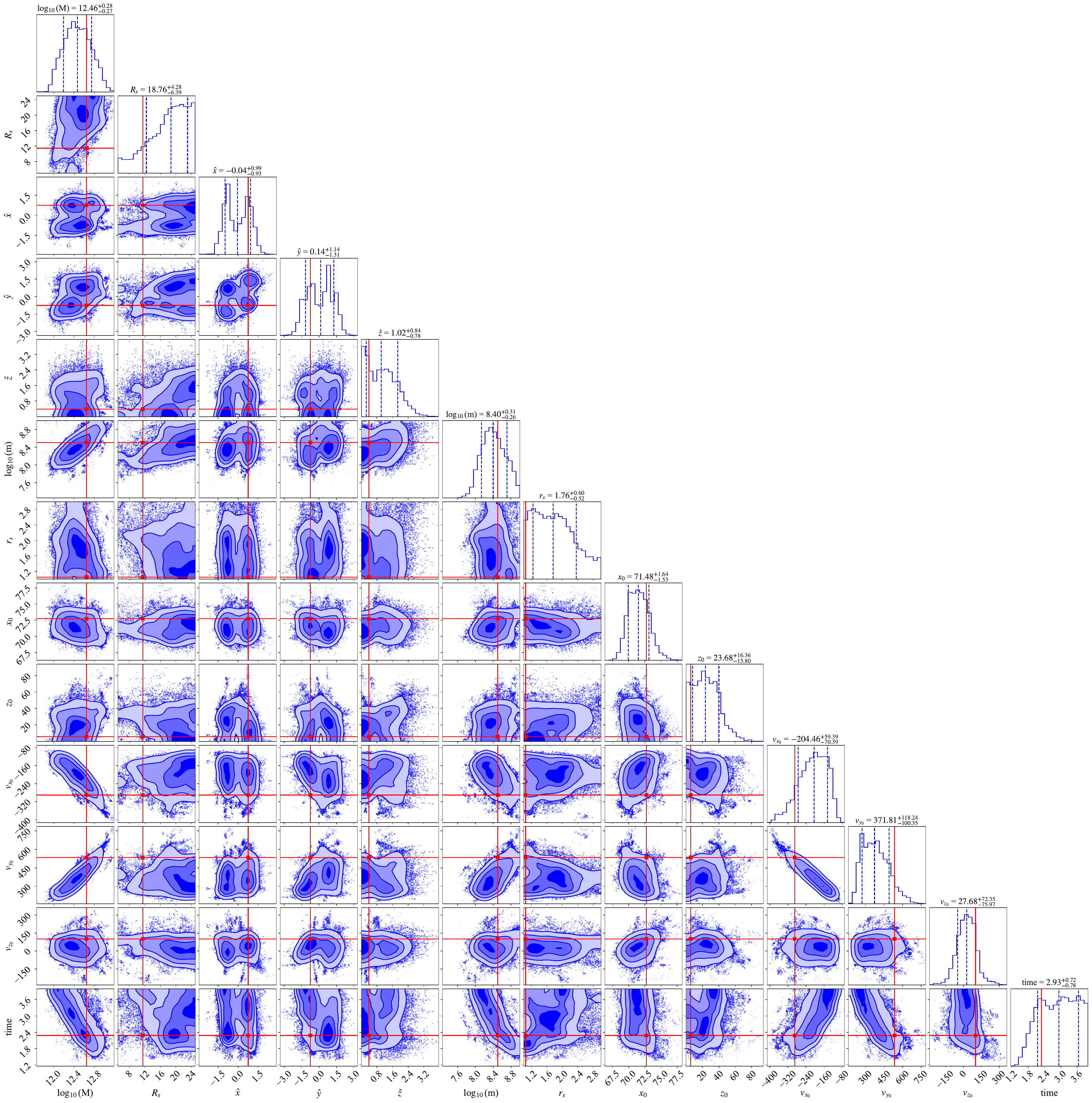}
    \caption{Corner plot for a single projected stream fit (13 parameters). Posteriors are shown in blue and the ground truth in red. The mock data include 2\% Gaussian radial noise. Overall, the posteriors recover the truth, though several degeneracies remain (see Section \ref{Degen}).}
    \label{fig:corner}
\end{figure*}

We begin by fitting each mock track individually. For this, we employ a Bayesian inference approach grounded in Bayes' theorem \citep{Bayes1763}. Let $D$ denote the data associated with a single stream (e.g.\ binned radii as a function of angle), and let $\theta$ denote the corresponding model parameters. We assume a specified forward model $M$ that maps $\theta$ to predicted observables. Bayes' theorem for continuous parameters reads

\begin{equation}
  p(\theta | D, M)
  = \frac{p(D | \theta, M)\,\pi(\theta | M)}{Z(M)},
  \label{eq:bayes_general}
\end{equation}

\noindent where:

\begin{itemize}
  \item $p(\theta | D, M)$ is the posterior density,
  \item $\pi(\theta | M)$ is the prior density on $\theta$ given model $M$,
  \item $p(D | \theta, M)$ is the likelihood, and
  \item the evidence is
    \begin{equation}
      Z(M) = p(D | M)
           = \int p(D | \theta, M)\,\pi(\theta | M)\,\mathrm{d}\theta.
           \label{eq:evidence}
    \end{equation}
\end{itemize}

In what follows we keep $M$ fixed and write $p(D|\theta)$ for brevity. Since we are not performing a model comparison, we drop the evidence term and work with the proportional form. For the inference, we adopt the same wide, weakly informative priors used in data generation (Table~\ref{table:prior_orbit}), but without the three selection cuts used to define mock-observation quality.

We assume that the observed projected track is binned in an angular coordinate $\phi$, yielding $N$ bins at angles $\{\phi_{i}\}_{i=1}^{N}$. In each bin we observe a projected radius $r^{\mathrm{data}}_{i}$ with an associated radial uncertainty $\sigma_{i} > 0$. The forward model $M$ predicts, for a given $\theta$, corresponding radii $r^{\mathrm{model}}_{i}(\theta)$ at the same angles.

We adopt a Gaussian noise model in radius:

\begin{equation}
  r^{\mathrm{data}}_{i}
  \sim \mathcal{N}\!\left(
    r^{\mathrm{model}}_{i}(\theta),\,
    \sigma_{i}^{2}
  \right),
  \qquad i = 1,\dots,N,
\end{equation}

\noindent with the bins conditionally independent given $\theta$. This yields the likelihood

\begin{equation}
  p(D | \theta)
  = \prod_{i=1}^{N}
    \frac{1}{\sqrt{2\pi}\,\sigma_{i}}
    \exp\!\left[
      -\frac{1}{2}
      \left(
        \frac{r^{\mathrm{data}}_{i} - r^{\mathrm{model}}_{i}(\theta)}{\sigma_{i}}
      \right)^{2}
    \right].
\end{equation}

\noindent For numerical work, we maximise or sample from the log-likelihood

\begin{equation}
  \log p(D | \theta)
  = -\frac{1}{2}
    \sum_{i=1}^{N}
    \left[
      \left(
        \frac{r^{\mathrm{model}}_{i}(\theta) - r^{\mathrm{data}}_{i}}{\sigma_{i}}
      \right)^{2}
      + \log\!\left(2\pi \sigma_{i}^{2}\right)
    \right].
  \label{eq:loglike}
\end{equation}

In the specific mock setup considered, each $\sigma_{i}$ is taken to be a fixed fraction of the observed radius,

\begin{equation}
  \sigma_{i} = \beta \,r^{\mathrm{data}}_{i},
\end{equation}

\noindent with $\beta$ a constant. Here, as mentioned in Section \ref{sec:track}, we added a 2\% Gaussian noise to the radius of the mock data based on STRRINGS \citep{Sola2025}. Therefore, for our Gaussian likelihood to be correct and completely unbiased, we set $\beta = 0.02$.

We note that this likelihood form involves several simplifying assumptions. In practice, the position of a stream track on the sky carries two dimensional uncertainties (in both the radial and angular directions), which would introduce correlated uncertainties in the one dimensional radial measurements across angular bins. We simplify this by binning in angle and treating each bin as independent with uncertainty only in $r$, which makes the likelihood tractable. Furthermore, we adopt a relative uncertainty rather than an absolute one. This choice, calibrated on the average noise level in STRRINGS, prevents bins at large radii from disproportionately dominating the fit: for example, an absolute uncertainty of 2\,kpc would represent a 20\% error at $r=10$\,kpc but only 2\% at $r=100$\,kpc, artificially up weighting the outer track. That said, a purely radial fractional uncertainty is itself a simplification, as real observational uncertainties depend not only on distance but also on local surface brightness, stream width, and background level, none of which are modelled here. A more complete treatment would require a two dimensional likelihood with a realistic, position dependent noise model, which we defer to future work with real data.

Models that do not cover all angular bins present in the data (for example, if the modelled track is too short in $\phi$) can either be discarded, $\log p(D | \theta) = -\infty$ for such $\theta$, or penalised by adding an explicit term to Eq.~\eqref{eq:loglike}. For instance, if $\Delta\phi_{\mathrm{data}}$ is the total angular extent of the data and $\Delta\phi_{\mathrm{overlap}}(\theta)$ is the angular overlap of the model on the data, one possible penalty is

\begin{equation}\label{eq:loglike_penalised}
  \log p_{\rm eff}(D \mid \theta)
  = \log p(D \mid \theta)
    - \lambda\,\left[1 - \frac{\Delta\phi_{\mathrm{overlap}}(\theta)}{\Delta\phi_{\rm data}}\right]_{+}^{2}
\end{equation}

\noindent where $[x]_{+} = \max(x,0)$ and $\lambda > 0$ is a tunable constant. In the main text of a results paper one should specify whether a hard rejection or an explicit penalty of the form \eqref{eq:loglike_penalised} is used. For this work, if the model's angular extend does not cover the data, we discard it, and if it extends beyond the range of the data, no penalty is applied. This reflects the assumption that limited observational coverage does not rule out intrinsically longer tracks. Indeed, we can only detect the portion of a stream that is bright enough to rise above the surface brightness limit of a given survey. Therefore, the fainter parts of a stream may remain undetected, even if the stream physically extends into those regions.

Given the dimensionality and complex posteriors, we use dynamic nested sampling with \texttt{Dynesty} \citep{Higson2019,Speagle2020,Koposov2022}, starting from 2000 live points and using the \textit{rslice} sampler. The initial number of live points was determined through trial runs, and we rely on \texttt{Dynesty}'s dynamic sampling to automatically allocate additional live points in regions of high posterior mass. Across our 105 individual mock-stream fits, the effective sample size ranges from $\sim$$54\,000$ to $\sim$$96\,000$ (median $\sim$$66\,000$), and the log-evidence uncertainty is $\Delta \log \mathcal{Z} \lesssim 0.14$ in all cases. Typical single-stream fits on a 44-core CPU take $\sim$10~hours. Although \texttt{StreaMAX} is fully differentiable, we currently use \texttt{JAX} primarily for computational speed rather than for gradient-based sampling. We have experimented with Hamiltonian Monte Carlo (HMC) but found that it struggles with the multi-modality present in our posteriors (see Section~\ref{Degen}). More modern approaches designed for multi-modal problems, such as parallel-tempered HMC or normalising-flow-based samplers \citep[e.g.\ \texttt{flowMC};][]{Wong2023}, are promising avenues for future work.

\begin{figure*}
    \centering
    \includegraphics[width=1.0\linewidth]{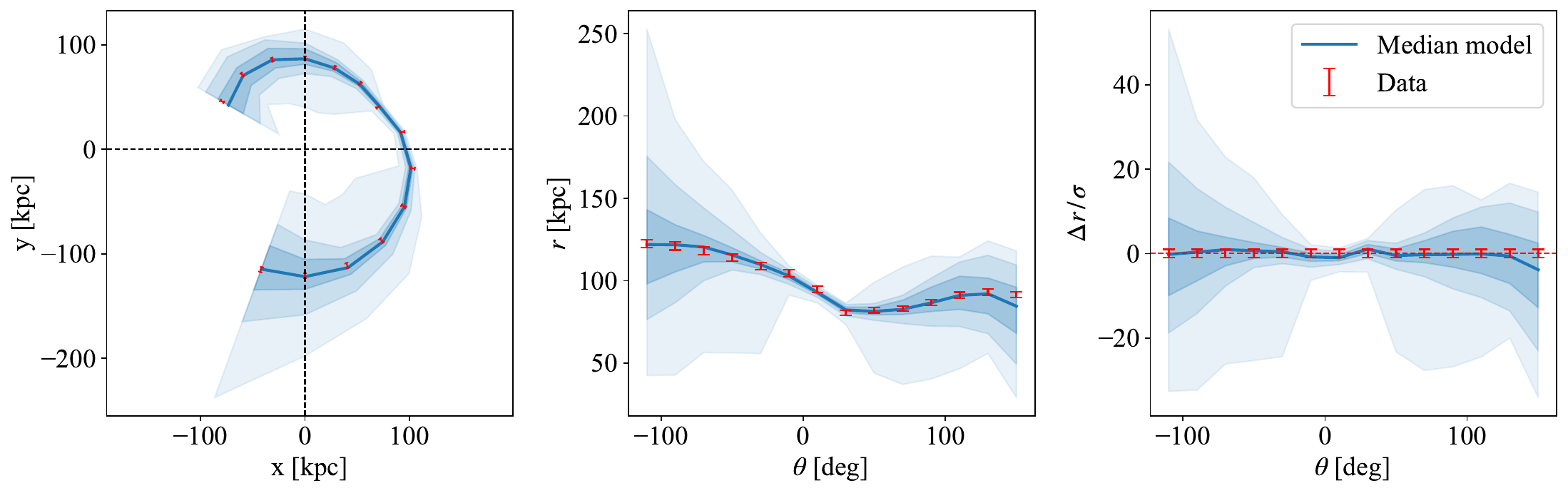}
    \caption{Summary of the 1000 samples drawn uniformly at random from the posterior (blue) compared to the ground truth (red) for a single stream. Shading indicates the 1, 2, and 3$\sigma$ drawn sample envelopes. \textbf{Left:} Comparison between the samples and the mock data on the $XY$ plane. \textbf{Middle:} Same comparison in the radial distance versus angle plane, where the likelihood is evaluated at fixed angle bins. \textbf{Right}: Signed normalised residuals, $(r_i^{\mathrm{model}}-r_i^{\mathrm{data}})/\sigma_i$, as a function of angle. Most bins are consistent with zero within $1\sigma$; the larger excursions near the ends of the stream reflect lower particle counts and increased sampling noise.}
    \label{fig:fits}
\end{figure*}

\subsection{Parameter Inference}\label{params_inf}

Figure~\ref{fig:corner} presents the full 13-parameter corner plot for one mock track. Despite our prior constraints and parameterisation choices, several degeneracies persist (see Section~\ref{Degen}). Nevertheless, most posteriors align well with the ground truth (red). Recall that all parameters except the halo flattening $q$ are treated as nuisance paramters. The orientation components $(\hat{x},\hat{y},\hat{z})$, from which $q$ is deterministically obtained, are comparatively well constrained, though they exhibit multi-modality among themselves. Importantly, they show no strong degeneracy with the remaining parameters (see Section~\ref{flatten} for some examples of flattening posteriors).

A notable feature is the marginalised posterior of the integration time. Low values are disfavoured up to $\sim 2$~Gyr, beyond which the posterior becomes broadly flat. This is a by-product of our conservative likelihood design: model tracks that are shorter than the observed angular coverage incur a strong penalty, whereas longer tracks are not penalised. As a result, the sampler is biased toward longer integration times, and once the integration time is sufficiently large for the model to span the observed range, the posterior plateaus. We discuss time-related degeneracies in Section~\ref{Degen}, but note here that this behaviour is driven by the likelihood construction rather than by information in the data.

To validate the fits, we draw 1000 posterior samples at random and overlay their modelled tracks on the mock observations in Figure~\ref{fig:fits}. The full posterior typically contains hundreds of millions of samples, but we downsample to avoid remodelling all of these tracks. Because the 1000 samples are selected uniformly at random, the displayed distribution is representative of the full posterior. The modelled tracks align closely with the data. Deviations are larger toward the ends of the tails, where fewer particles are present, leading to increased model deviation. The normalised residuals (right panel), defined as $(r_i^{\mathrm{model}}-r_i^{\mathrm{data}})/\sigma_i$ in the radius-angle space where the likelihood is evaluated, remain centred on zero and are consistent within $1\sigma$, with the largest excursions again confined to the stream ends. Together, these checks confirm that the inferred models are consistent with the mock data.

\subsection{Degeneracies}\label{Degen}

\begin{figure*}
    \centering
    
    \hspace{0.1\linewidth}
    \textbf{Progenitor Mass}
    \hspace{0.1\linewidth}
    \textbf{Progenitor Velocity}
    \hspace{0.1\linewidth}
    \textbf{Total Integration Time}
    \vspace{0.5em}
    
    \includegraphics[width=0.8\linewidth]{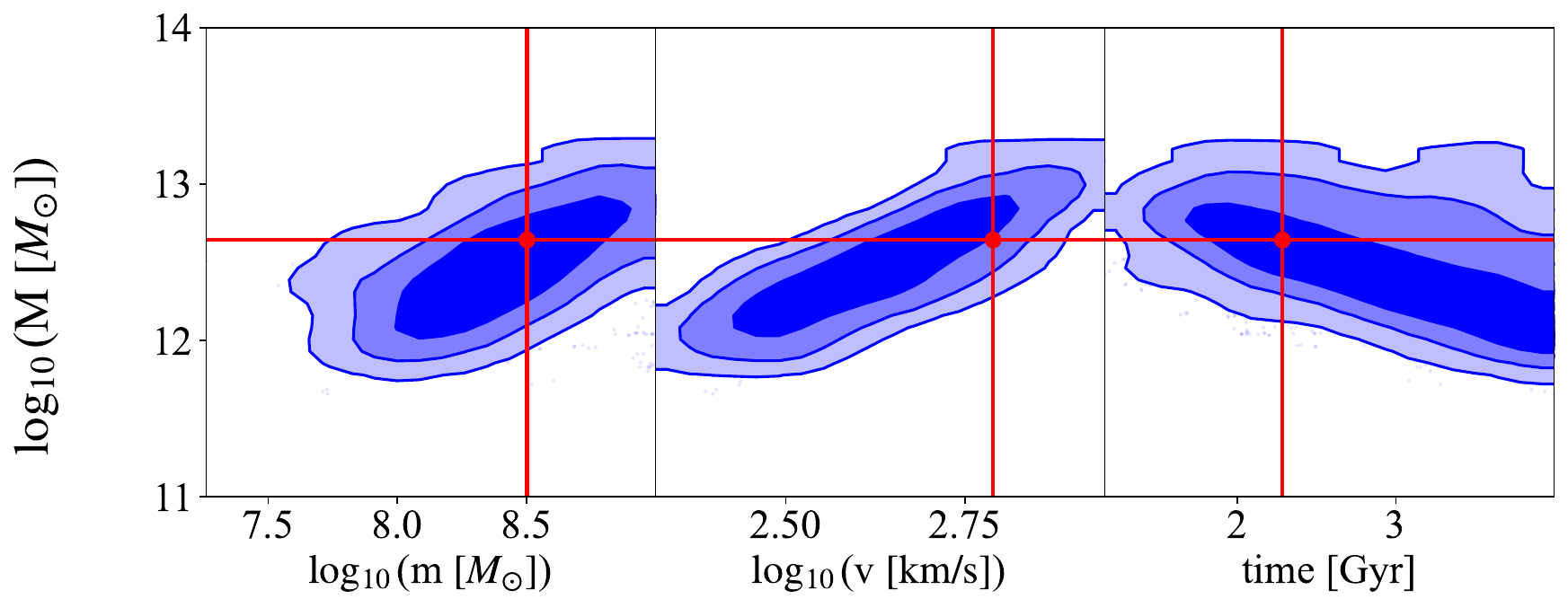}
    \caption{Joint posterior distributions showing the degeneracy of the mass of the halo ($\log_{10}(M \ [M_{\odot}])$) as a function of the mass of the progenitor ($\log_{10}(m \ [M_{\odot}])$), the velocity ($\log_{10}(\text{v} \ [\text{km/s}])$) and integration time. Ground truth is shown is red. Increasing the mass of the halo can be compensated by also increasing the mass and velocity of the progenitor while reducing the time. Without kinematic data these degeneracies cannot be broken.}
    \label{fig:degen}
\end{figure*}

In this work we fit only the projected track of modelled streams, with no velocity information. As shown by \citet{Pearson2022}, this implies a fundamental scale degeneracy between orbital velocity and halo mass: without kinematics it is not possible to fix the physical mass scale of the system. To visualise these relations we compute the present-day velocity $\text{v}=\sqrt{v_{x_0}^2+v_{y_0}^2+v_{z_0}^2}$ and examine the correlations between $\{\log_{10}(M),\log_{10}(m),\log_{10}(\text{v}), \text{time}\}$ (Figure~\ref{fig:degen}). Because $\text{v}^2\propto M$, $\log_{10}(M)$ and $\log_{10}(\text{v})$ are linearly correlated in log–log space. A second linear relation between $\log_{10}(M)$ and $\log_{10}(m)$ arises from the Lagrange-point scaling. In the rotating frame of the progenitor, the tidal radius $r_t$ is set by the balance between the progenitor’s self-gravity and the tidal force exerted by the host: $r_t \propto \left(\frac{m}{M}\right)^{\frac{1}{3}}$. If $m$ and $M$ are scaled proportionally, the ratio $m/M$ remains constant, and therefore the tidal radius (i.e. the location of the Lagrange points where particles are released) remains unchanged. Since our stream generation method releases debris precisely from these Lagrange points, a fixed $r_t$ produces an identical phase-space structure. The resulting streams are thus completely degenerate as long as $M/m$ is held constant. In logarithmic units, keeping this ratio fixed corresponds to a linear relation between $\log_{10}(M)$ and $\log_{10}(m)$. Finally, the integration time trades off against the mass–velocity scale: if $(M,\text{v})$ are scaled down, a longer integration time is required to match the observed length and vice-versa. From the inferred joint posteriors in Figure~\ref{fig:degen}, we see that although the parameters are strongly degenerate, the ground truth lies directly along the degeneracy. When looking only at the marginalized posteriors (Figure~\ref{fig:corner}), these parameters may appear well constrained. However, we stress that such constraints are prior-driven: likelihood values are nearly constant along the degeneracy, but the prior does not weight all regions equally, producing apparent preferences in the marginal distributions. Uniform priors on the parameters do not guarantee uniform priors on the degeneracies. In principle, one could reparameterise the problem to uniformly sample along the degeneracies; in practice, because enclosed mass varies with radius, the exact mass–velocity relation is position dependent. We therefore retain the raw parameters and explicitly note that these apparent constraints are prior dominated.

\begin{figure*}
    \centering
    \includegraphics[width=1.0\linewidth]{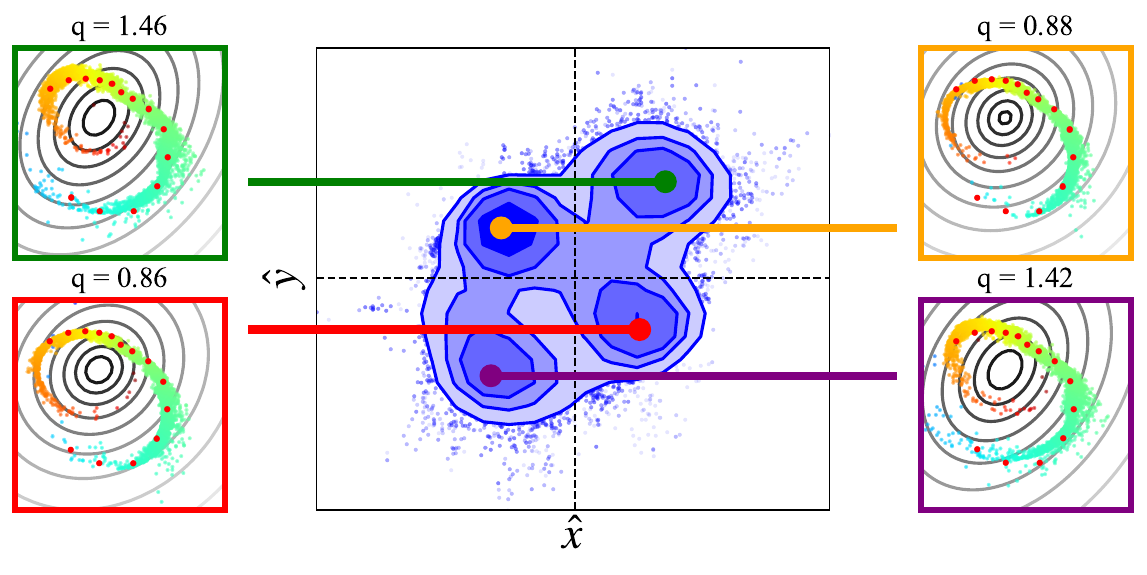}
    \caption{Zoom on the correlated posterior of the orientation from one mock stream track projected onto the $XY$-plane (the setting is the same than Figure \ref{fig:corner}). For each of the four visible modes we show the best-fit stream (rainbow ordering of the stream particles where bluer means further away from the progenitor clockwise and redder, anti-clockwise) atop the mock data (red) and the projected equipotentials as grey contours; the inferred flattening $q$ is listed above each panel. Modes correspond to projected rotations and oblate versus prolate swaps that give similar projected potential.}
    \label{fig:q_degen}
\end{figure*}

Figure~\ref{fig:corner} also reveals clear modalities between the different orientations that propagate to the flattening. Figure~\ref{fig:q_degen} zooms in on the $(\hat{x},\hat{y})$ posterior, showing four modes. For each mode we plot the best-fit stream with the rainbow colormap colour-coding the ordering of the stream particles over the mock data (red) and the projected equipotentials; the corresponding flattening $q$ is listed atop each panel. All four solutions fit the data and share nearly the same projected potential even though $(\hat{x},\hat{y})$ and $q$ differ. Looking at the two pairs, green-purple and red-orange, we see that in both cases the flattening is the same, either oblate or prolate. In the posterior, the modalities are equivalent to flipping the sign of both $\hat{x}$ and $\hat{y}$ which, in the projected plane, represents a simple 180$^\circ$ rotation. These modalities are due to the loss of information from working with projections. Indeed, the potentials' orientations are different in 3-dimensions but look similar when projected. Other posterior pairs, for example the green and orange solutions, differ in both orientation and flattening, yet still produce nearly identical projected tracks. In these cases, the inferred orientation flips the sign of $\hat{x}$ while $\hat{y}$ remains similar, which corresponds to a $\sim 90^\circ$ rotation of the halo in the projected plane. Combined with the change from an oblate ($q=0.86$) to a prolate ($q=1.46$) configuration, this transformation yields similar projected potentials. As a result, both 3D models fit the 2D data almost equally well. This demonstrates that when the projected potential matches the ground truth, distinct posterior modalities can appear. That being said, these modes are not perfect. Our results show that the correct solution is nonetheless favoured. This reinforces the idea that even in projection, stream tracks carry information about the underlying three dimensional potential. This phenomenon is in agreement with the results presented in \cite{Nibauer2023} showing that the projected curvature allows to constrain the morphology of the halo's potential. Notably, the best fit streams in Figure~\ref{fig:q_degen} differ in the regions not covered by the mock data, suggesting that longer streams (i.e. wider angular coverage) could help break this degeneracy. We return to these points in more detail in Section~\ref{flatten}.

Lastly, looking back at Figure \ref{fig:corner}, we see in the marginalized posterior of $\hat{z}$ two modes. Here, these modalities mostly dictated the length of the vector (see Equation \ref{eq:flattening_from_norm}) which maps to the flattening, allowing to vary the potential from oblate to prolate without affecting the projected orientation in the XY plane. This is how pairs of different flattening in Figure \ref{fig:q_degen} obtain different values for $q$ without varying the absolute size of $\hat{x}$ and $\hat{y}$.

Although the three-dimensional components defining the DM halo orientation ($\hat{x}$, $\hat{y}$, $\hat{z}$) are highly degenerate with one another, Figure~\ref{fig:corner} shows that the projected orientation is nonetheless well constrained. While this aspect is not explored further in the present work, such constraints could be compared to the observed orientation of the baryonic components in real galaxies, allowing the misalignment between the stellar (or stellar + gaseous) disk and the DM halo to be quantified statistically \citep[e.g.][]{Bailin2005, Bett2010, Debattista2013, Shao2016, Tenneti2016}. Measuring this misalignment across a large sample, which we are currently doing with \texttt{STRRINGS}, enables to study the population-level relationship between halo and disk orientations. Moreover, the inferred sky-projected orientations could be analysed to test whether haloes are randomly distributed or exhibit coherent alignments tracing the surrounding large-scale structure, as suggested by studies of the cosmic web \citep[e.g.][]{Tempel2013, Codis2015, Welker2020}.

\begin{figure}
    \centering
    \includegraphics[width=\columnwidth]{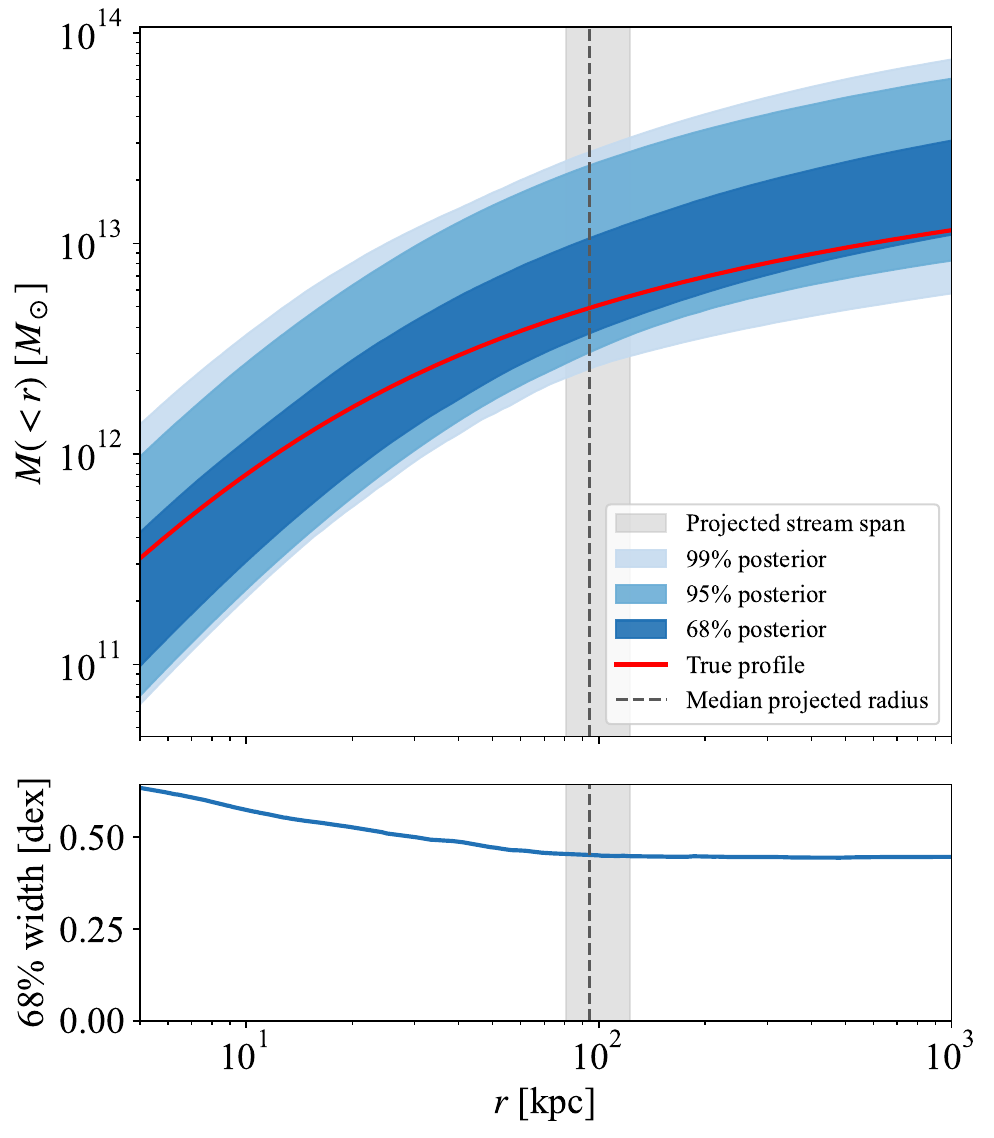}
    \caption{Enclosed-mass profile from a single projected-stream fit. \emph{Top:} the 68\% (dark blue), 95\% (medium blue), and 99\% (light blue) posterior ranges on the spherical-equivalent NFW enclosed mass $M(<r)$ compared to the true profile (red). The grey band marks the projected radial extent of the stream bins used in the likelihood, and the dashed black line shows the median projected radius. \emph{Bottom:} 68\% posterior width in $\log_{10} M(<r)$ as a function of radius. The broad, nearly flat uncertainty confirms that the enclosed mass is unconstrained by the projected track alone.}
    \label{fig:mass_enclosed}
\end{figure}

Figure~\ref{fig:mass_enclosed} shows the enclosed-mass profile $M(<r)$ implied by the posterior samples of $(\log M, R_s)$ for the same individual fit as Figure \ref{fig:corner} and \ref{fig:fits}. As expected from the mass--velocity degeneracy, the projected stream track alone does not constrain the absolute mass scale: the inferred enclosed mass spans several orders of magnitude at all radii and is not significantly tighter near the projected extent of the stream. This indicates that the apparent constraint on $M(<r)$ is prior dominated rather than localised around the stream radius. The true profile remains consistent with the posterior range, so the inference is not strongly biased, but it is largely uninformative for enclosed mass. By contrast, the halo flattening $q$ is still well recovered from the same data (Section~\ref{flatten}) because it is encoded in the shape of the projected track rather than its scale. In future work we will incorporate line-of-sight velocity information, which should break the mass--velocity degeneracy and permit genuinely constraining enclosed-mass measurements.

We note that the degeneracies discussed above are predominantly true, mathematical degeneracies arising from the loss of information in projection, rather than artefacts of finite measurement uncertainty. We have verified this by running fits with progressively smaller added Gaussian noise: the multi-modal structure in the posterior, in particular the oblate/prolate ambiguity, persists as the noise is reduced. While the individual modes do narrow, reducing the noise also makes the likelihood surface considerably spikier, requiring substantially more live points and likelihood evaluations to adequately explore the parameter space. In practice, the increased computational cost outweighs the marginal gain in precision, reinforcing that the dominant source of uncertainty in our projected-track fits is the intrinsic degeneracy of the problem rather than observational noise.

\begin{figure*}
    \centering
    \textbf{Oblate} \hspace{0.26\linewidth}
    \textbf{Spherical} \hspace{0.26\linewidth}
    \textbf{Prolate} \\
    \includegraphics[width=1.0\linewidth]{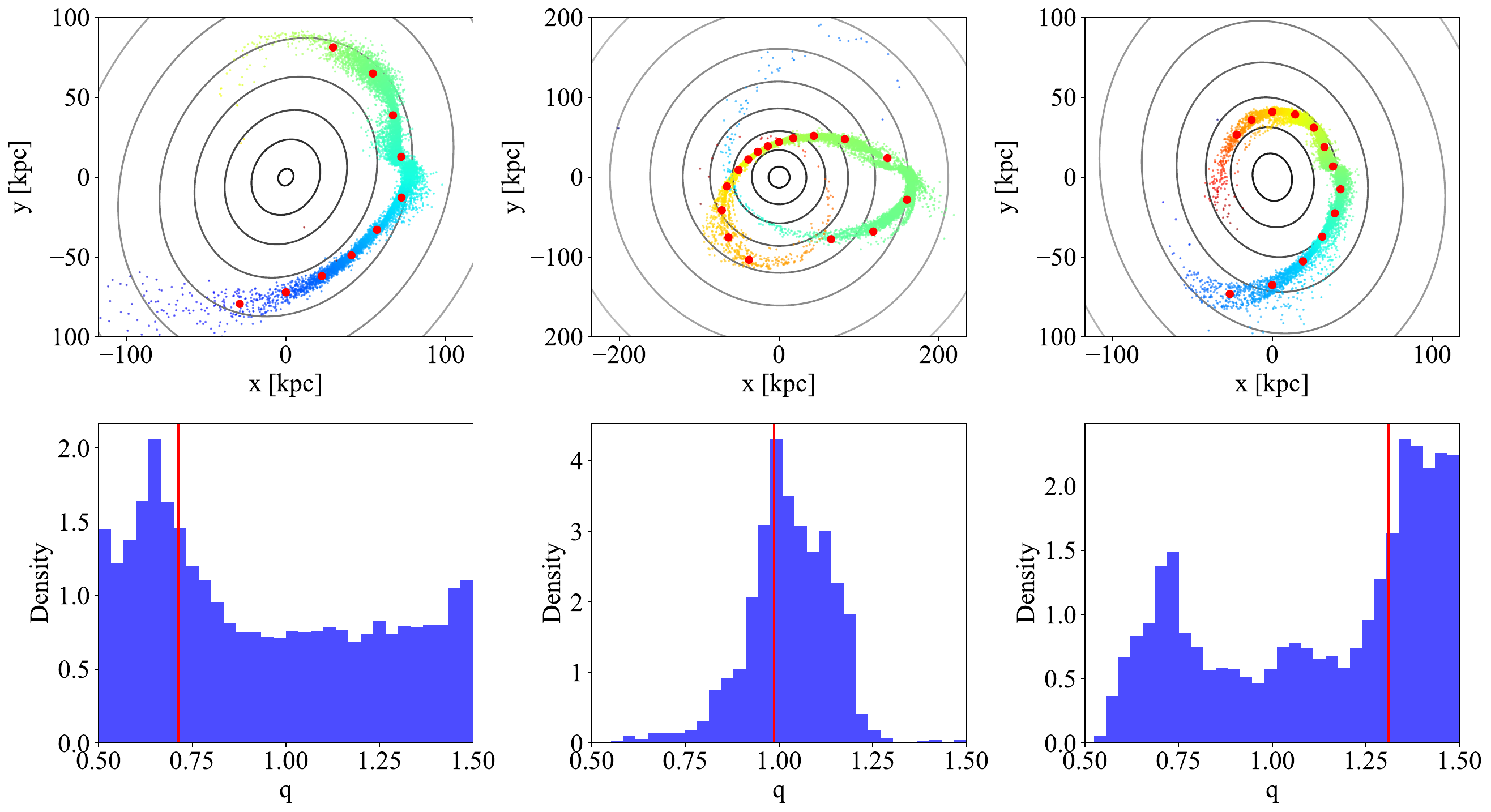}
    \caption{Examples of best-fit streams projected in the XY plane (top) and their posteriors for the flattening $q$ (bottom). From left to right: oblate, spherical, and prolate cases, with the ground truth in red. In all three cases, the fits recover the mock tracks and the posterior on the flattening shows a preference for values close to the ground truth. The secondary mode in the prolate case (bottom right) reflects projection symmetries discussed in Section \ref{Degen}.}
    \label{fig:q_fits}
\end{figure*}

\subsection{Flattening}\label{flatten}

We now run our inference pipeline for three different cases: ground truth oblate, spherical and prolate. With posteriors on $(\hat{x},\hat{y},\hat{z})$ from three different individual fits derived in Section \ref{params_inf}, we obtain the halo flattening using Equation~\ref{eq:flattening_from_norm}. Figure~\ref{fig:q_fits} shows the three examples respectively from left to right, each with the best-fit stream (top) and the corresponding posterior for $q$ (bottom). The ground truths are marked in red. In all cases the fits match the mock data well and the $q$ posteriors favour the correct values. For the spherical case, extreme oblate or prolate solutions are ruled out and the posterior peaks tightly at $q\simeq1$. For the oblate case, no value of $q$ is completely ruled out but the correct value is prefered. This is not always true for all oblate cases. Lastly, the prolate case also does not completely rule out any flattening possibility but does exhibit a secondary mode in the oblate region of the flattening. This is consistent with the projection induced modalities discussed earlier in Section \ref{Degen}. Nevertheless, the correct flattening value is clearly preferred. Because the projected track still encodes information about the underlying three-dimensional potential (see Appendix of \citealt{Nibauer2023}), the data retains discriminating power beyond simple sky projections. Our results show that even when working only with projected tracks, it is still possible to recover meaningful information about the full 3D potential. While being informative, the resulting posteriors are still relatively broad, which limits the strength of any single stream constraint on $q$. For this reason, in the next section we combine individual stream fits to obtain tighter constraints on the parameters of the underlying population distribution.

Before proceeding, it is important to acknowledge that our model isolates a single morphological parameter ($q$) within an idealised, axisymmetric potential. This simplification allows the population inference to focus exclusively on the halo flattening, neglecting all other inferred parameters. In reality, triaxiality, radially varying density profiles, and baryonic components (e.g. a disk) would introduce additional shape-related parameters and degeneracies, resulting in broader posteriors and a more complex population inference. These effects lie beyond the scope of the present work, which is intended to quantify the information content of the projected track in a controlled setting. Their impact will be explored in future studies.

\section{Population fit}\label{sec:hierarchical}

\subsection{Inference}

Beyond individual fits, our primary objective is to constrain the population distribution of halo flattening. While single stream fits may yield wide posteriors, we can combine them using hierarchical Bayesian inference to obtain robust constraints on population parameters and shed light on what is the morphology of DM haloes considering the ensemble \citep{Chua2019, Prada2019}.

We now consider a population of $N$ streams. For stream $n$ we denote its data by $d_{n}$ and its parameters by $\theta_{n}$. Let $\alpha$ be the hyperparameters that describe the population distribution of interest (here, the distribution of halo flattening $q$).

We assume that, conditioned on $\theta_{n}$, the data sets $\{d_{n}\}$ are independent, and that the $\theta_{n}$ are drawn independently from a population prior $\pi(\theta_{n} | \alpha)$. The hyperprior on $\alpha$ is denoted by $\pi(\alpha)$.

The hierarchical posterior for $\alpha$ given all data $\{d_{n}\}_{n=1}^{N}$ is then

\begin{equation}
  p(\alpha | \{d_{n}\})
  \propto \pi(\alpha)
  \prod_{n=1}^{N}
  \int
    p(d_{n} | \theta_{n})\,
    \pi(\theta_{n} | \alpha)\,
  \mathrm{d}\theta_{n},
  \label{eq:hier_general}
\end{equation}

\noindent where proportionality reflects omission of the global normalisation constant $p(\{d_{n}\})$. Evaluating this posterior directly would require refitting each individual model for every sample of \(\alpha\), yielding a computational cost of \(\mathcal{O}(N^2)\), which is infeasible given the cost of stream modeling and the large \(N\) required for tight constraints. Instead, we follow the reweighting approach introduced in \citet{Hogg2010}, which allows for a computationally efficient \(\mathcal{O}(N)\) solution. The key idea is to use the posterior samples from individual fits to recover the underlying likelihood.

\begin{figure*}
    \centering
    \textbf{Oblate} \hspace{0.23\linewidth}
    \textbf{Spherical} \hspace{0.23\linewidth}
    \textbf{Prolate} \\
    \begin{minipage}[c]{1.0\linewidth}
        \includegraphics[width=0.33\textwidth]{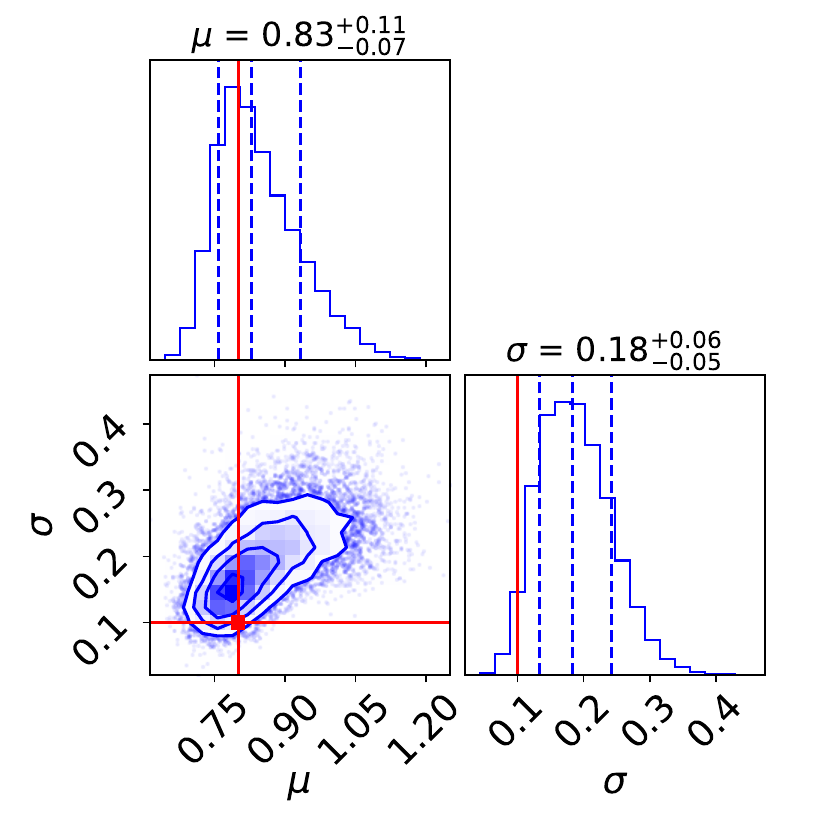}
        \includegraphics[width=0.33\textwidth]{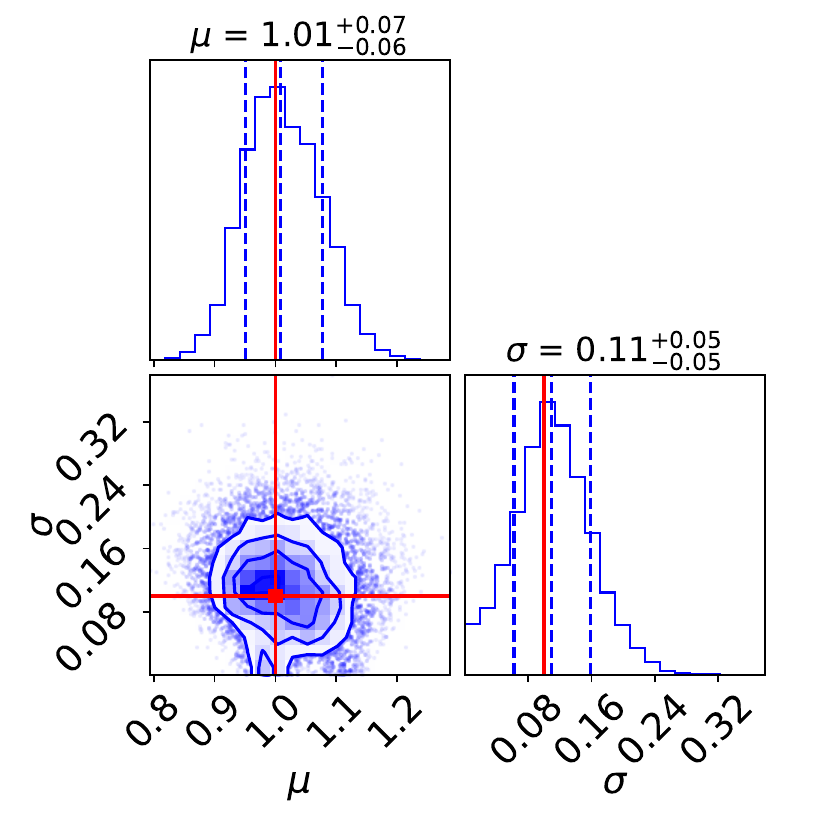}
        \includegraphics[width=0.33\textwidth]{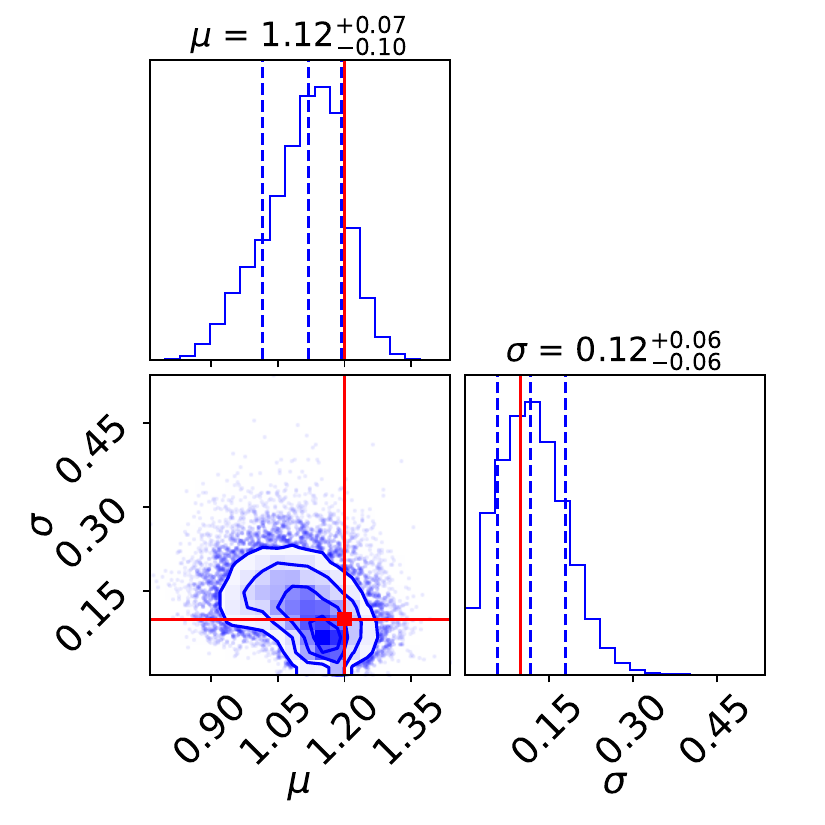}
    \end{minipage} \\
    \caption[]{Posteriors on the parameters of the underlying population distribution of flattening for three distinct dark matter halo morphologies: oblate (left), spherical (middle) and prolate (right). Blue dashed vertical lines indicate the 16th, 50th, and 84th percentiles, corresponding to the median and $\pm1\sigma$ intervals (values shown above each panel). These posteriors were obtained with three distinct populations of 35 streams each with the ground truth shown in red. We see a small bias towards more spherical haloes for both the oblate and prolate distributions, this bias is not present in the spherical case. The standard deviation of the distribution also shows this bias. For all three cases, the ground truth is within 2$\sigma$ of the infered posterior distributions.}
    \label{fig:population}
\end{figure*}

Suppose we have already performed individual Bayesian inferences for each stream, using some prior $\pi(\theta_{n})$ (not depending on $\alpha$), and obtained the posterior

\begin{equation}
  p(\theta_{n} | d_{n})
  = \frac{p(d_{n} | \theta_{n})\,\pi(\theta_{n})}{Z_{n}},
  \label{eq:single_posterior}
\end{equation}

\noindent with

\begin{equation}
  Z_{n}
  = p(d_{n})
  = \int p(d_{n} | \theta_{n})\,\pi(\theta_{n})\,\mathrm{d}\theta_{n}
\end{equation}

\noindent the evidence for stream $n$ under the individual prior.

\noindent Rearranging Eq.~\eqref{eq:single_posterior} gives

\begin{equation}
  p(d_{n} | \theta_{n})
  = \frac{p(\theta_{n} | d_{n})}{\pi(\theta_{n})}\,Z_{n}.
\end{equation}

\noindent Substituting into Eq.~\eqref{eq:hier_general},

\begin{align}
  p(\alpha | \{d_{n}\})
  &\propto \pi(\alpha)
    \prod_{n=1}^{N}
    \int
      \frac{p(\theta_{n} | d_{n})}{\pi(\theta_{n})}\,
      Z_{n}\,
      \pi(\theta_{n} | \alpha)\,
    \mathrm{d}\theta_{n} \\
  &\propto \pi(\alpha)
    \prod_{n=1}^{N}
    \int
      p(\theta_{n} | d_{n})\,
      \frac{\pi(\theta_{n} | \alpha)}{\pi(\theta_{n})}\,
    \mathrm{d}\theta_{n},
    \label{eq:hier_reweight_theta}
\end{align}

\noindent where we have dropped the product of evidences $\prod_{n} Z_{n}$, which does not depend on $\alpha$.

The integral in Eq.~\eqref{eq:hier_reweight_theta} is an expectation over the known posterior $p(\theta_{n} | d_{n})$ and can be approximated with posterior samples. Let $\{\theta_{ni}\}_{i=1}^{K_{n}}$ be $K_{n}$ samples from $p(\theta_{n}|d_{n})$. Then

\begin{equation}
  \int
    p(\theta_{n} | d_{n})\,
    \frac{\pi(\theta_{n} | \alpha)}{\pi(\theta_{n})}\,
  \mathrm{d}\theta_{n}
  \approx
  \frac{1}{K_{n}}
  \sum_{i=1}^{K_{n}}
    \frac{\pi(\theta_{ni} | \alpha)}{\pi(\theta_{ni})}.
\end{equation}

\noindent Thus the hierarchical posterior can be approximated as

\begin{equation}
  p(\alpha | \{d_{n}\})
  \propto \pi(\alpha)
  \prod_{n=1}^{N}
  \left[
    \frac{1}{K_{n}}
    \sum_{i=1}^{K_{n}}
      \frac{\pi(\theta_{ni} | \alpha)}{\pi(\theta_{ni})}
  \right].
  \label{eq:hier_reweight_MC_theta}
\end{equation}

In the application of interest, the population model acts only on the halo flattening $q$. We decompose the parameters as

\begin{equation}
  \theta_{n} = (q_{n}, \psi_{n}),
\end{equation}

\noindent where $\psi_{n}$ collects all nuisance parameters (e.g.\ halo mass, scale radius, progenitor phase--space coordinates). We assume that the individual prior factorises as

\begin{equation}
  \pi(\theta_{n}) = \pi(q_{n})\,\pi(\psi_{n}),
\end{equation}

\noindent and that the population prior has the form

\begin{equation}
  \pi(\theta_{n} | \alpha)
  = \pi(q_{n} | \alpha)\,\pi(\psi_{n}),
\end{equation}

\noindent with the same prior $\pi(\psi_{n})$ on the nuisance parameters. The ratio in Eq.~\eqref{eq:hier_reweight_MC_theta} then simplifies to

\begin{equation}
  \frac{\pi(\theta_{ni} | \alpha)}{\pi(\theta_{ni})}
  = \frac{\pi(q_{ni} | \alpha)}{\pi(q_{ni})}.
\end{equation}

\noindent If the individual prior on $q$ is uniform on $[q_{\min},q_{\max}]$,

\begin{equation}
  \pi(q_{n}) =
  \begin{cases}
    \dfrac{1}{q_{\max} - q_{\min}},
      & q_{\min} \le q_{n} \le q_{\max},\\[4pt]
    0, & \text{otherwise},
  \end{cases}
\end{equation}

\noindent then $\pi(q_{ni})$ is a constant for all posterior samples used in the reweighting. That constant can be absorbed into the overall normalisation of $p(\alpha | \{d_{n}\})$, and we obtain

\begin{equation}
  p(\alpha | \{d_{n}\})
  \propto \pi(\alpha)
  \prod_{n=1}^{N}
  \left[
    \frac{1}{K_{n}}
    \sum_{i=1}^{K_{n}}
      p(q_{ni} | d_{n}) \pi(q_{ni} | \alpha)
  \right].
  \label{eq:hier_reweight_final}
\end{equation}

Equation~\eqref{eq:hier_reweight_final} is the working expression used to evaluate the hierarchical likelihood for any proposed hyperparameters $\alpha$. This method is computationally efficient and scalable but relies on a critical condition: the conditional prior on the flatening, \(\pi(q|\alpha)\), must lie within the actual prior, \(\pi(q)\), use for the individual inference. Since no new samples are generated in this second-stage inference, the individual posteriors must already sufficiently cover the relevant parameter space. This is the case here since our individual posteriors on flattening are usually quite broad and very rarely completely rule out a region of parameter space.

As a concrete example, we consider a population model in which the physical flattening $q$ is distributed as a Gaussian with mean $\mu_{\mathrm{pop}}$ and standard deviation $\sigma_{\mathrm{pop}}$, truncated to the same support as the individual prior, $[q_{\min},q_{\max}]$. The hyperparameters are

\begin{equation}
  \alpha = (\mu_{\mathrm{pop}}, \sigma_{\mathrm{pop}}).
\end{equation}

\noindent The (unnormalised) Gaussian density is

\begin{equation}
  \phi(q | \mu_{\mathrm{pop}}, \sigma_{\mathrm{pop}})
  = \frac{1}{\sqrt{2\pi}\,\sigma_{\mathrm{pop}}}
    \exp\!\left[
      -\frac{1}{2}
      \left(
        \frac{q - \mu_{\mathrm{pop}}}{\sigma_{\mathrm{pop}}}
      \right)^{2}
    \right].
\end{equation}

\noindent The truncated density is

\begin{equation}
  \pi(q | \alpha)
  =
  \begin{cases}
    \dfrac{\phi(q | \mu_{\mathrm{pop}}, \sigma_{\mathrm{pop}})}
          {\Phi\!\left(\dfrac{q_{\max}-\mu_{\mathrm{pop}}}{\sigma_{\mathrm{pop}}}\right)
           - \Phi\!\left(\dfrac{q_{\min}-\mu_{\mathrm{pop}}}{\sigma_{\mathrm{pop}}}\right)},
       & q_{\min} \le q \le q_{\max},\\[10pt]
    0, & \text{otherwise},
  \end{cases}
  \label{eq:truncated_gaussian}
\end{equation}

\noindent where $\Phi$ is the standard normal cdf. In practice, when evaluating the reweighting factor $\pi(q_{ni}|\alpha)$ inside Eq.~\eqref{eq:hier_reweight_final}, it is sufficient to use the properly normalised truncated density \eqref{eq:truncated_gaussian}. The hyperprior $\pi(\alpha)$ can be chosen, for example, as a product of broad uniform priors on $\mu_{\mathrm{pop}}$ and $\sigma_{\mathrm{pop}}$.

To initialize the different and seperate population distributions, we generate \(N\) streams for each, drawing true values of \(q\) from one of three underlying population distributions assuming that all haloes are drawn from a single Gaussian population:
\begin{enumerate}
    \item Oblate: \(\mu_{\mathrm{pop}} = 0.8\), \(\sigma_{\mathrm{pop}} = 0.1\),
    \item Spherical: \(\mu_{\mathrm{pop}} = 1.0\), \(\sigma_{\mathrm{pop}} = 0.1\),
    \item Prolate: \(\mu_{\mathrm{pop}} = 1.2\), \(\sigma_{\mathrm{pop}} = 0.1\).
\end{enumerate}

For every data point in the dataset which represent a modelled stream, we first obtain the posterior on the flattening by performing N individual fits. In all the following examples, we set $N=35$ motivated by STRRINGS \citep{Sola2025}. From there, we infer the population parameters \((\mu_{\mathrm{pop}}, \sigma_{\mathrm{pop}})\) through our Bayesian hierarchical inference pipeline using the same dynamic nested sampling implementation from \texttt{dynesty}, this time with 500 live points, which is sufficient given the low dimensionality (two parameters) and the absence of multi-modality at the population level. Given the low dimensionality, we use the uniform sampler, which helps accelerate the inference. The priors are uniform for both and range from [0.5, 1.5] for $\mu_{\mathrm{pop}}$ and [0, 1] for $\sigma_{\mathrm{pop}}$.

Figure \ref{fig:population} shows the posteriors of the average and standard deviation of three different populations of modelled projected tracks. From left to right, the population distributions represent the situation (i), (ii) and (iii) as shown from the ground truth in red. Each population comprises 35 streams, chosen to mimic the \texttt{STRRINGS} catalogue. Even with our highly optimised population inference pipeline, obtaining posterior distributions for a single population requires roughly one week of computation. We therefore adopt a sample size large enough to constrain the parameters while keeping the computational cost manageable. Increasing the number of streams generally improves the precision and reduces bias in the inferred parameters, though this effect is not strictly linear. In practice, the total information content varies significantly between individual streams, so population level constraints depend on both the number and quality of the streams.

\begin{figure}
    \centering
    \includegraphics[width=\columnwidth]{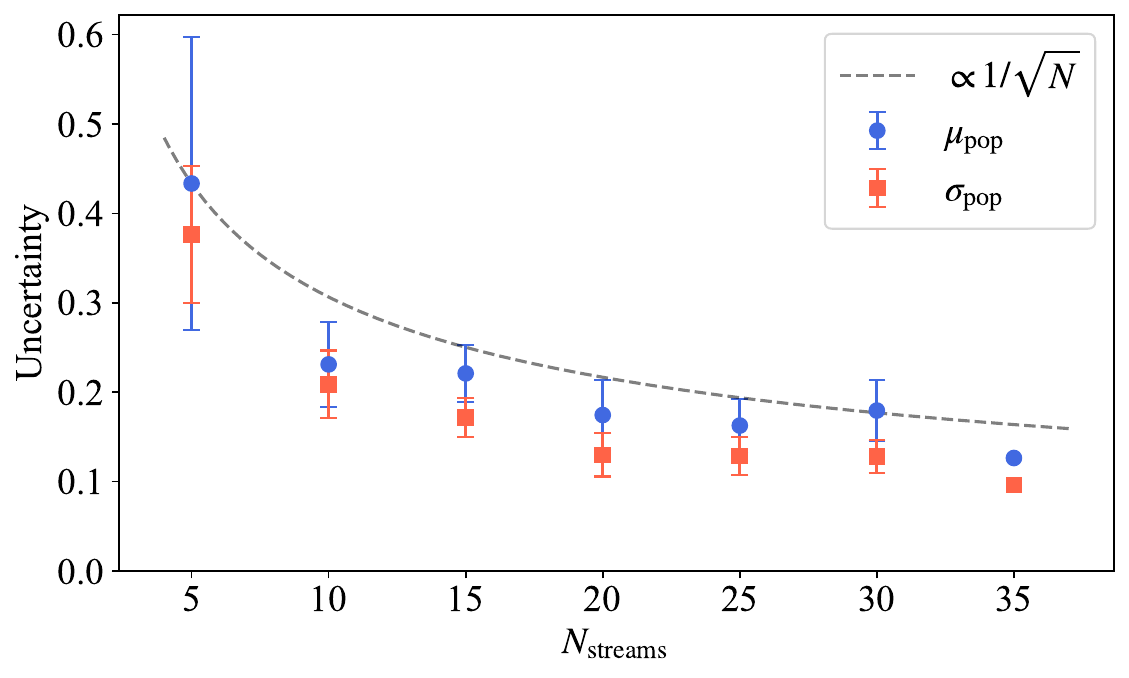}
    \caption{Population uncertainty as a function of the number of streams $N_{streams}$, where the uncertainty is defined as the 68\% posterior width. Blue circles show the uncertainty on the population mean $\mu_{\mathrm{pop}}$, and orange squares show the uncertainty on the population standard deviation $\sigma_{\mathrm{pop}}$. For $N \leq 30$, each point is the mean over 10 bootstrap iterations and the error bars show the standard deviation across bootstrap samples. The $N=35$ point uses the full 35-stream catalogue once and therefore has no error bar. The dashed line shows the expected $1/\sqrt{N}$ scaling. The uncertainty decreases broadly as $1/\sqrt{N}$, with faster than expected improvement at small $N$.}
    \label{fig:scaling_N}
\end{figure}

To explore this, we examined the absolute mean error as a function of stream length (both physical and angular), mean radial distance, and radial variation, finding no clear correlation (see Appendix \ref{app:A} for more information). Although, since length typically increases the available phase-space information \citep{Bonaca2018}, individual streams are expected to yield better fits when longer, this effect does not translate straightforwardly to populations. As shown by \citet{Nibauer2023}, the most constraining features in stream tracks for halo morphology are the straight segments. Because longer streams do not necessarily contain more of these segments, their contribution to the overall population level inference is not systematically stronger. Moreover, since we are constraining the halo flattening, an azimuthal property, the radial distance and variation of the track have little impact on the quality of the fits in contrast to \citet{Nibauer2025b}, who found a radial dependence when fitting for the host halo’s radial density profile.

We see that in all three cases, the parameters are all confidently within $2\sigma$. For the spherical case (middle pannel), there is no bias in the posterior distribution of the mean position, $\mu$, whereas for the oblate (left) and prolate (right) case there is a clear bias (larger tail) pointing towards the spherical case. We explain this from two phenomena. Firstly, we remark that the correct axis ratio is always the most extreme flattening. Indeed, any small deviations in orientation or measurement error tend to make the system appear more spherical. Secondly, the modalities between oblate and prolate discussed in Section \ref{Degen} can also explain the bias towards  the other end of the flattening space. Importantly, the multimodalities seen in some individual posteriors between oblate and prolate are completely absent at the population level. Just like $\mu$, the posteriors on the standard deviations, $\sigma$, are confidently constrained but also exhibits bias towards larger deviation which also related to the two aformentioned reasons. A larger standard deviation for the population model helps better fit the outskirts of the distribution. Overall, we demonstrate that while individual projected stream tracks are weakly informative, fitting as few as 35 streams allows us to confidently constrain the population-level shape of dark matter haloes.

\subsection{Scaling with sample size}\label{sec:scaling_N}

To assess how population constraints improve with the number of streams, we repeat the spherical population fit ($\mu_{\mathrm{pop}}=1.0$, $\sigma_{\mathrm{pop}}=0.1$) for N = 5, 10, 15, 20, 25 and 30 streams. For $N \leq 30$, we perform 10 bootstrap iterations, each drawing a random subset of $N$ streams from the full 35-stream mock catalogue and running the hierarchical inference independently. For $N=35$ we use the full catalogue once, so there is no bootstrap scatter associated with that point. Figure~\ref{fig:scaling_N} shows the uncertainty on $\mu_{\mathrm{pop}}$ and $\sigma_{\mathrm{pop}}$ as a function of $N$, where we define the uncertainty as the 68\% posterior width. In both cases, the overall trend is broadly consistent with the $1/\sqrt{N}$ scaling expected for independent Gaussian measurements. At small $N$, the improvement is somewhat faster than $1/\sqrt{N}$, which we attribute to the rapid breaking of the oblate/prolate degeneracy as streams with diverse orientations are combined.

\subsection{Sensitivity to population-model mismatch}\label{sec:mismatch_pop}

\begin{figure}
    \centering
    \includegraphics[width=\columnwidth]{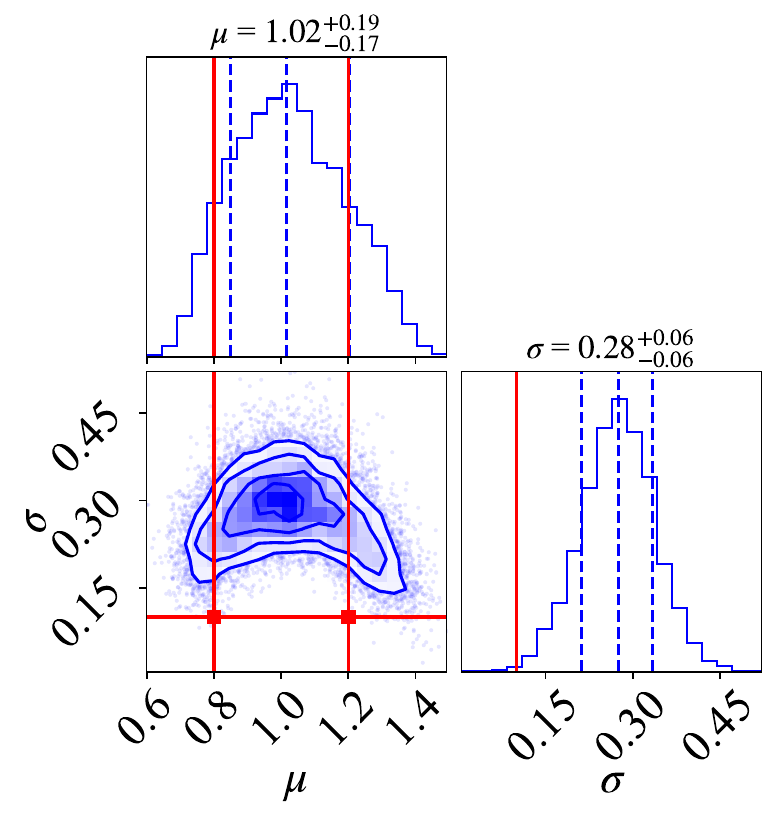}
    \caption{Population level model mismatch test. A bimodal mock distribution with peaks at $q=0.8$ and $q=1.2$ is fit with a single Gaussian. Red guides indicate the two true component means. The recovered posterior prefers an intermediate $\mu_{\mathrm{pop}}$ with broad uncertainty and an inflated $\sigma_{\mathrm{pop}}$, while the tails extend toward the two true peaks without reproducing the genuine bimodal structure.}
    \label{fig:population_mismatch}
\end{figure}

The tests above are intentionally unbiased: the true population distribution and the fitted popualtion model are both Gaussian. To assess the impact of model mismatch, we generated a bimodal mock population with two components centred at $q=0.8$ and $q=1.2$, each with intrinsic width $\sigma=0.1$, and then fit that sample with the same single Gaussian population model used above. Figure~\ref{fig:population_mismatch} shows the result. The recovered posterior favours an intermediate value of $\mu_{\mathrm{pop}}$, with a broad uncertainty, while $\sigma_{\mathrm{pop}}$ becomes substantially inflated in order to accommodate both components. The posterior on $\mu_{\mathrm{pop}}$ also develops two tails pointing toward the locations of the true peaks, but it does not recover the genuine bimodal structure or extend significantly beyond those peaks, as expected when a unimodal model is forced to approximate a bounded bimodal truth. This demonstrates that the inferred hyperparameters can be systematically biased when the assumed population family is incorrect, even if the individual stream fits are themselves well behaved.

The encouraging aspect is that the second-stage hierarchical fit is computationally inexpensive, typically taking only a few minutes, so alternative population models and stress tests can readily be explored in practice. Nevertheless, the inference remains conditional on the chosen model. Cosmological simulations can provide physically motivated guidance for plausible models and priors \citep{JingSuto2002, Allgood2006, Bett2007, VeraCiro2011, Velliscig2015, Butsky2016, Chua2019}, but they do not remove the need to test multiple models. Allowing mixtures or other more flexible population models is therefore an important next step for real applications.

\section{Discussion and Conclusions}\label{sec:discussion}

We developed a linearly scalable, \texttt{JAX}-accelerated hierarchical Bayesian framework to infer the population-level morphology of DM haloes using tracks of projected stellar streams. Our stream generator, \href{https://github.com/David-Chemaly/StreaMAX}{\texttt{StreaMAX} \textsc{\Large{\scalebox{0.8}{\faGithub}}}}, achieves substantial speedups via compilation and vectorised parallelism. Tailored to extragalactic applications, our method extracts information from photometry alone, with no kinematic inputs. First, we fit streams individually leading to broad posterior on the halo flattening, with each fit taking $\sim$10~hours on a 44-core CPU. Second, by treating the individual fits as a population, we combine individual posteriors and obtain tight constraints on the population level parameters.

Our main results are summarized as follows:
\begin{enumerate}
\item Forward modelling of stellar streams with a \texttt{JAX}-compiled particle-spray and dynamic nested sampling reliably recovers halo flattening $q$ across oblate, spherical, and prolate morphologies, while limiting strong couplings with other (nuisance) parameters.
\item Individual fits sometimes exhibit projection-induced multi-modality between oblate and prolate solutions; at the population level these modes manifest as potential biases.
\item Combining posteriors from 35 simulated streams accurately reconstructs the input shape distribution, demonstrating clear discrimination among oblate, spherical, and prolate populations.
\item Population posteriors show a mild tendency toward rounder ($q\!\approx\!1$) shapes and slightly inflated standard deviations, attributable to projection effects and sampling inefficiencies.
\end{enumerate}

This study serves a as proof of concept for utilizing extragalactic streams for population inference of DM haloes. Here, we address some of the limitations of this work given our assumptions:

\begin{enumerate}
\item We fit only the ridgeline (track) in projection, discarding width and surface density information. Using these observables will require a more realistic stream model (e.g.\ mass loss and variable spray emission) and a likelihood that models not only position but also width and density.
\item One projected progenitor coordinate is fixed, effectively assuming a known on-track position. A more general parameterisation would allow the progenitor to slide freely along the track.
\item The host potential is intentionally simplified (axisymmetric NFW, no baryons) making flattening the only parameter dictating the shape of the halo. This allowed us to constrain population parameters without having to worry about degeneracies amongst morphological parameters. Allowing triaxiality, radial profile variations, and baryonic components will introduce additional, physically relevant degeneracies.
\item All tests are closed loop: mocks and fits share the same forward model and noise, and the population prior and likelihood models match. This allowed testing the capabilities of our pipeline without worrying about biases, but we stress that real data will inevitably violate these assumptions, and model mismatch can bias inferences. We verified this explicitly in Section~\ref{sec:mismatch_pop}: fitting a bimodal population with a single Gaussian pulls the inferred mean toward an intermediate value and inflates the inferred dispersion. Because the population fit itself is fast, such stress tests can and should be repeated for multiple plausible models in real applications.
\end{enumerate}

Our computationally efficient framework is well suited to upcoming wide-field imaging from Euclid and Rubin/LSST, which will vastly expand the census of extragalactic streams, creating unprecedented opportunities for population-level analyses. Looking ahead, we plan to refine our approach by incorporating additional stream properties, such as widths of the stream along the track, and allow for the progenitor to move completely freely in the projected plan. Furthermore, we are working on more accurately modeling the host's potential by also fitting for the baryonic matter of the galaxy. We are also exploring more complex dark matter halo distributions and the effect of assuming the wrong underlying population distribution on our retrieval. 

In a future work, we will apply this methodology to observations using the \texttt{STRRINGS} catalogue \citep{Sola2025} of 35 faint, curved, and extended streams identified in the Dark Energy Spectroscopic Instrument Legacy Imaging Surveys \citep[DESI–LS;][]{Dey2019}. To our knowledge, this will provide the first population scale constraints on dark matter halo shapes derived from stellar streams in external galaxies.

\section*{Acknowledgements}

The authors would like to thank Tom Hehir, Chris Moore and Kaisey Mandel for their valuable insight. DC acknowledges funding from the Harding Distinguished Postgraduate Scholars Program. ES is grateful to the Leverhulme Trust for funding under the grant number RPG-2021-205. SK acknowledges support from the Science \& Technology Facilities Council (STFC) grant ST/Y001001/1.

\section*{Data Availability}

All data used in this work were generated through simulations using code developed by the authors. The model used to simulate stream tracks is publicly available as a \texttt{Python} package on \href{https://github.com/David-Chemaly/StreaMAX}{\texttt{StreaMAX} \textsc{\Large{\scalebox{0.8}{\faGithub}}}}. The full fitting pipeline, along with scripts to reproduce all figures and results, is available upon request.



\bibliographystyle{mnras}
\bibliography{references} 

@ARTICLE{Hendel2015,
       author = {{Hendel}, David and {Johnston}, Kathryn V.},
        title = "{Tidal debris morphology and the orbits of satellite galaxies}",
      journal = {\mnras},
         year = 2015,
        month = dec,
       volume = {454},
       number = {3},
        pages = {2472-2485},
          doi = {10.1093/mnras/stv2035},
archivePrefix = {arXiv},
       eprint = {1509.06369},
 primaryClass = {astro-ph.GA},
       adsurl = {https://ui.adsabs.harvard.edu/abs/2015MNRAS.454.2472H}
}

@ARTICLE{Sanders2014,
       author = {{Sanders}, Jason L.},
        title = "{Probabilistic model for constraining the Galactic potential using tidal streams}",
      journal = {\mnras},
         year = 2014,
        month = sep,
       volume = {443},
       number = {1},
        pages = {423-431},
          doi = {10.1093/mnras/stu1159},
archivePrefix = {arXiv},
       eprint = {1401.7602},
 primaryClass = {astro-ph.GA},
       adsurl = {https://ui.adsabs.harvard.edu/abs/2014MNRAS.443..423S}
}

@ARTICLE{Eyre2011,
       author = {{Eyre}, Andy and {Binney}, James},
        title = "{The mechanics of tidal streams}",
      journal = {\mnras},
         year = 2011,
        month = may,
       volume = {413},
       number = {3},
        pages = {1852-1874},
          doi = {10.1111/j.1365-2966.2011.18270.x},
archivePrefix = {arXiv},
       eprint = {1011.3672},
 primaryClass = {astro-ph.GA},
       adsurl = {https://ui.adsabs.harvard.edu/abs/2011MNRAS.413.1852E}
}

@ARTICLE{Bovy2014,
       author = {{Bovy}, Jo},
        title = "{Dynamical Modeling of Tidal Streams}",
      journal = {\apj},
         year = 2014,
        month = nov,
       volume = {795},
       number = {1},
          eid = {95},
        pages = {95},
          doi = {10.1088/0004-637X/795/1/95},
archivePrefix = {arXiv},
       eprint = {1401.2985},
 primaryClass = {astro-ph.GA},
       adsurl = {https://ui.adsabs.harvard.edu/abs/2014ApJ...795...95B}
}

@ARTICLE{Binney2008streams,
       author = {{Binney}, James},
        title = "{Fitting orbits to tidal streams}",
      journal = {\mnras},
         year = 2008,
        month = may,
       volume = {386},
       number = {1},
        pages = {L47-L51},
          doi = {10.1111/j.1745-3933.2008.00458.x},
archivePrefix = {arXiv},
       eprint = {0802.1485},
 primaryClass = {astro-ph},
       adsurl = {https://ui.adsabs.harvard.edu/abs/2008MNRAS.386L..47B}
}

@ARTICLE{Amorisco2015,
       author = {{Amorisco}, N.~C.},
        title = "{On feathers, bifurcations and shells: the dynamics of tidal streams across the mass scale}",
      journal = {\mnras},
         year = 2015,
        month = jun,
       volume = {450},
       number = {1},
        pages = {575-591},
          doi = {10.1093/mnras/stv648},
archivePrefix = {arXiv},
       eprint = {1410.0360},
 primaryClass = {astro-ph.GA},
       adsurl = {https://ui.adsabs.harvard.edu/abs/2015MNRAS.450..575A}
}

@ARTICLE{LSST2009,
       author = {{LSST Science Collaboration} and {Abell}, Paul A. and {Allison}, Julius and {Anderson}, Scott F. and {Andrew}, John R. and {Angel}, J. Roger P. and {Armus}, Lee and {Arnett}, David and {Asztalos}, S.~J. and {Axelrod}, Tim S. and {Bailey}, Stephen and {Ballantyne}, D.~R. and {Bankert}, Justin R. and {Barkhouse}, Wayne A. and {Barr}, Jeffrey D. and {Barrientos}, L. Felipe and {Barth}, Aaron J. and {Bartlett}, James G. and {Becker}, Andrew C. and {Becla}, Jacek and {Beers}, Timothy C. and {Bernstein}, Joseph P. and {Biswas}, Rahul and {Blanton}, Michael R. and {Bloom}, Joshua S. and {Bochanski}, John J. and {Boeshaar}, Pat and {Borne}, Kirk D. and {Bradac}, Marusa and {Brandt}, W.~N. and {Bridge}, Carrie R. and {Brown}, Michael E. and {Brunner}, Robert J. and {Bullock}, James S. and {Burgasser}, Adam J. and {Burge}, James H. and {Burke}, David L. and {Cargile}, Phillip A. and {Chandrasekharan}, Srinivasan and {Chartas}, George and {Chesley}, Steven R. and {Chu}, You-Hua and {Cinabro}, David and {Claire}, Mark W. and {Claver}, Charles F. and {Clowe}, Douglas and {Connolly}, A.~J. and {Cook}, Kem H. and {Cooke}, Jeff and {Cooray}, Asantha and {Covey}, Kevin R. and {Culliton}, Christopher S. and {de Jong}, Roelof and {de Vries}, Willem H. and {Debattista}, Victor P. and {Delgado}, Francisco and {Dell'Antonio}, Ian P. and {Dhital}, Saurav and {Di Stefano}, Rosanne and {Dickinson}, Mark and {Dilday}, Benjamin and {Djorgovski}, S.~G. and {Dobler}, Gregory and {Donalek}, Ciro and {Dubois-Felsmann}, Gregory and {Durech}, Josef and {Eliasdottir}, Ardis and {Eracleous}, Michael and {Eyer}, Laurent and {Falco}, Emilio E. and {Fan}, Xiaohui and {Fassnacht}, Christopher D. and {Ferguson}, Harry C. and {Fernandez}, Yanga R. and {Fields}, Brian D. and {Finkbeiner}, Douglas and {Figueroa}, Eduardo E. and {Fox}, Derek B. and {Francke}, Harold and {Frank}, James S. and {Frieman}, Josh and {Fromenteau}, Sebastien and {Furqan}, Muhammad and {Galaz}, Gaspar and {Gal-Yam}, A. and {Garnavich}, Peter and {Gawiser}, Eric and {Geary}, John and {Gee}, Perry and {Gibson}, Robert R. and {Gilmore}, Kirk and {Grace}, Emily A. and {Green}, Richard F. and {Gressler}, William J. and {Grillmair}, Carl J. and {Habib}, Salman and {Haggerty}, J.~S. and {Hamuy}, Mario and {Harris}, Alan W. and {Hawley}, Suzanne L. and {Heavens}, Alan F. and {Hebb}, Leslie and {Henry}, Todd J. and {Hileman}, Edward and {Hilton}, Eric J. and {Hoadley}, Keri and {Holberg}, J.~B. and {Holman}, Matt J. and {Howell}, Steve B. and {Infante}, Leopoldo and {Ivezic}, Zeljko and {Jacoby}, Suzanne H. and {Jain}, Bhuvnesh and {R} and {Jedicke} and {Jee}, M. James and {Garrett Jernigan}, J. and {Jha}, Saurabh W. and {Johnston}, Kathryn V. and {Jones}, R. Lynne and {Juric}, Mario and {Kaasalainen}, Mikko and {Styliani} and {Kafka} and {Kahn}, Steven M. and {Kaib}, Nathan A. and {Kalirai}, Jason and {Kantor}, Jeff and {Kasliwal}, Mansi M. and {Keeton}, Charles R. and {Kessler}, Richard and {Knezevic}, Zoran and {Kowalski}, Adam and {Krabbendam}, Victor L. and {Krughoff}, K. Simon and {Kulkarni}, Shrinivas and {Kuhlman}, Stephen and {Lacy}, Mark and {Lepine}, Sebastien and {Liang}, Ming and {Lien}, Amy and {Lira}, Paulina and {Long}, Knox S. and {Lorenz}, Suzanne and {Lotz}, Jennifer M. and {Lupton}, R.~H. and {Lutz}, Julie and {Macri}, Lucas M. and {Mahabal}, Ashish A. and {Mandelbaum}, Rachel and {Marshall}, Phil and {May}, Morgan and {McGehee}, Peregrine M. and {Meadows}, Brian T. and {Meert}, Alan and {Milani}, Andrea and {Miller}, Christopher J. and {Miller}, Michelle and {Mills}, David and {Minniti}, Dante and {Monet}, David and {Mukadam}, Anjum S. and {Nakar}, Ehud and {Neill}, Douglas R. and {Newman}, Jeffrey A. and {Nikolaev}, Sergei and {Nordby}, Martin and {O'Connor}, Paul and {Oguri}, Masamune and {Oliver}, John and {Olivier}, Scot S. and {Olsen}, Julia K. and {Olsen}, Knut and {Olszewski}, Edward W. and {Oluseyi}, Hakeem and {Padilla}, Nelson D. and {Parker}, Alex and {Pepper}, Joshua and {Peterson}, John R. and {Petry}, Catherine and {Pinto}, Philip A. and {Pizagno}, James L. and {Popescu}, Bogdan and {Prsa}, Andrej and {Radcka}, Veljko and {Raddick}, M. Jordan and {Rasmussen}, Andrew and {Rau}, Arne and {Rho}, Jeonghee and {Rhoads}, James E. and {Richards}, Gordon T. and {Ridgway}, Stephen T. and {Robertson}, Brant E. and {Roskar}, Rok and {Saha}, Abhijit and {Sarajedini}, Ata and {Scannapieco}, Evan and {Schalk}, Terry and {Schindler}, Rafe and {Schmidt}, Samuel},
        title = "{LSST Science Book, Version 2.0}",
      journal = {arXiv e-prints},
         year = 2009,
        month = dec,
          eid = {arXiv:0912.0201},
        pages = {arXiv:0912.0201},
          doi = {10.48550/arXiv.0912.0201},
archivePrefix = {arXiv},
       eprint = {0912.0201},
 primaryClass = {astro-ph.IM},
       adsurl = {https://ui.adsabs.harvard.edu/abs/2009arXiv0912.0201L}
}

@ARTICLE{Laine2018,
       author = {{Laine}, Seppo and {Martinez-Delgado}, David and {Trujillo}, Ignacio and {Duc}, Pierre-Alain and {Grillmair}, Carl J. and {Frenk}, Carlos S. and {Hendel}, David and {Johnston}, Kathryn V. and {Mihos}, J. Chris and {Moustakas}, John and {Beaton}, Rachael L. and {Romanowsky}, Aaron J. and {Greco}, Johnny and {Erkal}, Denis},
        title = "{LSST Cadence Optimization White Paper in Support of Observations of Unresolved Tidal Stellar Streams in Galaxies beyond the Local Group}",
      journal = {arXiv e-prints},
         year = 2018,
        month = dec,
          eid = {arXiv:1812.04897},
        pages = {arXiv:1812.04897},
          doi = {10.48550/arXiv.1812.04897},
archivePrefix = {arXiv},
       eprint = {1812.04897},
 primaryClass = {astro-ph.GA},
       adsurl = {https://ui.adsabs.harvard.edu/abs/2018arXiv181204897L}
}

@ARTICLE{Pearson2024,
       author = {{Pearson}, Sarah and {Bonaca}, Ana and {Chen}, Yingtian and {Gnedin}, Oleg Y.},
        title = "{Forecasting the Population of Globular Cluster Streams in Milky Way{\textendash}type Galaxies}",
      journal = {\apj},
         year = 2024,
        month = nov,
       volume = {976},
       number = {1},
          eid = {54},
        pages = {54},
          doi = {10.3847/1538-4357/ad8348},
archivePrefix = {arXiv},
       eprint = {2405.15851},
 primaryClass = {astro-ph.GA},
       adsurl = {https://ui.adsabs.harvard.edu/abs/2024ApJ...976...54P}
}

@ARTICLE{Pearson2019,
       author = {{Pearson}, Sarah and {Starkenburg}, Tjitske K. and {Johnston}, Kathryn V. and {Williams}, Benjamin F. and {Ibata}, Rodrigo A. and {Khan}, Rubab},
        title = "{Detecting Thin Stellar Streams in External Galaxies: Resolved Stars and Integrated Light}",
      journal = {\apj},
         year = 2019,
        month = sep,
       volume = {883},
       number = {1},
          eid = {87},
        pages = {87},
          doi = {10.3847/1538-4357/ab3e06},
archivePrefix = {arXiv},
       eprint = {1906.03264},
 primaryClass = {astro-ph.GA},
       adsurl = {https://ui.adsabs.harvard.edu/abs/2019ApJ...883...87P}
}

@ARTICLE{Mouhcine2010,
       author = {{Mouhcine}, M. and {Ibata}, R. and {Rejkuba}, M.},
        title = "{A Panoramic View of the Milky Way Analog NGC 891}",
      journal = {\apjl},
         year = 2010,
        month = may,
       volume = {714},
       number = {1},
        pages = {L12-L15},
          doi = {10.1088/2041-8205/714/1/L12},
archivePrefix = {arXiv},
       eprint = {1002.0461},
 primaryClass = {astro-ph.GA},
       adsurl = {https://ui.adsabs.harvard.edu/abs/2010ApJ...714L..12M}
}

@INPROCEEDINGS{Carlin2016,
       author = {{Carlin}, Jeffrey L. and {Beaton}, Rachael L. and {Mart{\'\i}nez-Delgado}, David and {Gabany}, R. Jay},
        title = "{Stellar Tidal Streams in External Galaxies}",
    booktitle = {Tidal Streams in the Local Group and Beyond},
         year = 2016,
       editor = {{Newberg}, Heidi Jo and {Carlin}, Jeffrey L.},
       series = {Astrophysics and Space Science Library},
       volume = {420},
        month = jan,
        pages = {219},
          doi = {10.1007/978-3-319-19336-6_9},
archivePrefix = {arXiv},
       eprint = {1603.04656},
 primaryClass = {astro-ph.GA},
       adsurl = {https://ui.adsabs.harvard.edu/abs/2016ASSL..420..219C}
}

@ARTICLE{Yavetz2023,
       author = {{Yavetz}, Tomer D. and {Johnston}, Kathryn V. and {Pearson}, Sarah and {Price-Whelan}, Adrian M. and {Hamilton}, Chris},
        title = "{Stream Fanning and Bifurcations: Observable Signatures of Resonances in Stellar Stream Morphology}",
      journal = {\apj},
         year = 2023,
        month = sep,
       volume = {954},
       number = {2},
          eid = {215},
        pages = {215},
          doi = {10.3847/1538-4357/ace7b9},
archivePrefix = {arXiv},
       eprint = {2212.11006},
 primaryClass = {astro-ph.GA},
       adsurl = {https://ui.adsabs.harvard.edu/abs/2023ApJ...954..215Y}
}

@ARTICLE{Weerasooriya2025,
       author = {{Weerasooriya}, Sachi and {Starkenburg}, Tjitske and {Cunningham}, Emily C. and {Johnston}, Kathryn V},
        title = "{Dancing Streams In Merging Halos: Stellar Streams in a MW--LMC-like merger}",
      journal = {arXiv e-prints},
         year = 2025,
        month = may,
          eid = {arXiv:2505.14792},
        pages = {arXiv:2505.14792},
          doi = {10.48550/arXiv.2505.14792},
archivePrefix = {arXiv},
       eprint = {2505.14792},
 primaryClass = {astro-ph.GA},
       adsurl = {https://ui.adsabs.harvard.edu/abs/2025arXiv250514792W}
}

@ARTICLE{Dillamore2022,
       author = {{Dillamore}, Adam M. and {Belokurov}, Vasily and {Evans}, N. Wyn and {Price-Whelan}, Adrian M.},
        title = "{The impact of a massive Sagittarius dSph on GD-1-like streams}",
      journal = {\mnras},
         year = 2022,
        month = oct,
       volume = {516},
       number = {2},
        pages = {1685-1703},
          doi = {10.1093/mnras/stac2311},
archivePrefix = {arXiv},
       eprint = {2205.13547},
 primaryClass = {astro-ph.GA},
       adsurl = {https://ui.adsabs.harvard.edu/abs/2022MNRAS.516.1685D}
}

@ARTICLE{Valluri2025,
       author = {{Valluri}, M. and {Fagrelius}, P. and {Koposov}, S.~E. and {Li}, T.~S. and {Gnedin}, Oleg Y. and {Bell}, E.~F. and {Carlberg}, R.~G. and {Cooper}, A.~P. and {Aguilar}, J. and {Ahlen}, S. and {Allende Prieto}, C. and {Belokurov}, V. and {Beraldo e Silva}, L. and {Brooks}, D. and {Bystr{\"o}m}, A. and {Claybaugh}, T. and {Dawson}, K. and {Dey}, A. and {Doel}, P. and {Forero-Romero}, J.~E. and {Gazta{\~n}aga}, E. and {Gontcho A Gontcho}, S. and {Han}, J. and {Honscheid}, K. and {Kisner}, T. and {Kremin}, A. and {Lambert}, A. and {Landriau}, M. and {Le Guillou}, L. and {Levi}, M.~E. and {de la Macorra}, A. and {Manera}, M. and {Martini}, P. and {Medina}, G.~E. and {Meisner}, A. and {Miquel}, R. and {Moustakas}, J. and {Myers}, A.~D. and {Najita}, J. and {Poppett}, C. and {Prada}, F. and {Rezaie}, M. and {Rossi}, G. and {Riley}, A.~H. and {Sanchez}, E. and {Schlegel}, D. and {Schubnell}, M. and {Sprayberry}, D. and {Tarl{\'e}}, G. and {Thomas}, G. and {Weaver}, B.~A. and {Wechsler}, R.~H. and {Zhou}, R. and {Zou}, H.},
        title = "{GD-1 Stellar Stream and Cocoon in the DESI Early Data Release}",
      journal = {\apj},
         year = 2025,
        month = feb,
       volume = {980},
       number = {1},
          eid = {71},
        pages = {71},
          doi = {10.3847/1538-4357/ada690},
archivePrefix = {arXiv},
       eprint = {2407.06336},
 primaryClass = {astro-ph.GA},
       adsurl = {https://ui.adsabs.harvard.edu/abs/2025ApJ...980...71V}
}

@ARTICLE{deBoer2020,
       author = {{de Boer}, T.~J.~L. and {Erkal}, D. and {Gieles}, M.},
        title = "{A closer look at the spur, blob, wiggle, and gaps in GD-1}",
      journal = {\mnras},
         year = 2020,
        month = jun,
       volume = {494},
       number = {4},
        pages = {5315-5332},
          doi = {10.1093/mnras/staa917},
archivePrefix = {arXiv},
       eprint = {1911.05745},
 primaryClass = {astro-ph.GA},
       adsurl = {https://ui.adsabs.harvard.edu/abs/2020MNRAS.494.5315D}
}

@ARTICLE{deBoer2018,
       author = {{de Boer}, T.~J.~L. and {Belokurov}, V. and {Koposov}, S.~E. and {Ferrarese}, L. and {Erkal}, D. and {C{\^o}t{\'e}}, P. and {Navarro}, J.~F.},
        title = "{A deeper look at the GD1 stream: density variations and wiggles}",
      journal = {\mnras},
         year = 2018,
        month = jun,
       volume = {477},
       number = {2},
        pages = {1893-1902},
          doi = {10.1093/mnras/sty677},
archivePrefix = {arXiv},
       eprint = {1801.08948},
 primaryClass = {astro-ph.GA},
       adsurl = {https://ui.adsabs.harvard.edu/abs/2018MNRAS.477.1893D}
}

@ARTICLE{Nibauer2025a,
       author = {{Nibauer}, Jacob and {Bonaca}, Ana},
        title = "{Galactic Accelerations from the GD-1 Stream Suggest a Tilted Dark Matter Halo}",
      journal = {\apjl},
         year = 2025,
        month = may,
       volume = {985},
       number = {1},
          eid = {L22},
        pages = {L22},
          doi = {10.3847/2041-8213/add0a9},
archivePrefix = {arXiv},
       eprint = {2504.07187},
 primaryClass = {astro-ph.GA},
       adsurl = {https://ui.adsabs.harvard.edu/abs/2025ApJ...985L..22N}
}

@ARTICLE{Palau2023,
       author = {{Palau}, Carles G. and {Miralda-Escud{\'e}}, Jordi},
        title = "{The oblateness of the Milky Way dark matter halo from the stellar streams of NGC 3201, M68, and Palomar 5}",
      journal = {\mnras},
         year = 2023,
        month = sep,
       volume = {524},
       number = {2},
        pages = {2124-2147},
          doi = {10.1093/mnras/stad1930},
archivePrefix = {arXiv},
       eprint = {2212.03587},
 primaryClass = {astro-ph.GA},
       adsurl = {https://ui.adsabs.harvard.edu/abs/2023MNRAS.524.2124P}
}

@ARTICLE{Bowden2015,
       author = {{Bowden}, A. and {Belokurov}, V. and {Evans}, N.~W.},
        title = "{Dipping our toes in the water: first models of GD-1 as a stream}",
      journal = {\mnras},
         year = 2015,
        month = may,
       volume = {449},
       number = {2},
        pages = {1391-1400},
          doi = {10.1093/mnras/stv285},
archivePrefix = {arXiv},
       eprint = {1502.00484},
 primaryClass = {astro-ph.GA},
       adsurl = {https://ui.adsabs.harvard.edu/abs/2015MNRAS.449.1391B}
}

@ARTICLE{Johnston2005,
       author = {{Johnston}, Kathryn V. and {Law}, David R. and {Majewski}, Steven R.},
        title = "{A Two Micron All Sky Survey View of the Sagittarius Dwarf Galaxy. III. Constraints on the Flattening of the Galactic Halo}",
      journal = {\apj},
         year = 2005,
        month = feb,
       volume = {619},
       number = {2},
        pages = {800-806},
          doi = {10.1086/426777},
archivePrefix = {arXiv},
       eprint = {astro-ph/0407565},
 primaryClass = {astro-ph},
       adsurl = {https://ui.adsabs.harvard.edu/abs/2005ApJ...619..800J}
}

@ARTICLE{Erkal2015,
       author = {{Erkal}, Denis and {Belokurov}, Vasily},
        title = "{Forensics of subhalo-stream encounters: the three phases of gap growth}",
      journal = {\mnras},
         year = 2015,
        month = jun,
       volume = {450},
       number = {1},
        pages = {1136-1149},
          doi = {10.1093/mnras/stv655},
archivePrefix = {arXiv},
       eprint = {1412.6035},
 primaryClass = {astro-ph.GA},
       adsurl = {https://ui.adsabs.harvard.edu/abs/2015MNRAS.450.1136E}
}

@ARTICLE{Yoon2011,
       author = {{Yoon}, Joo Heon and {Johnston}, Kathryn V. and {Hogg}, David W.},
        title = "{Clumpy Streams from Clumpy Halos: Detecting Missing Satellites with Cold Stellar Structures}",
      journal = {\apj},
         year = 2011,
        month = apr,
       volume = {731},
       number = {1},
          eid = {58},
        pages = {58},
          doi = {10.1088/0004-637X/731/1/58},
archivePrefix = {arXiv},
       eprint = {1012.2884},
 primaryClass = {astro-ph.GA},
       adsurl = {https://ui.adsabs.harvard.edu/abs/2011ApJ...731...58Y}
}

@ARTICLE{Li2022,
       author = {{Li}, Ting S. and {Ji}, Alexander P. and {Pace}, Andrew B. and {Erkal}, Denis and {Koposov}, Sergey E. and {Shipp}, Nora and {Da Costa}, Gary S. and {Cullinane}, Lara R. and {Kuehn}, Kyler and {Lewis}, Geraint F. and {Mackey}, Dougal and {Simpson}, Jeffrey D. and {Zucker}, Daniel B. and {Ferguson}, Peter S. and {Martell}, Sarah L. and {Bland-Hawthorn}, Joss and {Balbinot}, Eduardo and {Tavangar}, Kiyan and {Drlica-Wagner}, Alex and {De Silva}, Gayandhi M. and {Simon}, Joshua D.},
        title = "{S $^{5}$: The Orbital and Chemical Properties of One Dozen Stellar Streams}",
      journal = {\apj},
         year = 2022,
        month = mar,
       volume = {928},
       number = {1},
          eid = {30},
        pages = {30},
          doi = {10.3847/1538-4357/ac46d3},
archivePrefix = {arXiv},
       eprint = {2110.06950},
 primaryClass = {astro-ph.GA},
       adsurl = {https://ui.adsabs.harvard.edu/abs/2022ApJ...928...30L}
}

@ARTICLE{Shipp2018,
       author = {{Shipp}, N. and {Drlica-Wagner}, A. and {Balbinot}, E. and {Ferguson}, P. and {Erkal}, D. and {Li}, T.~S. and {Bechtol}, K. and {Belokurov}, V. and {Buncher}, B. and {Carollo}, D. and {Carrasco Kind}, M. and {Kuehn}, K. and {Marshall}, J.~L. and {Pace}, A.~B. and {Rykoff}, E.~S. and {Sevilla-Noarbe}, I. and {Sheldon}, E. and {Strigari}, L. and {Vivas}, A.~K. and {Yanny}, B. and {Zenteno}, A. and {Abbott}, T.~M.~C. and {Abdalla}, F.~B. and {Allam}, S. and {Avila}, S. and {Bertin}, E. and {Brooks}, D. and {Burke}, D.~L. and {Carretero}, J. and {Castander}, F.~J. and {Cawthon}, R. and {Crocce}, M. and {Cunha}, C.~E. and {D'Andrea}, C.~B. and {da Costa}, L.~N. and {Davis}, C. and {De Vicente}, J. and {Desai}, S. and {Diehl}, H.~T. and {Doel}, P. and {Evrard}, A.~E. and {Flaugher}, B. and {Fosalba}, P. and {Frieman}, J. and {Garc{\'\i}a-Bellido}, J. and {Gaztanaga}, E. and {Gerdes}, D.~W. and {Gruen}, D. and {Gruendl}, R.~A. and {Gschwend}, J. and {Gutierrez}, G. and {Hartley}, W. and {Honscheid}, K. and {Hoyle}, B. and {James}, D.~J. and {Johnson}, M.~D. and {Krause}, E. and {Kuropatkin}, N. and {Lahav}, O. and {Lin}, H. and {Maia}, M.~A.~G. and {March}, M. and {Martini}, P. and {Menanteau}, F. and {Miller}, C.~J. and {Miquel}, R. and {Nichol}, R.~C. and {Plazas}, A.~A. and {Romer}, A.~K. and {Sako}, M. and {Sanchez}, E. and {Santiago}, B. and {Scarpine}, V. and {Schindler}, R. and {Schubnell}, M. and {Smith}, M. and {Smith}, R.~C. and {Sobreira}, F. and {Suchyta}, E. and {Swanson}, M.~E.~C. and {Tarle}, G. and {Thomas}, D. and {Tucker}, D.~L. and {Walker}, A.~R. and {Wechsler}, R.~H. and {DES Collaboration}},
        title = "{Stellar Streams Discovered in the Dark Energy Survey}",
      journal = {\apj},
         year = 2018,
        month = aug,
       volume = {862},
       number = {2},
          eid = {114},
        pages = {114},
          doi = {10.3847/1538-4357/aacdab},
archivePrefix = {arXiv},
       eprint = {1801.03097},
 primaryClass = {astro-ph.GA},
       adsurl = {https://ui.adsabs.harvard.edu/abs/2018ApJ...862..114S}
}

@ARTICLE{Malhan2018,
       author = {{Malhan}, Khyati and {Ibata}, Rodrigo A. and {Martin}, Nicolas F.},
        title = "{Ghostly tributaries to the Milky Way: charting the halo's stellar streams with the Gaia DR2 catalogue}",
      journal = {\mnras},
         year = 2018,
        month = dec,
       volume = {481},
       number = {3},
        pages = {3442-3455},
          doi = {10.1093/mnras/sty2474},
archivePrefix = {arXiv},
       eprint = {1804.11339},
 primaryClass = {astro-ph.GA},
       adsurl = {https://ui.adsabs.harvard.edu/abs/2018MNRAS.481.3442M}
}

@ARTICLE{Price-Whelan2018,
       author = {{Price-Whelan}, Adrian M. and {Bonaca}, Ana},
        title = "{Off the Beaten Path: Gaia Reveals GD-1 Stars outside of the Main Stream}",
      journal = {\apjl},
         year = 2018,
        month = aug,
       volume = {863},
       number = {2},
          eid = {L20},
        pages = {L20},
          doi = {10.3847/2041-8213/aad7b5},
archivePrefix = {arXiv},
       eprint = {1805.00425},
 primaryClass = {astro-ph.GA},
       adsurl = {https://ui.adsabs.harvard.edu/abs/2018ApJ...863L..20P}
}

@ARTICLE{Grillmair2006,
       author = {{Grillmair}, C.~J. and {Dionatos}, O.},
        title = "{Detection of a 63{\textdegree} Cold Stellar Stream in the Sloan Digital Sky Survey}",
      journal = {\apjl},
         year = 2006,
        month = may,
       volume = {643},
       number = {1},
        pages = {L17-L20},
          doi = {10.1086/505111},
archivePrefix = {arXiv},
       eprint = {astro-ph/0604332},
 primaryClass = {astro-ph},
       adsurl = {https://ui.adsabs.harvard.edu/abs/2006ApJ...643L..17G}
}

@ARTICLE{Deason2024,
       author = {{Deason}, Alis J. and {Belokurov}, Vasily},
        title = "{Galactic Archaeology with Gaia}",
      journal = {\nar},
         year = 2024,
        month = dec,
       volume = {99},
          eid = {101706},
        pages = {101706},
          doi = {10.1016/j.newar.2024.101706},
archivePrefix = {arXiv},
       eprint = {2402.12443},
 primaryClass = {astro-ph.GA},
       adsurl = {https://ui.adsabs.harvard.edu/abs/2024NewAR..9901706D}
}

@ARTICLE{Bonaca2025,
       author = {{Bonaca}, Ana and {Price-Whelan}, Adrian M.},
        title = "{Stellar streams in the Gaia era}",
      journal = {\nar},
         year = 2025,
        month = jun,
       volume = {100},
          eid = {101713},
        pages = {101713},
          doi = {10.1016/j.newar.2024.101713},
archivePrefix = {arXiv},
       eprint = {2405.19410},
 primaryClass = {astro-ph.GA},
       adsurl = {https://ui.adsabs.harvard.edu/abs/2025NewAR.10001713B}
}

@article{Price2017,
  doi = {10.21105/joss.00388},
  url = {https://doi.org/10.21105%2Fjoss.00388},
  year = 2017,
  month = {oct},
  publisher = {The Open Journal},
  volume = {2},
  number = {18},
  author = {Adrian M. Price-Whelan},
  title = {Gala: A Python package for galactic dynamics},
  journal = {The Journal of Open Source Software}
}

@ARTICLE{Erkal2017,
       author = {{Erkal}, Denis and {Koposov}, Sergey E. and {Belokurov}, Vasily},
        title = "{A sharper view of Pal 5's tails: discovery of stream perturbations with a novel non-parametric technique}",
      journal = {\mnras},
         year = 2017,
        month = sep,
       volume = {470},
       number = {1},
        pages = {60-84},
          doi = {10.1093/mnras/stx1208},
archivePrefix = {arXiv},
       eprint = {1609.01282},
 primaryClass = {astro-ph.GA},
       adsurl = {https://ui.adsabs.harvard.edu/abs/2017MNRAS.470...60E}
}

@ARTICLE{Walder2024,
       author = {{Walder}, Madison and {Erkal}, Denis and {Collins}, Michelle and {Martinez-Delgado}, David},
        title = "{Probing the dark matter haloes of external galaxies with stellar streams}",
      journal = {arXiv e-prints},
         year = 2024,
        month = feb,
          eid = {arXiv:2402.13314},
        pages = {arXiv:2402.13314},
          doi = {10.48550/arXiv.2402.13314},
archivePrefix = {arXiv},
       eprint = {2402.13314},
 primaryClass = {astro-ph.GA},
       adsurl = {https://ui.adsabs.harvard.edu/abs/2024arXiv240213314W}
}

@ARTICLE{Gibbons2014,
       author = {{Gibbons}, S.~L.~J. and {Belokurov}, V. and {Evans}, N.~W.},
        title = "{`Skinny Milky Way please', says Sagittarius}",
      journal = {\mnras},
         year = 2014,
        month = dec,
       volume = {445},
       number = {4},
        pages = {3788-3802},
          doi = {10.1093/mnras/stu1986},
archivePrefix = {arXiv},
       eprint = {1406.2243},
 primaryClass = {astro-ph.GA},
       adsurl = {https://ui.adsabs.harvard.edu/abs/2014MNRAS.445.3788G}
}

@ARTICLE{Higson2019,
       author = {{Higson}, Edward and {Handley}, Will and {Hobson}, Mike and {Lasenby}, Anthony},
        title = "{Dynamic nested sampling: an improved algorithm for parameter estimation and evidence calculation}",
      journal = {Statistics and Computing},
         year = 2019,
        month = sep,
       volume = {29},
       number = {5},
        pages = {891-913},
          doi = {10.1007/s11222-018-9844-0},
archivePrefix = {arXiv},
       eprint = {1704.03459},
 primaryClass = {stat.CO},
       adsurl = {https://ui.adsabs.harvard.edu/abs/2019S&C....29..891H}
}

@software{Koposov2022,
       author = {{Koposov}, Sergey and {Speagle}, Josh and {Barbary}, Kyle and {Ashton}, Gregory and {Bennett}, Ed and {Buchner}, Johannes and {Scheffler}, Carl and {Cook}, Ben and {Talbot}, Colm and {Guillochon}, James and {Cubillos}, Patricio and {Asensio Ramos}, Andr{\'e}s and {Johnson}, Ben and {Lang}, Dustin and {Ilya} and {Dartiailh}, Matthieu and {Nitz}, Alex and {McCluskey}, Andrew and {Archibald}, Anne and {Deil}, Christoph and {Foreman-Mackey}, Dan and {Goldstein}, Danny and {Tollerud}, Erik and {Leja}, Joel and {Kirk}, Matthew and {Pitkin}, Matt and {Sheehan}, Patrick and {Cargile}, Phillip and {Ruskin23} and {Angus}, Ruth},
        title = "{joshspeagle/dynesty: v2.0.3}",
         year = 2022,
        month = dec,
          eid = {10.5281/zenodo.7388523},
          doi = {10.5281/zenodo.7388523},
      version = {v2.0.3},
    publisher = {Zenodo},
       adsurl = {https://ui.adsabs.harvard.edu/abs/2022zndo...7388523K}
}

@ARTICLE{Speagle2020,
       author = {{Speagle}, Joshua S.},
        title = "{DYNESTY: a dynamic nested sampling package for estimating Bayesian posteriors and evidences}",
      journal = {\mnras},
         year = 2020,
        month = apr,
       volume = {493},
       number = {3},
        pages = {3132-3158},
          doi = {10.1093/mnras/staa278},
archivePrefix = {arXiv},
       eprint = {1904.02180},
 primaryClass = {astro-ph.IM},
       adsurl = {https://ui.adsabs.harvard.edu/abs/2020MNRAS.493.3132S}
}

@ARTICLE{Nibauer2023,
       author = {{Nibauer}, Jacob and {Bonaca}, Ana and {Johnston}, Kathryn V.},
        title = "{Constraining the Gravitational Potential from the Projected Morphology of Extragalactic Tidal Streams}",
      journal = {\apj},
         year = 2023,
        month = sep,
       volume = {954},
       number = {2},
          eid = {195},
        pages = {195},
          doi = {10.3847/1538-4357/ace9bc},
archivePrefix = {arXiv},
       eprint = {2303.17406},
 primaryClass = {astro-ph.GA},
       adsurl = {https://ui.adsabs.harvard.edu/abs/2023ApJ...954..195N}
}

@ARTICLE{Sanders2013,
       author = {{Sanders}, Jason L. and {Binney}, James},
        title = "{Stream-orbit misalignment - I. The dangers of orbit-fitting}",
      journal = {\mnras},
         year = 2013,
        month = aug,
       volume = {433},
       number = {3},
        pages = {1813-1825},
          doi = {10.1093/mnras/stt806},
archivePrefix = {arXiv},
       eprint = {1305.1935},
 primaryClass = {astro-ph.GA},
       adsurl = {https://ui.adsabs.harvard.edu/abs/2013MNRAS.433.1813S}
}

@ARTICLE{Pearson2022,
       author = {{Pearson}, Sarah and {Price-Whelan}, Adrian M. and {Hogg}, David W. and {Seth}, Anil C. and {Sand}, David J. and {Hunt}, Jason A.~S. and {Crnojevi{\'c}}, Denija},
        title = "{Mapping Dark Matter with Extragalactic Stellar Streams: The Case of Centaurus A}",
      journal = {\apj},
         year = 2022,
        month = dec,
       volume = {941},
       number = {1},
          eid = {19},
        pages = {19},
          doi = {10.3847/1538-4357/ac9bfb},
archivePrefix = {arXiv},
       eprint = {2205.12277},
 primaryClass = {astro-ph.GA},
       adsurl = {https://ui.adsabs.harvard.edu/abs/2022ApJ...941...19P}
}

@ARTICLE{Johnston1995,
       author = {{Johnston}, Kathryn V. and {Spergel}, David N. and {Hernquist}, Lars},
        title = "{The Disruption of the Sagittarius Dwarf Galaxy}",
      journal = {\apj},
         year = 1995,
        month = oct,
       volume = {451},
        pages = {598},
          doi = {10.1086/176247},
archivePrefix = {arXiv},
       eprint = {astro-ph/9502005},
 primaryClass = {astro-ph},
       adsurl = {https://ui.adsabs.harvard.edu/abs/1995ApJ...451..598J}
}

@ARTICLE{Koposov2010,
       author = {{Koposov}, Sergey E. and {Rix}, Hans-Walter and {Hogg}, David W.},
        title = "{Constraining the Milky Way Potential with a Six-Dimensional Phase-Space Map of the GD-1 Stellar Stream}",
      journal = {\apj},
         year = 2010,
        month = mar,
       volume = {712},
       number = {1},
        pages = {260-273},
          doi = {10.1088/0004-637X/712/1/260},
archivePrefix = {arXiv},
       eprint = {0907.1085},
 primaryClass = {astro-ph.GA},
       adsurl = {https://ui.adsabs.harvard.edu/abs/2010ApJ...712..260K}
}

@ARTICLE{Helmi1999,
       author = {{Helmi}, Amina and {White}, Simon D.~M.},
        title = "{Building up the stellar halo of the Galaxy}",
      journal = {\mnras},
         year = 1999,
        month = aug,
       volume = {307},
       number = {3},
        pages = {495-517},
          doi = {10.1046/j.1365-8711.1999.02616.x},
archivePrefix = {arXiv},
       eprint = {astro-ph/9901102},
 primaryClass = {astro-ph},
       adsurl = {https://ui.adsabs.harvard.edu/abs/1999MNRAS.307..495H}
}

@ARTICLE{Bovy2016,
       author = {{Bovy}, Jo and {Bahmanyar}, Anita and {Fritz}, Tobias K. and {Kallivayalil}, Nitya},
        title = "{The Shape of the Inner Milky Way Halo from Observations of the Pal 5 and GD--1 Stellar Streams}",
      journal = {\apj},
         year = 2016,
        month = dec,
       volume = {833},
       number = {1},
          eid = {31},
        pages = {31},
          doi = {10.3847/1538-4357/833/1/31},
archivePrefix = {arXiv},
       eprint = {1609.01298},
 primaryClass = {astro-ph.GA},
       adsurl = {https://ui.adsabs.harvard.edu/abs/2016ApJ...833...31B}
}

@ARTICLE{Pearson2015,
       author = {{Pearson}, Sarah and {K{\"u}pper}, Andreas H.~W. and {Johnston}, Kathryn V. and {Price-Whelan}, Adrian M.},
        title = "{Tidal Stream Morphology as an Indicator of Dark Matter Halo Geometry: The Case of Palomar 5}",
      journal = {\apj},
         year = 2015,
        month = jan,
       volume = {799},
       number = {1},
          eid = {28},
        pages = {28},
          doi = {10.1088/0004-637X/799/1/28},
archivePrefix = {arXiv},
       eprint = {1410.3477},
 primaryClass = {astro-ph.GA},
       adsurl = {https://ui.adsabs.harvard.edu/abs/2015ApJ...799...28P}
}

@ARTICLE{Kupper2015,
       author = {{K{\"u}pper}, Andreas H.~W. and {Balbinot}, Eduardo and {Bonaca}, Ana and {Johnston}, Kathryn V. and {Hogg}, David W. and {Kroupa}, Pavel and {Santiago}, Basilio X.},
        title = "{Globular Cluster Streams as Galactic High-Precision Scales{\textemdash}the Poster Child Palomar 5}",
      journal = {\apj},
         year = 2015,
        month = apr,
       volume = {803},
       number = {2},
          eid = {80},
        pages = {80},
          doi = {10.1088/0004-637X/803/2/80},
archivePrefix = {arXiv},
       eprint = {1502.02658},
 primaryClass = {astro-ph.GA},
       adsurl = {https://ui.adsabs.harvard.edu/abs/2015ApJ...803...80K}
}

@ARTICLE{Vasiliev2021,
       author = {{Vasiliev}, Eugene and {Belokurov}, Vasily and {Erkal}, Denis},
        title = "{Tango for three: Sagittarius, LMC, and the Milky Way}",
      journal = {\mnras},
         year = 2021,
        month = feb,
       volume = {501},
       number = {2},
        pages = {2279-2304},
          doi = {10.1093/mnras/staa3673},
archivePrefix = {arXiv},
       eprint = {2009.10726},
 primaryClass = {astro-ph.GA},
       adsurl = {https://ui.adsabs.harvard.edu/abs/2021MNRAS.501.2279V}
}

@ARTICLE{Koposov2023,
       author = {{Koposov}, Sergey E. and {Erkal}, Denis and {Li}, Ting S. and {Da Costa}, Gary S. and {Cullinane}, Lara R. and {Ji}, Alexander P. and {Kuehn}, Kyler and {Lewis}, Geraint F. and {Pace}, Andrew B. and {Shipp}, Nora and {Zucker}, Daniel B. and {Bland-Hawthorn}, Joss and {Lilleengen}, Sophia and {Martell}, Sarah L. and {S5 Collaboration}},
        title = "{S $^{5}$: Probing the Milky Way and Magellanic Clouds potentials with the 6D map of the Orphan-Chenab stream}",
      journal = {\mnras},
         year = 2023,
        month = jun,
       volume = {521},
       number = {4},
        pages = {4936-4962},
          doi = {10.1093/mnras/stad551},
archivePrefix = {arXiv},
       eprint = {2211.04495},
 primaryClass = {astro-ph.GA},
       adsurl = {https://ui.adsabs.harvard.edu/abs/2023MNRAS.521.4936K}
}

@ARTICLE{Ibata2002,
       author = {{Ibata}, R.~A. and {Lewis}, G.~F. and {Irwin}, M.~J. and {Quinn}, T.},
        title = "{Uncovering cold dark matter halo substructure with tidal streams}",
      journal = {\mnras},
         year = 2002,
        month = jun,
       volume = {332},
       number = {4},
        pages = {915-920},
          doi = {10.1046/j.1365-8711.2002.05358.x},
archivePrefix = {arXiv},
       eprint = {astro-ph/0110690},
 primaryClass = {astro-ph},
       adsurl = {https://ui.adsabs.harvard.edu/abs/2002MNRAS.332..915I}
}

@ARTICLE{Bonaca2014,
       author = {{Bonaca}, Ana and {Geha}, Marla and {K{\"u}pper}, Andreas H.~W. and {Diemand}, J{\"u}rg and {Johnston}, Kathryn V. and {Hogg}, David W.},
        title = "{Milky Way Mass and Potential Recovery Using Tidal Streams in a Realistic Halo}",
      journal = {\apj},
         year = 2014,
        month = nov,
       volume = {795},
       number = {1},
          eid = {94},
        pages = {94},
          doi = {10.1088/0004-637X/795/1/94},
archivePrefix = {arXiv},
       eprint = {1406.6063},
 primaryClass = {astro-ph.GA},
       adsurl = {https://ui.adsabs.harvard.edu/abs/2014ApJ...795...94B}
}

@ARTICLE{Erkal2016,
       author = {{Erkal}, Denis and {Belokurov}, Vasily and {Bovy}, Jo and {Sanders}, Jason L.},
        title = "{The number and size of subhalo-induced gaps in stellar streams}",
      journal = {\mnras},
         year = 2016,
        month = nov,
       volume = {463},
       number = {1},
        pages = {102-119},
          doi = {10.1093/mnras/stw1957},
archivePrefix = {arXiv},
       eprint = {1606.04946},
 primaryClass = {astro-ph.GA},
       adsurl = {https://ui.adsabs.harvard.edu/abs/2016MNRAS.463..102E}
}

@ARTICLE{MartinezDelgado2010,
       author = {{Mart{\'\i}nez-Delgado}, David and {Gabany}, R. Jay and {Crawford}, Ken and {Zibetti}, Stefano and {Majewski}, Steven R. and {Rix}, Hans-Walter and {Fliri}, J{\"u}rgen and {Carballo-Bello}, Julio A. and {Bardalez-Gagliuffi}, Daniella C. and {Pe{\~n}arrubia}, Jorge and {Chonis}, Taylor S. and {Madore}, Barry and {Trujillo}, Ignacio and {Schirmer}, Mischa and {McDavid}, David A.},
        title = "{Stellar Tidal Streams in Spiral Galaxies of the Local Volume: A Pilot Survey with Modest Aperture Telescopes}",
      journal = {\aj},
         year = 2010,
        month = oct,
       volume = {140},
       number = {4},
        pages = {962-967},
          doi = {10.1088/0004-6256/140/4/962},
archivePrefix = {arXiv},
       eprint = {1003.4860},
 primaryClass = {astro-ph.CO},
       adsurl = {https://ui.adsabs.harvard.edu/abs/2010AJ....140..962M}
}

@ARTICLE{MartinezDelgado2023,
       author = {{Mart{\'\i}nez-Delgado}, David and {Cooper}, Andrew P. and {Rom{\'a}n}, Javier and {Pillepich}, Annalisa and {Erkal}, Denis and {Pearson}, Sarah and {Moustakas}, John and {Laporte}, Chervin F.~P. and {Laine}, Seppo and {Akhlaghi}, Mohammad and {Lang}, Dustin and {Makarov}, Dmitry and {Borlaff}, Alejandro S. and {Donatiello}, Giuseppe and {Pearson}, William J. and {Mir{\'o}-Carretero}, Juan and {Cuillandre}, Jean-Charles and {Dom{\'\i}nguez}, Helena and {Roca-F{\`a}brega}, Santi and {Frenk}, Carlos S. and {Schmidt}, Judy and {G{\'o}mez-Flechoso}, Mar{\'\i}a A. and {Guzman}, Rafael and {Libeskind}, Noam I. and {Dey}, Arjun and {Weaver}, Benjamin A. and {Schlegel}, David and {Myers}, Adam D. and {Valdes}, Frank G.},
        title = "{Hidden depths in the local Universe: The Stellar Stream Legacy Survey}",
      journal = {\aap},
         year = 2023,
        month = mar,
       volume = {671},
          eid = {A141},
        pages = {A141},
          doi = {10.1051/0004-6361/202245011},
archivePrefix = {arXiv},
       eprint = {2104.06071},
 primaryClass = {astro-ph.GA},
       adsurl = {https://ui.adsabs.harvard.edu/abs/2023A&A...671A.141M}
}

@ARTICLE{Crnojevic2016,
       author = {{Crnojevi{\'c}}, D. and {Sand}, D.~J. and {Spekkens}, K. and {Caldwell}, N. and {Guhathakurta}, P. and {McLeod}, B. and {Seth}, A. and {Simon}, J.~D. and {Strader}, J. and {Toloba}, E.},
        title = "{The Extended Halo of Centaurus A: Uncovering Satellites, Streams, and Substructures}",
      journal = {\apj},
         year = 2016,
        month = may,
       volume = {823},
       number = {1},
          eid = {19},
        pages = {19},
          doi = {10.3847/0004-637X/823/1/19},
archivePrefix = {arXiv},
       eprint = {1512.05366},
 primaryClass = {astro-ph.GA},
       adsurl = {https://ui.adsabs.harvard.edu/abs/2016ApJ...823...19C}
}

@ARTICLE{vanDokkum2019,
       author = {{van Dokkum}, Pieter and {Gilhuly}, Colleen and {Bonaca}, Ana and {Merritt}, Allison and {Danieli}, Shany and {Lokhorst}, Deborah and {Abraham}, Roberto and {Conroy}, Charlie and {Greco}, Johnny P.},
        title = "{Dragonfly Imaging of the Galaxy NGC 5907: A Different View of the Iconic Stellar Stream}",
      journal = {\apjl},
         year = 2019,
        month = oct,
       volume = {883},
       number = {2},
          eid = {L32},
        pages = {L32},
          doi = {10.3847/2041-8213/ab40c9},
archivePrefix = {arXiv},
       eprint = {1906.11260},
 primaryClass = {astro-ph.GA},
       adsurl = {https://ui.adsabs.harvard.edu/abs/2019ApJ...883L..32V}
}

@ARTICLE{Ivezic2019,
       author = {{Ivezi{\'c}}, {\v{Z}}eljko and {Kahn}, Steven M. and {Tyson}, J. Anthony and {Abel}, Bob and {Acosta}, Emily and {Allsman}, Robyn and {Alonso}, David and {AlSayyad}, Yusra and {Anderson}, Scott F. and {Andrew}, John and {Angel}, James Roger P. and {Angeli}, George Z. and {Ansari}, Reza and {Antilogus}, Pierre and {Araujo}, Constanza and {Armstrong}, Robert and {Arndt}, Kirk T. and {Astier}, Pierre and {Aubourg}, {\'E}ric and {Auza}, Nicole and {Axelrod}, Tim S. and {Bard}, Deborah J. and {Barr}, Jeff D. and {Barrau}, Aurelian and {Bartlett}, James G. and {Bauer}, Amanda E. and {Bauman}, Brian J. and {Baumont}, Sylvain and {Bechtol}, Ellen and {Bechtol}, Keith and {Becker}, Andrew C. and {Becla}, Jacek and {Beldica}, Cristina and {Bellavia}, Steve and {Bianco}, Federica B. and {Biswas}, Rahul and {Blanc}, Guillaume and {Blazek}, Jonathan and {Blandford}, Roger D. and {Bloom}, Josh S. and {Bogart}, Joanne and {Bond}, Tim W. and {Booth}, Michael T. and {Borgland}, Anders W. and {Borne}, Kirk and {Bosch}, James F. and {Boutigny}, Dominique and {Brackett}, Craig A. and {Bradshaw}, Andrew and {Brandt}, William Nielsen and {Brown}, Michael E. and {Bullock}, James S. and {Burchat}, Patricia and {Burke}, David L. and {Cagnoli}, Gianpietro and {Calabrese}, Daniel and {Callahan}, Shawn and {Callen}, Alice L. and {Carlin}, Jeffrey L. and {Carlson}, Erin L. and {Chandrasekharan}, Srinivasan and {Charles-Emerson}, Glenaver and {Chesley}, Steve and {Cheu}, Elliott C. and {Chiang}, Hsin-Fang and {Chiang}, James and {Chirino}, Carol and {Chow}, Derek and {Ciardi}, David R. and {Claver}, Charles F. and {Cohen-Tanugi}, Johann and {Cockrum}, Joseph J. and {Coles}, Rebecca and {Connolly}, Andrew J. and {Cook}, Kem H. and {Cooray}, Asantha and {Covey}, Kevin R. and {Cribbs}, Chris and {Cui}, Wei and {Cutri}, Roc and {Daly}, Philip N. and {Daniel}, Scott F. and {Daruich}, Felipe and {Daubard}, Guillaume and {Daues}, Greg and {Dawson}, William and {Delgado}, Francisco and {Dellapenna}, Alfred and {de Peyster}, Robert and {de Val-Borro}, Miguel and {Digel}, Seth W. and {Doherty}, Peter and {Dubois}, Richard and {Dubois-Felsmann}, Gregory P. and {Durech}, Josef and {Economou}, Frossie and {Eifler}, Tim and {Eracleous}, Michael and {Emmons}, Benjamin L. and {Fausti Neto}, Angelo and {Ferguson}, Henry and {Figueroa}, Enrique and {Fisher-Levine}, Merlin and {Focke}, Warren and {Foss}, Michael D. and {Frank}, James and {Freemon}, Michael D. and {Gangler}, Emmanuel and {Gawiser}, Eric and {Geary}, John C. and {Gee}, Perry and {Geha}, Marla and {Gessner}, Charles J.~B. and {Gibson}, Robert R. and {Gilmore}, D. Kirk and {Glanzman}, Thomas and {Glick}, William and {Goldina}, Tatiana and {Goldstein}, Daniel A. and {Goodenow}, Iain and {Graham}, Melissa L. and {Gressler}, William J. and {Gris}, Philippe and {Guy}, Leanne P. and {Guyonnet}, Augustin and {Haller}, Gunther and {Harris}, Ron and {Hascall}, Patrick A. and {Haupt}, Justine and {Hernandez}, Fabio and {Herrmann}, Sven and {Hileman}, Edward and {Hoblitt}, Joshua and {Hodgson}, John A. and {Hogan}, Craig and {Howard}, James D. and {Huang}, Dajun and {Huffer}, Michael E. and {Ingraham}, Patrick and {Innes}, Walter R. and {Jacoby}, Suzanne H. and {Jain}, Bhuvnesh and {Jammes}, Fabrice and {Jee}, M. James and {Jenness}, Tim and {Jernigan}, Garrett and {Jevremovi{\'c}}, Darko and {Johns}, Kenneth and {Johnson}, Anthony S. and {Johnson}, Margaret W.~G. and {Jones}, R. Lynne and {Juramy-Gilles}, Claire and {Juri{\'c}}, Mario and {Kalirai}, Jason S. and {Kallivayalil}, Nitya J. and {Kalmbach}, Bryce and {Kantor}, Jeffrey P. and {Karst}, Pierre and {Kasliwal}, Mansi M. and {Kelly}, Heather and {Kessler}, Richard and {Kinnison}, Veronica and {Kirkby}, David and {Knox}, Lloyd and {Kotov}, Ivan V. and {Krabbendam}, Victor L. and {Krughoff}, K. Simon and {Kub{\'a}nek}, Petr and {Kuczewski}, John and {Kulkarni}, Shri and {Ku}, John and {Kurita}, Nadine R. and {Lage}, Craig S. and {Lambert}, Ron and {Lange}, Travis and {Langton}, J. Brian and {Le Guillou}, Laurent and {Levine}, Deborah and {Liang}, Ming and {Lim}, Kian-Tat and {Lintott}, Chris J. and {Long}, Kevin E. and {Lopez}, Margaux and {Lotz}, Paul J. and {Lupton}, Robert H. and {Lust}, Nate B. and {MacArthur}, Lauren A. and {Mahabal}, Ashish and {Mandelbaum}, Rachel and {Markiewicz}, Thomas W. and {Marsh}, Darren S. and {Marshall}, Philip J. and {Marshall}, Stuart and {May}, Morgan and {McKercher}, Robert and {McQueen}, Michelle and {Meyers}, Joshua and {Migliore}, Myriam and {Miller}, Michelle and {Mills}, David J.},
        title = "{LSST: From Science Drivers to Reference Design and Anticipated Data Products}",
      journal = {\apj},
         year = 2019,
        month = mar,
       volume = {873},
       number = {2},
          eid = {111},
        pages = {111},
          doi = {10.3847/1538-4357/ab042c},
archivePrefix = {arXiv},
       eprint = {0805.2366},
 primaryClass = {astro-ph},
       adsurl = {https://ui.adsabs.harvard.edu/abs/2019ApJ...873..111I}
}

@ARTICLE{Laureijs2011,
       author = {{Laureijs}, R. and {Amiaux}, J. and {Arduini}, S. and {Augu{\`e}res}, J. -L. and {Brinchmann}, J. and {Cole}, R. and {Cropper}, M. and {Dabin}, C. and {Duvet}, L. and {Ealet}, A. and {Garilli}, B. and {Gondoin}, P. and {Guzzo}, L. and {Hoar}, J. and {Hoekstra}, H. and {Holmes}, R. and {Kitching}, T. and {Maciaszek}, T. and {Mellier}, Y. and {Pasian}, F. and {Percival}, W. and {Rhodes}, J. and {Saavedra Criado}, G. and {Sauvage}, M. and {Scaramella}, R. and {Valenziano}, L. and {Warren}, S. and {Bender}, R. and {Castander}, F. and {Cimatti}, A. and {Le F{\`e}vre}, O. and {Kurki-Suonio}, H. and {Levi}, M. and {Lilje}, P. and {Meylan}, G. and {Nichol}, R. and {Pedersen}, K. and {Popa}, V. and {Rebolo Lopez}, R. and {Rix}, H. -W. and {Rottgering}, H. and {Zeilinger}, W. and {Grupp}, F. and {Hudelot}, P. and {Massey}, R. and {Meneghetti}, M. and {Miller}, L. and {Paltani}, S. and {Paulin-Henriksson}, S. and {Pires}, S. and {Saxton}, C. and {Schrabback}, T. and {Seidel}, G. and {Walsh}, J. and {Aghanim}, N. and {Amendola}, L. and {Bartlett}, J. and {Baccigalupi}, C. and {Beaulieu}, J. -P. and {Benabed}, K. and {Cuby}, J. -G. and {Elbaz}, D. and {Fosalba}, P. and {Gavazzi}, G. and {Helmi}, A. and {Hook}, I. and {Irwin}, M. and {Kneib}, J. -P. and {Kunz}, M. and {Mannucci}, F. and {Moscardini}, L. and {Tao}, C. and {Teyssier}, R. and {Weller}, J. and {Zamorani}, G. and {Zapatero Osorio}, M.~R. and {Boulade}, O. and {Foumond}, J.~J. and {Di Giorgio}, A. and {Guttridge}, P. and {James}, A. and {Kemp}, M. and {Martignac}, J. and {Spencer}, A. and {Walton}, D. and {Bl{\"u}mchen}, T. and {Bonoli}, C. and {Bortoletto}, F. and {Cerna}, C. and {Corcione}, L. and {Fabron}, C. and {Jahnke}, K. and {Ligori}, S. and {Madrid}, F. and {Martin}, L. and {Morgante}, G. and {Pamplona}, T. and {Prieto}, E. and {Riva}, M. and {Toledo}, R. and {Trifoglio}, M. and {Zerbi}, F. and {Abdalla}, F. and {Douspis}, M. and {Grenet}, C. and {Borgani}, S. and {Bouwens}, R. and {Courbin}, F. and {Delouis}, J. -M. and {Dubath}, P. and {Fontana}, A. and {Frailis}, M. and {Grazian}, A. and {Koppenh{\"o}fer}, J. and {Mansutti}, O. and {Melchior}, M. and {Mignoli}, M. and {Mohr}, J. and {Neissner}, C. and {Noddle}, K. and {Poncet}, M. and {Scodeggio}, M. and {Serrano}, S. and {Shane}, N. and {Starck}, J. -L. and {Surace}, C. and {Taylor}, A. and {Verdoes-Kleijn}, G. and {Vuerli}, C. and {Williams}, O.~R. and {Zacchei}, A. and {Altieri}, B. and {Escudero Sanz}, I. and {Kohley}, R. and {Oosterbroek}, T. and {Astier}, P. and {Bacon}, D. and {Bardelli}, S. and {Baugh}, C. and {Bellagamba}, F. and {Benoist}, C. and {Bianchi}, D. and {Biviano}, A. and {Branchini}, E. and {Carbone}, C. and {Cardone}, V. and {Clements}, D. and {Colombi}, S. and {Conselice}, C. and {Cresci}, G. and {Deacon}, N. and {Dunlop}, J. and {Fedeli}, C. and {Fontanot}, F. and {Franzetti}, P. and {Giocoli}, C. and {Garcia-Bellido}, J. and {Gow}, J. and {Heavens}, A. and {Hewett}, P. and {Heymans}, C. and {Holland}, A. and {Huang}, Z. and {Ilbert}, O. and {Joachimi}, B. and {Jennins}, E. and {Kerins}, E. and {Kiessling}, A. and {Kirk}, D. and {Kotak}, R. and {Krause}, O. and {Lahav}, O. and {van Leeuwen}, F. and {Lesgourgues}, J. and {Lombardi}, M. and {Magliocchetti}, M. and {Maguire}, K. and {Majerotto}, E. and {Maoli}, R. and {Marulli}, F. and {Maurogordato}, S. and {McCracken}, H. and {McLure}, R. and {Melchiorri}, A. and {Merson}, A. and {Moresco}, M. and {Nonino}, M. and {Norberg}, P. and {Peacock}, J. and {Pello}, R. and {Penny}, M. and {Pettorino}, V. and {Di Porto}, C. and {Pozzetti}, L. and {Quercellini}, C. and {Radovich}, M. and {Rassat}, A. and {Roche}, N. and {Ronayette}, S. and {Rossetti}, E.},
        title = "{Euclid Definition Study Report}",
      journal = {arXiv e-prints},
         year = 2011,
        month = oct,
          eid = {arXiv:1110.3193},
        pages = {arXiv:1110.3193},
          doi = {10.48550/arXiv.1110.3193},
archivePrefix = {arXiv},
       eprint = {1110.3193},
 primaryClass = {astro-ph.CO},
       adsurl = {https://ui.adsabs.harvard.edu/abs/2011arXiv1110.3193L}
}

@ARTICLE{Vasiliev2019,
       author = {{Vasiliev}, Eugene},
        title = "{AGAMA: action-based galaxy modelling architecture}",
      journal = {\mnras},
         year = 2019,
        month = jan,
       volume = {482},
       number = {2},
        pages = {1525-1544},
          doi = {10.1093/mnras/sty2672},
archivePrefix = {arXiv},
       eprint = {1802.08239},
 primaryClass = {astro-ph.GA},
       adsurl = {https://ui.adsabs.harvard.edu/abs/2019MNRAS.482.1525V}
}

@ARTICLE{Fardal2015,
       author = {{Fardal}, Mark A. and {Huang}, Shuiyao and {Weinberg}, Martin D.},
        title = "{Generation of mock tidal streams}",
      journal = {\mnras},
         year = 2015,
        month = sep,
       volume = {452},
       number = {1},
        pages = {301-319},
          doi = {10.1093/mnras/stv1198},
archivePrefix = {arXiv},
       eprint = {1410.1861},
 primaryClass = {astro-ph.GA},
       adsurl = {https://ui.adsabs.harvard.edu/abs/2015MNRAS.452..301F}
}

@BOOK{Hockney1988,
       author = {{Hockney}, R.~W. and {Eastwood}, J.~W.},
        title = "{Computer simulation using particles}",
         year = 1988,
       adsurl = {https://ui.adsabs.harvard.edu/abs/1988csup.book.....H}
}

@ARTICLE{Bayes1763,
       author = {{Bayes}, Mr. and {Price}, Mr.},
        title = "{An Essay towards Solving a Problem in the Doctrine of Chances. By the Late Rev. Mr. Bayes, F. R. S. Communicated by Mr. Price, in a Letter to John Canton, A. M. F. R. S.}",
      journal = {Philosophical Transactions of the Royal Society of London Series I},
         year = 1763,
        month = jan,
       volume = {53},
        pages = {370-418},
       adsurl = {https://ui.adsabs.harvard.edu/abs/1763RSPT...53..370B}
}

@ARTICLE{Bullock2005,
       author = {{Bullock}, James S. and {Johnston}, Kathryn V.},
        title = "{Tracing Galaxy Formation with Stellar Halos. I. Methods}",
      journal = {\apj},
         year = 2005,
        month = dec,
       volume = {635},
       number = {2},
        pages = {931-949},
          doi = {10.1086/497422},
archivePrefix = {arXiv},
       eprint = {astro-ph/0506467},
 primaryClass = {astro-ph},
       adsurl = {https://ui.adsabs.harvard.edu/abs/2005ApJ...635..931B}
}

@ARTICLE{Ibata2001b,
       author = {{Ibata}, Rodrigo and {Lewis}, Geraint F. and {Irwin}, Michael and {Totten}, Edward and {Quinn}, Thomas},
        title = "{Great Circle Tidal Streams: Evidence for a Nearly Spherical Massive Dark Halo around the Milky Way}",
      journal = {\apj},
         year = 2001,
        month = apr,
       volume = {551},
       number = {1},
        pages = {294-311},
          doi = {10.1086/320060},
archivePrefix = {arXiv},
       eprint = {astro-ph/0004011},
 primaryClass = {astro-ph},
       adsurl = {https://ui.adsabs.harvard.edu/abs/2001ApJ...551..294I}
}

@ARTICLE{Johnston2001,
       author = {{Johnston}, Kathryn V. and {Sackett}, Penny D. and {Bullock}, James S.},
        title = "{Interpreting Debris from Satellite Disruption in External Galaxies}",
      journal = {\apj},
         year = 2001,
        month = aug,
       volume = {557},
       number = {1},
        pages = {137-149},
          doi = {10.1086/321644},
archivePrefix = {arXiv},
       eprint = {astro-ph/0101543},
 primaryClass = {astro-ph},
       adsurl = {https://ui.adsabs.harvard.edu/abs/2001ApJ...557..137J}
}

@ARTICLE{Carlberg2012,
       author = {{Carlberg}, R.~G. and {Grillmair}, C.~J. and {Hetherington}, Nathan},
        title = "{The Pal 5 Star Stream Gaps}",
      journal = {\apj},
         year = 2012,
        month = nov,
       volume = {760},
       number = {1},
          eid = {75},
        pages = {75},
          doi = {10.1088/0004-637X/760/1/75},
archivePrefix = {arXiv},
       eprint = {1209.1741},
 primaryClass = {astro-ph.CO},
       adsurl = {https://ui.adsabs.harvard.edu/abs/2012ApJ...760...75C}
}

@ARTICLE{Banik2019,
       author = {{Banik}, Nilanjan and {Bovy}, Jo},
        title = "{Effects of baryonic and dark matter substructure on the Pal 5 stream}",
      journal = {\mnras},
         year = 2019,
        month = apr,
       volume = {484},
       number = {2},
        pages = {2009-2020},
          doi = {10.1093/mnras/stz142},
archivePrefix = {arXiv},
       eprint = {1809.09640},
 primaryClass = {astro-ph.GA},
       adsurl = {https://ui.adsabs.harvard.edu/abs/2019MNRAS.484.2009B}
}

@ARTICLE{GaiaCollaboration2018,
       author = {{Gaia Collaboration} and {Brown}, A.~G.~A. and {Vallenari}, A. and {Prusti}, T. and {de Bruijne}, J.~H.~J. and {Babusiaux}, C. and {Bailer-Jones}, C.~A.~L. and {Biermann}, M. and {Evans}, D.~W. and {Eyer}, L. and {Jansen}, F. and {Jordi}, C. and {Klioner}, S.~A. and {Lammers}, U. and {Lindegren}, L. and {Luri}, X. and {Mignard}, F. and {Panem}, C. and {Pourbaix}, D. and {Randich}, S. and {Sartoretti}, P. and {Siddiqui}, H.~I. and {Soubiran}, C. and {van Leeuwen}, F. and {Walton}, N.~A. and {Arenou}, F. and {Bastian}, U. and {Cropper}, M. and {Drimmel}, R. and {Katz}, D. and {Lattanzi}, M.~G. and {Bakker}, J. and {Cacciari}, C. and {Casta{\~n}eda}, J. and {Chaoul}, L. and {Cheek}, N. and {De Angeli}, F. and {Fabricius}, C. and {Guerra}, R. and {Holl}, B. and {Masana}, E. and {Messineo}, R. and {Mowlavi}, N. and {Nienartowicz}, K. and {Panuzzo}, P. and {Portell}, J. and {Riello}, M. and {Seabroke}, G.~M. and {Tanga}, P. and {Th{\'e}venin}, F. and {Gracia-Abril}, G. and {Comoretto}, G. and {Garcia-Reinaldos}, M. and {Teyssier}, D. and {Altmann}, M. and {Andrae}, R. and {Audard}, M. and {Bellas-Velidis}, I. and {Benson}, K. and {Berthier}, J. and {Blomme}, R. and {Burgess}, P. and {Busso}, G. and {Carry}, B. and {Cellino}, A. and {Clementini}, G. and {Clotet}, M. and {Creevey}, O. and {Davidson}, M. and {De Ridder}, J. and {Delchambre}, L. and {Dell'Oro}, A. and {Ducourant}, C. and {Fern{\'a}ndez-Hern{\'a}ndez}, J. and {Fouesneau}, M. and {Fr{\'e}mat}, Y. and {Galluccio}, L. and {Garc{\'\i}a-Torres}, M. and {Gonz{\'a}lez-N{\'u}{\~n}ez}, J. and {Gonz{\'a}lez-Vidal}, J.~J. and {Gosset}, E. and {Guy}, L.~P. and {Halbwachs}, J. -L. and {Hambly}, N.~C. and {Harrison}, D.~L. and {Hern{\'a}ndez}, J. and {Hestroffer}, D. and {Hodgkin}, S.~T. and {Hutton}, A. and {Jasniewicz}, G. and {Jean-Antoine-Piccolo}, A. and {Jordan}, S. and {Korn}, A.~J. and {Krone-Martins}, A. and {Lanzafame}, A.~C. and {Lebzelter}, T. and {L{\"o}ffler}, W. and {Manteiga}, M. and {Marrese}, P.~M. and {Mart{\'\i}n-Fleitas}, J.~M. and {Moitinho}, A. and {Mora}, A. and {Muinonen}, K. and {Osinde}, J. and {Pancino}, E. and {Pauwels}, T. and {Petit}, J. -M. and {Recio-Blanco}, A. and {Richards}, P.~J. and {Rimoldini}, L. and {Robin}, A.~C. and {Sarro}, L.~M. and {Siopis}, C. and {Smith}, M. and {Sozzetti}, A. and {S{\"u}veges}, M. and {Torra}, J. and {van Reeven}, W. and {Abbas}, U. and {Abreu Aramburu}, A. and {Accart}, S. and {Aerts}, C. and {Altavilla}, G. and {{\'A}lvarez}, M.~A. and {Alvarez}, R. and {Alves}, J. and {Anderson}, R.~I. and {Andrei}, A.~H. and {Anglada Varela}, E. and {Antiche}, E. and {Antoja}, T. and {Arcay}, B. and {Astraatmadja}, T.~L. and {Bach}, N. and {Baker}, S.~G. and {Balaguer-N{\'u}{\~n}ez}, L. and {Balm}, P. and {Barache}, C. and {Barata}, C. and {Barbato}, D. and {Barblan}, F. and {Barklem}, P.~S. and {Barrado}, D. and {Barros}, M. and {Barstow}, M.~A. and {Bartholom{\'e} Mu{\~n}oz}, S. and {Bassilana}, J. -L. and {Becciani}, U. and {Bellazzini}, M. and {Berihuete}, A. and {Bertone}, S. and {Bianchi}, L. and {Bienaym{\'e}}, O. and {Blanco-Cuaresma}, S. and {Boch}, T. and {Boeche}, C. and {Bombrun}, A. and {Borrachero}, R. and {Bossini}, D. and {Bouquillon}, S. and {Bourda}, G. and {Bragaglia}, A. and {Bramante}, L. and {Breddels}, M.~A. and {Bressan}, A. and {Brouillet}, N. and {Br{\"u}semeister}, T. and {Brugaletta}, E. and {Bucciarelli}, B. and {Burlacu}, A. and {Busonero}, D. and {Butkevich}, A.~G. and {Buzzi}, R. and {Caffau}, E. and {Cancelliere}, R. and {Cannizzaro}, G. and {Cantat-Gaudin}, T. and {Carballo}, R. and {Carlucci}, T. and {Carrasco}, J.~M. and {Casamiquela}, L. and {Castellani}, M. and {Castro-Ginard}, A. and {Charlot}, P. and {Chemin}, L. and {Chiavassa}, A. and {Cocozza}, G. and {Costigan}, G. and {Cowell}, S. and {Crifo}, F. and {Crosta}, M. and {Crowley}, C. and {Cuypers}, J. and {Dafonte}, C. and {Damerdji}, Y. and {Dapergolas}, A. and {David}, P. and {David}, M. and {de Laverny}, P. and {De Luise}, F.},
        title = "{Gaia Data Release 2. Summary of the contents and survey properties}",
      journal = {\aap},
         year = 2018,
        month = aug,
       volume = {616},
          eid = {A1},
        pages = {A1},
          doi = {10.1051/0004-6361/201833051},
archivePrefix = {arXiv},
       eprint = {1804.09365},
 primaryClass = {astro-ph.GA},
       adsurl = {https://ui.adsabs.harvard.edu/abs/2018A&A...616A...1G}
}

@ARTICLE{GaiaCollaboration2021,
       author = {{Gaia Collaboration} and {Brown}, A.~G.~A. and {Vallenari}, A. and {Prusti}, T. and {de Bruijne}, J.~H.~J. and {Babusiaux}, C. and {Biermann}, M. and {Creevey}, O.~L. and {Evans}, D.~W. and {Eyer}, L. and {Hutton}, A. and {Jansen}, F. and {Jordi}, C. and {Klioner}, S.~A. and {Lammers}, U. and {Lindegren}, L. and {Luri}, X. and {Mignard}, F. and {Panem}, C. and {Pourbaix}, D. and {Randich}, S. and {Sartoretti}, P. and {Soubiran}, C. and {Walton}, N.~A. and {Arenou}, F. and {Bailer-Jones}, C.~A.~L. and {Bastian}, U. and {Cropper}, M. and {Drimmel}, R. and {Katz}, D. and {Lattanzi}, M.~G. and {van Leeuwen}, F. and {Bakker}, J. and {Cacciari}, C. and {Casta{\~n}eda}, J. and {De Angeli}, F. and {Ducourant}, C. and {Fabricius}, C. and {Fouesneau}, M. and {Fr{\'e}mat}, Y. and {Guerra}, R. and {Guerrier}, A. and {Guiraud}, J. and {Jean-Antoine Piccolo}, A. and {Masana}, E. and {Messineo}, R. and {Mowlavi}, N. and {Nicolas}, C. and {Nienartowicz}, K. and {Pailler}, F. and {Panuzzo}, P. and {Riclet}, F. and {Roux}, W. and {Seabroke}, G.~M. and {Sordo}, R. and {Tanga}, P. and {Th{\'e}venin}, F. and {Gracia-Abril}, G. and {Portell}, J. and {Teyssier}, D. and {Altmann}, M. and {Andrae}, R. and {Bellas-Velidis}, I. and {Benson}, K. and {Berthier}, J. and {Blomme}, R. and {Brugaletta}, E. and {Burgess}, P.~W. and {Busso}, G. and {Carry}, B. and {Cellino}, A. and {Cheek}, N. and {Clementini}, G. and {Damerdji}, Y. and {Davidson}, M. and {Delchambre}, L. and {Dell'Oro}, A. and {Fern{\'a}ndez-Hern{\'a}ndez}, J. and {Galluccio}, L. and {Garc{\'\i}a-Lario}, P. and {Garcia-Reinaldos}, M. and {Gonz{\'a}lez-N{\'u}{\~n}ez}, J. and {Gosset}, E. and {Haigron}, R. and {Halbwachs}, J. -L. and {Hambly}, N.~C. and {Harrison}, D.~L. and {Hatzidimitriou}, D. and {Heiter}, U. and {Hern{\'a}ndez}, J. and {Hestroffer}, D. and {Hodgkin}, S.~T. and {Holl}, B. and {Jan{\ss}en}, K. and {Jevardat de Fombelle}, G. and {Jordan}, S. and {Krone-Martins}, A. and {Lanzafame}, A.~C. and {L{\"o}ffler}, W. and {Lorca}, A. and {Manteiga}, M. and {Marchal}, O. and {Marrese}, P.~M. and {Moitinho}, A. and {Mora}, A. and {Muinonen}, K. and {Osborne}, P. and {Pancino}, E. and {Pauwels}, T. and {Petit}, J. -M. and {Recio-Blanco}, A. and {Richards}, P.~J. and {Riello}, M. and {Rimoldini}, L. and {Robin}, A.~C. and {Roegiers}, T. and {Rybizki}, J. and {Sarro}, L.~M. and {Siopis}, C. and {Smith}, M. and {Sozzetti}, A. and {Ulla}, A. and {Utrilla}, E. and {van Leeuwen}, M. and {van Reeven}, W. and {Abbas}, U. and {Abreu Aramburu}, A. and {Accart}, S. and {Aerts}, C. and {Aguado}, J.~J. and {Ajaj}, M. and {Altavilla}, G. and {{\'A}lvarez}, M.~A. and {{\'A}lvarez Cid-Fuentes}, J. and {Alves}, J. and {Anderson}, R.~I. and {Anglada Varela}, E. and {Antoja}, T. and {Audard}, M. and {Baines}, D. and {Baker}, S.~G. and {Balaguer-N{\'u}{\~n}ez}, L. and {Balbinot}, E. and {Balog}, Z. and {Barache}, C. and {Barbato}, D. and {Barros}, M. and {Barstow}, M.~A. and {Bartolom{\'e}}, S. and {Bassilana}, J. -L. and {Bauchet}, N. and {Baudesson-Stella}, A. and {Becciani}, U. and {Bellazzini}, M. and {Bernet}, M. and {Bertone}, S. and {Bianchi}, L. and {Blanco-Cuaresma}, S. and {Boch}, T. and {Bombrun}, A. and {Bossini}, D. and {Bouquillon}, S. and {Bragaglia}, A. and {Bramante}, L. and {Breedt}, E. and {Bressan}, A. and {Brouillet}, N. and {Bucciarelli}, B. and {Burlacu}, A. and {Busonero}, D. and {Butkevich}, A.~G. and {Buzzi}, R. and {Caffau}, E. and {Cancelliere}, R. and {C{\'a}novas}, H. and {Cantat-Gaudin}, T. and {Carballo}, R. and {Carlucci}, T. and {Carnerero}, M.~I. and {Carrasco}, J.~M. and {Casamiquela}, L. and {Castellani}, M. and {Castro-Ginard}, A. and {Castro Sampol}, P. and {Chaoul}, L. and {Charlot}, P. and {Chemin}, L. and {Chiavassa}, A. and {Cioni}, M. -R.~L. and {Comoretto}, G. and {Cooper}, W.~J. and {Cornez}, T. and {Cowell}, S. and {Crifo}, F. and {Crosta}, M. and {Crowley}, C. and {Dafonte}, C. and {Dapergolas}, A. and {David}, M. and {David}, P.},
        title = "{Gaia Early Data Release 3. Summary of the contents and survey properties}",
      journal = {\aap},
         year = 2021,
        month = may,
       volume = {649},
          eid = {A1},
        pages = {A1},
          doi = {10.1051/0004-6361/202039657},
archivePrefix = {arXiv},
       eprint = {2012.01533},
 primaryClass = {astro-ph.GA},
       adsurl = {https://ui.adsabs.harvard.edu/abs/2021A&A...649A...1G}
}

@ARTICLE{York2000
,
       author = {{York}, Donald G. and {Adelman}, J. and {Anderson}, Jr., John E. and {Anderson}, Scott F. and {Annis}, James and {Bahcall}, Neta A. and {Bakken}, J.~A. and {Barkhouser}, Robert and {Bastian}, Steven and {Berman}, Eileen and {Boroski}, William N. and {Bracker}, Steve and {Briegel}, Charlie and {Briggs}, John W. and {Brinkmann}, J. and {Brunner}, Robert and {Burles}, Scott and {Carey}, Larry and {Carr}, Michael A. and {Castander}, Francisco J. and {Chen}, Bing and {Colestock}, Patrick L. and {Connolly}, A.~J. and {Crocker}, J.~H. and {Csabai}, Istv{\'a}n and {Czarapata}, Paul C. and {Davis}, John Eric and {Doi}, Mamoru and {Dombeck}, Tom and {Eisenstein}, Daniel and {Ellman}, Nancy and {Elms}, Brian R. and {Evans}, Michael L. and {Fan}, Xiaohui and {Federwitz}, Glenn R. and {Fiscelli}, Larry and {Friedman}, Scott and {Frieman}, Joshua A. and {Fukugita}, Masataka and {Gillespie}, Bruce and {Gunn}, James E. and {Gurbani}, Vijay K. and {de Haas}, Ernst and {Haldeman}, Merle and {Harris}, Frederick H. and {Hayes}, J. and {Heckman}, Timothy M. and {Hennessy}, G.~S. and {Hindsley}, Robert B. and {Holm}, Scott and {Holmgren}, Donald J. and {Huang}, Chi-hao and {Hull}, Charles and {Husby}, Don and {Ichikawa}, Shin-Ichi and {Ichikawa}, Takashi and {Ivezi{\'c}}, {\v{Z}}eljko and {Kent}, Stephen and {Kim}, Rita S.~J. and {Kinney}, E. and {Klaene}, Mark and {Kleinman}, A.~N. and {Kleinman}, S. and {Knapp}, G.~R. and {Korienek}, John and {Kron}, Richard G. and {Kunszt}, Peter Z. and {Lamb}, D.~Q. and {Lee}, B. and {Leger}, R. French and {Limmongkol}, Siriluk and {Lindenmeyer}, Carl and {Long}, Daniel C. and {Loomis}, Craig and {Loveday}, Jon and {Lucinio}, Rich and {Lupton}, Robert H. and {MacKinnon}, Bryan and {Mannery}, Edward J. and {Mantsch}, P.~M. and {Margon}, Bruce and {McGehee}, Peregrine and {McKay}, Timothy A. and {Meiksin}, Avery and {Merelli}, Aronne and {Monet}, David G. and {Munn}, Jeffrey A. and {Narayanan}, Vijay K. and {Nash}, Thomas and {Neilsen}, Eric and {Neswold}, Rich and {Newberg}, Heidi Jo and {Nichol}, R.~C. and {Nicinski}, Tom and {Nonino}, Mario and {Okada}, Norio and {Okamura}, Sadanori and {Ostriker}, Jeremiah P. and {Owen}, Russell and {Pauls}, A. George and {Peoples}, John and {Peterson}, R.~L. and {Petravick}, Donald and {Pier}, Jeffrey R. and {Pope}, Adrian and {Pordes}, Ruth and {Prosapio}, Angela and {Rechenmacher}, Ron and {Quinn}, Thomas R. and {Richards}, Gordon T. and {Richmond}, Michael W. and {Rivetta}, Claudio H. and {Rockosi}, Constance M. and {Ruthmansdorfer}, Kurt and {Sandford}, Dale and {Schlegel}, David J. and {Schneider}, Donald P. and {Sekiguchi}, Maki and {Sergey}, Gary and {Shimasaku}, Kazuhiro and {Siegmund}, Walter A. and {Smee}, Stephen and {Smith}, J. Allyn and {Snedden}, S. and {Stone}, R. and {Stoughton}, Chris and {Strauss}, Michael A. and {Stubbs}, Christopher and {SubbaRao}, Mark and {Szalay}, Alexander S. and {Szapudi}, Istvan and {Szokoly}, Gyula P. and {Thakar}, Anirudda R. and {Tremonti}, Christy and {Tucker}, Douglas L. and {Uomoto}, Alan and {Vanden Berk}, Dan and {Vogeley}, Michael S. and {Waddell}, Patrick and {Wang}, Shu-i. and {Watanabe}, Masaru and {Weinberg}, David H. and {Yanny}, Brian and {Yasuda}, Naoki and {SDSS Collaboration}},
        title = "{The Sloan Digital Sky Survey: Technical Summary}",
      journal = {\aj},
         year = 2000,
        month = sep,
       volume = {120},
       number = {3},
        pages = {1579-1587},
          doi = {10.1086/301513},
archivePrefix = {arXiv},
       eprint = {astro-ph/0006396},
 primaryClass = {astro-ph},
       adsurl = {https://ui.adsabs.harvard.edu/abs/2000AJ....120.1579Y}
}

@ARTICLE{DESCollaboration2018,
       author = {{Abbott}, T.~M.~C. and {Abdalla}, F.~B. and {Allam}, S. and {Amara}, A. and {Annis}, J. and {Asorey}, J. and {Avila}, S. and {Ballester}, O. and {Banerji}, M. and {Barkhouse}, W. and {Baruah}, L. and {Baumer}, M. and {Bechtol}, K. and {Becker}, M.~R. and {Benoit-L{\'e}vy}, A. and {Bernstein}, G.~M. and {Bertin}, E. and {Blazek}, J. and {Bocquet}, S. and {Brooks}, D. and {Brout}, D. and {Buckley-Geer}, E. and {Burke}, D.~L. and {Busti}, V. and {Campisano}, R. and {Cardiel-Sas}, L. and {Carnero Rosell}, A. and {Carrasco Kind}, M. and {Carretero}, J. and {Castander}, F.~J. and {Cawthon}, R. and {Chang}, C. and {Chen}, X. and {Conselice}, C. and {Costa}, G. and {Crocce}, M. and {Cunha}, C.~E. and {D'Andrea}, C.~B. and {da Costa}, L.~N. and {Das}, R. and {Daues}, G. and {Davis}, T.~M. and {Davis}, C. and {De Vicente}, J. and {DePoy}, D.~L. and {DeRose}, J. and {Desai}, S. and {Diehl}, H.~T. and {Dietrich}, J.~P. and {Dodelson}, S. and {Doel}, P. and {Drlica-Wagner}, A. and {Eifler}, T.~F. and {Elliott}, A.~E. and {Evrard}, A.~E. and {Farahi}, A. and {Fausti Neto}, A. and {Fernandez}, E. and {Finley}, D.~A. and {Flaugher}, B. and {Foley}, R.~J. and {Fosalba}, P. and {Friedel}, D.~N. and {Frieman}, J. and {Garc{\'\i}a-Bellido}, J. and {Gaztanaga}, E. and {Gerdes}, D.~W. and {Giannantonio}, T. and {Gill}, M.~S.~S. and {Glazebrook}, K. and {Goldstein}, D.~A. and {Gower}, M. and {Gruen}, D. and {Gruendl}, R.~A. and {Gschwend}, J. and {Gupta}, R.~R. and {Gutierrez}, G. and {Hamilton}, S. and {Hartley}, W.~G. and {Hinton}, S.~R. and {Hislop}, J.~M. and {Hollowood}, D. and {Honscheid}, K. and {Hoyle}, B. and {Huterer}, D. and {Jain}, B. and {James}, D.~J. and {Jeltema}, T. and {Johnson}, M.~W.~G. and {Johnson}, M.~D. and {Kacprzak}, T. and {Kent}, S. and {Khullar}, G. and {Klein}, M. and {Kovacs}, A. and {Koziol}, A.~M.~G. and {Krause}, E. and {Kremin}, A. and {Kron}, R. and {Kuehn}, K. and {Kuhlmann}, S. and {Kuropatkin}, N. and {Lahav}, O. and {Lasker}, J. and {Li}, T.~S. and {Li}, R.~T. and {Liddle}, A.~R. and {Lima}, M. and {Lin}, H. and {L{\'o}pez-Reyes}, P. and {MacCrann}, N. and {Maia}, M.~A.~G. and {Maloney}, J.~D. and {Manera}, M. and {March}, M. and {Marriner}, J. and {Marshall}, J.~L. and {Martini}, P. and {McClintock}, T. and {McKay}, T. and {McMahon}, R.~G. and {Melchior}, P. and {Menanteau}, F. and {Miller}, C.~J. and {Miquel}, R. and {Mohr}, J.~J. and {Morganson}, E. and {Mould}, J. and {Neilsen}, E. and {Nichol}, R.~C. and {Nogueira}, F. and {Nord}, B. and {Nugent}, P. and {Nunes}, L. and {Ogando}, R.~L.~C. and {Old}, L. and {Pace}, A.~B. and {Palmese}, A. and {Paz-Chinch{\'o}n}, F. and {Peiris}, H.~V. and {Percival}, W.~J. and {Petravick}, D. and {Plazas}, A.~A. and {Poh}, J. and {Pond}, C. and {Porredon}, A. and {Pujol}, A. and {Refregier}, A. and {Reil}, K. and {Ricker}, P.~M. and {Rollins}, R.~P. and {Romer}, A.~K. and {Roodman}, A. and {Rooney}, P. and {Ross}, A.~J. and {Rykoff}, E.~S. and {Sako}, M. and {Sanchez}, M.~L. and {Sanchez}, E. and {Santiago}, B. and {Saro}, A. and {Scarpine}, V. and {Scolnic}, D. and {Serrano}, S. and {Sevilla-Noarbe}, I. and {Sheldon}, E. and {Shipp}, N. and {Silveira}, M.~L. and {Smith}, M. and {Smith}, R.~C. and {Smith}, J.~A. and {Soares-Santos}, M. and {Sobreira}, F. and {Song}, J. and {Stebbins}, A. and {Suchyta}, E. and {Sullivan}, M. and {Swanson}, M.~E.~C. and {Tarle}, G. and {Thaler}, J. and {Thomas}, D. and {Thomas}, R.~C. and {Troxel}, M.~A. and {Tucker}, D.~L. and {Vikram}, V. and {Vivas}, A.~K. and {Walker}, A.~R. and {Wechsler}, R.~H. and {Weller}, J. and {Wester}, W. and {Wolf}, R.~C. and {Wu}, H. and {Yanny}, B. and {Zenteno}, A. and {Zhang}, Y. and {Zuntz}, J. and {DES Collaboration} and {Juneau}, S. and {Fitzpatrick}, M. and {Nikutta}, R.},
        title = "{The Dark Energy Survey: Data Release 1}",
      journal = {\apjs},
         year = 2018,
        month = dec,
       volume = {239},
       number = {2},
          eid = {18},
        pages = {18},
          doi = {10.3847/1538-4365/aae9f0},
archivePrefix = {arXiv},
       eprint = {1801.03181},
 primaryClass = {astro-ph.IM},
       adsurl = {https://ui.adsabs.harvard.edu/abs/2018ApJS..239...18A}
}

@ARTICLE{Cui2012,
       author = {{Cui}, Xiang-Qun and {Zhao}, Yong-Heng and {Chu}, Yao-Quan and {Li}, Guo-Ping and {Li}, Qi and {Zhang}, Li-Ping and {Su}, Hong-Jun and {Yao}, Zheng-Qiu and {Wang}, Ya-Nan and {Xing}, Xiao-Zheng and {Li}, Xin-Nan and {Zhu}, Yong-Tian and {Wang}, Gang and {Gu}, Bo-Zhong and {Luo}, A. -Li and {Xu}, Xin-Qi and {Zhang}, Zhen-Chao and {Liu}, Gen-Rong and {Zhang}, Hao-Tong and {Yang}, De-Hua and {Cao}, Shu-Yun and {Chen}, Hai-Yuan and {Chen}, Jian-Jun and {Chen}, Kun-Xin and {Chen}, Ying and {Chu}, Jia-Ru and {Feng}, Lei and {Gong}, Xue-Fei and {Hou}, Yong-Hui and {Hu}, Hong-Zhuan and {Hu}, Ning-Sheng and {Hu}, Zhong-Wen and {Jia}, Lei and {Jiang}, Fang-Hua and {Jiang}, Xiang and {Jiang}, Zi-Bo and {Jin}, Ge and {Li}, Ai-Hua and {Li}, Yan and {Li}, Ye-Ping and {Liu}, Guan-Qun and {Liu}, Zhi-Gang and {Lu}, Wen-Zhi and {Mao}, Yin-Dun and {Men}, Li and {Qi}, Yong-Jun and {Qi}, Zhao-Xiang and {Shi}, Huo-Ming and {Tang}, Zheng-Hong and {Tao}, Qing-Sheng and {Wang}, Da-Qi and {Wang}, Dan and {Wang}, Guo-Min and {Wang}, Hai and {Wang}, Jia-Ning and {Wang}, Jian and {Wang}, Jian-Ling and {Wang}, Jian-Ping and {Wang}, Lei and {Wang}, Shu-Qing and {Wang}, You and {Wang}, Yue-Fei and {Xu}, Ling-Zhe and {Xu}, Yan and {Yang}, Shi-Hai and {Yu}, Yong and {Yuan}, Hui and {Yuan}, Xiang-Yan and {Zhai}, Chao and {Zhang}, Jing and {Zhang}, Yan-Xia and {Zhang}, Yong and {Zhao}, Ming and {Zhou}, Fang and {Zhou}, Guo-Hua and {Zhu}, Jie and {Zou}, Si-Cheng},
        title = "{The Large Sky Area Multi-Object Fiber Spectroscopic Telescope (LAMOST)}",
      journal = {Research in Astronomy and Astrophysics},
         year = 2012,
        month = sep,
       volume = {12},
       number = {9},
        pages = {1197-1242},
          doi = {10.1088/1674-4527/12/9/003},
       adsurl = {https://ui.adsabs.harvard.edu/abs/2012RAA....12.1197C}
}

@ARTICLE{Majewski2017,
       author = {{Majewski}, Steven R. and {Schiavon}, Ricardo P. and {Frinchaboy}, Peter M. and {Allende Prieto}, Carlos and {Barkhouser}, Robert and {Bizyaev}, Dmitry and {Blank}, Basil and {Brunner}, Sophia and {Burton}, Adam and {Carrera}, Ricardo and {Chojnowski}, S. Drew and {Cunha}, K{\'a}tia and {Epstein}, Courtney and {Fitzgerald}, Greg and {Garc{\'\i}a P{\'e}rez}, Ana E. and {Hearty}, Fred R. and {Henderson}, Chuck and {Holtzman}, Jon A. and {Johnson}, Jennifer A. and {Lam}, Charles R. and {Lawler}, James E. and {Maseman}, Paul and {M{\'e}sz{\'a}ros}, Szabolcs and {Nelson}, Matthew and {Nguyen}, Duy Coung and {Nidever}, David L. and {Pinsonneault}, Marc and {Shetrone}, Matthew and {Smee}, Stephen and {Smith}, Verne V. and {Stolberg}, Todd and {Skrutskie}, Michael F. and {Walker}, Eric and {Wilson}, John C. and {Zasowski}, Gail and {Anders}, Friedrich and {Basu}, Sarbani and {Beland}, Stephane and {Blanton}, Michael R. and {Bovy}, Jo and {Brownstein}, Joel R. and {Carlberg}, Joleen and {Chaplin}, William and {Chiappini}, Cristina and {Eisenstein}, Daniel J. and {Elsworth}, Yvonne and {Feuillet}, Diane and {Fleming}, Scott W. and {Galbraith-Frew}, Jessica and {Garc{\'\i}a}, Rafael A. and {Garc{\'\i}a-Hern{\'a}ndez}, D. An{\'\i}bal and {Gillespie}, Bruce A. and {Girardi}, L{\'e}o and {Gunn}, James E. and {Hasselquist}, Sten and {Hayden}, Michael R. and {Hekker}, Saskia and {Ivans}, Inese and {Kinemuchi}, Karen and {Klaene}, Mark and {Mahadevan}, Suvrath and {Mathur}, Savita and {Mosser}, Beno{\^\i}t and {Muna}, Demitri and {Munn}, Jeffrey A. and {Nichol}, Robert C. and {O'Connell}, Robert W. and {Parejko}, John K. and {Robin}, A.~C. and {Rocha-Pinto}, Helio and {Schultheis}, Matthias and {Serenelli}, Aldo M. and {Shane}, Neville and {Silva Aguirre}, Victor and {Sobeck}, Jennifer S. and {Thompson}, Benjamin and {Troup}, Nicholas W. and {Weinberg}, David H. and {Zamora}, Olga},
        title = "{The Apache Point Observatory Galactic Evolution Experiment (APOGEE)}",
      journal = {\aj},
         year = 2017,
        month = sep,
       volume = {154},
       number = {3},
          eid = {94},
        pages = {94},
          doi = {10.3847/1538-3881/aa784d},
archivePrefix = {arXiv},
       eprint = {1509.05420},
 primaryClass = {astro-ph.IM},
       adsurl = {https://ui.adsabs.harvard.edu/abs/2017AJ....154...94M}
}

@ARTICLE{VeraCiro2013,
       author = {{Vera-Ciro}, Carlos and {Helmi}, Amina},
        title = "{Constraints on the Shape of the Milky Way Dark Matter Halo from the Sagittarius Stream}",
      journal = {\apjl},
         year = 2013,
        month = aug,
       volume = {773},
       number = {1},
          eid = {L4},
        pages = {L4},
          doi = {10.1088/2041-8205/773/1/L4},
archivePrefix = {arXiv},
       eprint = {1304.4646},
 primaryClass = {astro-ph.GA},
       adsurl = {https://ui.adsabs.harvard.edu/abs/2013ApJ...773L...4V}
}

@ARTICLE{Bonaca2019,
       author = {{Bonaca}, Ana and {Hogg}, David W. and {Price-Whelan}, Adrian M. and {Conroy}, Charlie},
        title = "{The Spur and the Gap in GD-1: Dynamical Evidence for a Dark Substructure in the Milky Way Halo}",
      journal = {\apj},
         year = 2019,
        month = jul,
       volume = {880},
       number = {1},
          eid = {38},
        pages = {38},
          doi = {10.3847/1538-4357/ab2873},
archivePrefix = {arXiv},
       eprint = {1811.03631},
 primaryClass = {astro-ph.GA},
       adsurl = {https://ui.adsabs.harvard.edu/abs/2019ApJ...880...38B}
}

@ARTICLE{Sola2025,
       author = {{Sola}, E. and {Chemaly}, D. and {Belokurov}, V. and {M{\"u}ller}, O. and {Ardern-Arentsen}, A. and {Davies}, E.~Y. and {Laguna-Miralles}, J. and {Myeong}, G. and {Panagiotakis}, K. and {Zhang}, H. and {Erkal}, D. and {Koposov}, S.~E. and {Lang}, D. and {Nibauer}, J.},
        title = "{STRRINGS: STReams in Residual Images of Nearby GalaxieS}",
      journal = {arXiv e-prints},
         year = 2025,
        month = aug,
          eid = {arXiv:2508.02154},
        pages = {arXiv:2508.02154},
          doi = {10.48550/arXiv.2508.02154},
archivePrefix = {arXiv},
       eprint = {2508.02154},
 primaryClass = {astro-ph.GA},
       adsurl = {https://ui.adsabs.harvard.edu/abs/2025arXiv250802154S}
}

@ARTICLE{Tal2009,
       author = {{Tal}, Tomer and {van Dokkum}, Pieter G. and {Nelan}, Jenica and {Bezanson}, Rachel},
        title = "{The Frequency of Tidal Features Associated with Nearby Luminous Elliptical Galaxies From a Statistically Complete Sample}",
      journal = {\aj},
         year = 2009,
        month = nov,
       volume = {138},
       number = {5},
        pages = {1417-1427},
          doi = {10.1088/0004-6256/138/5/1417},
archivePrefix = {arXiv},
       eprint = {0908.1382},
 primaryClass = {astro-ph.CO},
       adsurl = {https://ui.adsabs.harvard.edu/abs/2009AJ....138.1417T}
}

@ARTICLE{Atkinson2013,
       author = {{Atkinson}, Adam M. and {Abraham}, Roberto G. and {Ferguson}, Annette M.~N.},
        title = "{Faint Tidal Features in Galaxies within the Canada-France-Hawaii Telescope Legacy Survey Wide Fields}",
      journal = {\apj},
         year = 2013,
        month = mar,
       volume = {765},
       number = {1},
          eid = {28},
        pages = {28},
          doi = {10.1088/0004-637X/765/1/28},
archivePrefix = {arXiv},
       eprint = {1301.4275},
 primaryClass = {astro-ph.CO},
       adsurl = {https://ui.adsabs.harvard.edu/abs/2013ApJ...765...28A}
}

@ARTICLE{Hood2018,
       author = {{Hood}, Callie E. and {Kannappan}, Sheila J. and {Stark}, David V. and {Dell'Antonio}, Ian P. and {Moffett}, Amanda J. and {Eckert}, Kathleen D. and {Norris}, Mark A. and {Hendel}, David},
        title = "{The Origin of Faint Tidal Features around Galaxies in the RESOLVE Survey}",
      journal = {\apj},
         year = 2018,
        month = apr,
       volume = {857},
       number = {2},
          eid = {144},
        pages = {144},
          doi = {10.3847/1538-4357/aab719},
archivePrefix = {arXiv},
       eprint = {1803.05447},
 primaryClass = {astro-ph.GA},
       adsurl = {https://ui.adsabs.harvard.edu/abs/2018ApJ...857..144H}
}

@ARTICLE{Bilek2020,
       author = {{B{\'\i}lek}, Michal and {Duc}, Pierre-Alain and {Cuillandre}, Jean-Charles and {Gwyn}, Stephen and {Cappellari}, Michele and {Bekaert}, David V. and {Bonfini}, Paolo and {Bitsakis}, Theodoros and {Paudel}, Sanjaya and {Krajnovi{\'c}}, Davor and {Durrell}, Patrick R. and {Marleau}, Francine},
        title = "{Census and classification of low-surface-brightness structures in nearby early-type galaxies from the MATLAS survey}",
      journal = {\mnras},
         year = 2020,
        month = oct,
       volume = {498},
       number = {2},
        pages = {2138-2166},
          doi = {10.1093/mnras/staa2248},
archivePrefix = {arXiv},
       eprint = {2007.13772},
 primaryClass = {astro-ph.GA},
       adsurl = {https://ui.adsabs.harvard.edu/abs/2020MNRAS.498.2138B}
}

@ARTICLE{Sola2022,
       author = {{Sola}, Elisabeth and {Duc}, Pierre-Alain and {Richards}, Felix and {Paiement}, Adeline and {Urbano}, Mathias and {Klehammer}, Julie and {B{\'\i}lek}, Michal and {Cuillandre}, Jean-Charles and {Gwyn}, Stephen and {McConnachie}, Alan},
        title = "{Characterization of low surface brightness structures in annotated deep images}",
      journal = {\aap},
         year = 2022,
        month = jun,
       volume = {662},
          eid = {A124},
        pages = {A124},
          doi = {10.1051/0004-6361/202142675},
archivePrefix = {arXiv},
       eprint = {2203.03973},
 primaryClass = {astro-ph.GA},
       adsurl = {https://ui.adsabs.harvard.edu/abs/2022A&A...662A.124S}
}

@ARTICLE{Yoon2024,
       author = {{Yoon}, Yongmin and {Ko}, Jongwan and {Chung}, Haeun and {Byun}, Woowon and {Chun}, Kyungwon},
        title = "{Shell-type Tidal Features Are More Frequently Detected in Slowly Rotating Early-type Galaxies than Stream- and Tail-type Features}",
      journal = {\apj},
         year = 2024,
        month = apr,
       volume = {965},
       number = {2},
          eid = {158},
        pages = {158},
          doi = {10.3847/1538-4357/ad34ad},
archivePrefix = {arXiv},
       eprint = {2404.03459},
 primaryClass = {astro-ph.GA},
       adsurl = {https://ui.adsabs.harvard.edu/abs/2024ApJ...965..158Y}
}

@ARTICLE{Johnston1998,
       author = {{Johnston}, Kathryn V.},
        title = "{A Prescription for Building the Milky Way's Halo from Disrupted Satellites}",
      journal = {\apj},
         year = 1998,
        month = mar,
       volume = {495},
       number = {1},
        pages = {297-308},
          doi = {10.1086/305273},
archivePrefix = {arXiv},
       eprint = {astro-ph/9710007},
 primaryClass = {astro-ph},
       adsurl = {https://ui.adsabs.harvard.edu/abs/1998ApJ...495..297J}
}

@ARTICLE{Kuepper2012,
       author = {{K{\"u}pper}, Andreas H.~W. and {Lane}, Richard R. and {Heggie}, Douglas C.},
        title = "{More on the structure of tidal tails}",
      journal = {\mnras},
         year = 2012,
        month = mar,
       volume = {420},
       number = {3},
        pages = {2700-2714},
          doi = {10.1111/j.1365-2966.2011.20242.x},
archivePrefix = {arXiv},
       eprint = {1111.5013},
 primaryClass = {astro-ph.GA},
       adsurl = {https://ui.adsabs.harvard.edu/abs/2012MNRAS.420.2700K}
}

@ARTICLE{Chen2025,
       author = {{Chen}, Yingtian and {Valluri}, Monica and {Gnedin}, Oleg Y. and {Ash}, Neil},
        title = "{Improved Particle Spray Algorithm for Modeling Globular Cluster Streams}",
      journal = {\apjs},
         year = 2025,
        month = feb,
       volume = {276},
       number = {2},
          eid = {32},
        pages = {32},
          doi = {10.3847/1538-4365/ad9904},
archivePrefix = {arXiv},
       eprint = {2408.01496},
 primaryClass = {astro-ph.GA},
       adsurl = {https://ui.adsabs.harvard.edu/abs/2025ApJS..276...32C}
}

@ARTICLE{ChenLi2025,
       author = {{Chen}, Yingtian and {Li}, Hui and {Gnedin}, Oleg Y.},
        title = "{Stellar Streams Reveal the Mass Loss of Globular Clusters}",
      journal = {\apjl},
         year = 2025,
        month = feb,
       volume = {980},
          eid = {L18},
        pages = {L18},
          doi = {10.3847/2041-8213/adaf93},
archivePrefix = {arXiv},
       eprint = {2411.19899},
 primaryClass = {astro-ph.GA},
       adsurl = {https://ui.adsabs.harvard.edu/abs/2025ApJ...980L..18C}
}

@ARTICLE{Tavangar2025,
       author = {{Tavangar}, Kiyan and {Price-Whelan}, Adrian M.},
        title = "{Inferring the Density and Membership of Stellar Streams with Flexible Models: The GD-1 Stream in Gaia Data Release 3}",
      journal = {\apj},
         year = 2025,
        month = jul,
       volume = {988},
          eid = {45},
        pages = {45},
          doi = {10.3847/1538-4357/addd1c},
archivePrefix = {arXiv},
       eprint = {2502.13236},
 primaryClass = {astro-ph.GA},
       adsurl = {https://ui.adsabs.harvard.edu/abs/2025ApJ...988...45T}
}

@ARTICLE{Wong2023,
       author = {{Wong}, Kaze W.~K. and {Gabri{\'e}}, Marylou and {Foreman-Mackey}, Daniel},
        title = "{flowMC: Normalizing flow enhanced sampling package for probabilistic inference in JAX}",
      journal = {The Journal of Open Source Software},
         year = 2023,
        month = mar,
       volume = {8},
        pages = {5021},
          doi = {10.21105/joss.05021},
archivePrefix = {arXiv},
       eprint = {2211.06397},
 primaryClass = {astro-ph.IM},
       adsurl = {https://ui.adsabs.harvard.edu/abs/2023JOSS....8.5021W}
}

@software{jax2018github,
  author = {James Bradbury and Roy Frostig and Peter Hawkins and Matthew James Johnson and Chris Leary and Dougal Maclaurin and George Necula and Adam Paszke and Jake Vander{P}las and Skye Wanderman-{M}ilne and Qiao Zhang},
  title = {{JAX}: composable transformations of {P}ython+{N}um{P}y programs},
  url = {http://github.com/jax-ml/jax},
  version = {0.3.13},
  year = {2018},
}

@ARTICLE{Dey2019,
       author = {{Dey}, Arjun and {Schlegel}, David J. and {Lang}, Dustin and {Blum}, Robert and {Burleigh}, Kaylan and {Fan}, Xiaohui and {Findlay}, Joseph R. and {Finkbeiner}, Doug and {Herrera}, David and {Juneau}, St{\'e}phanie and {Landriau}, Martin and {Levi}, Michael and {McGreer}, Ian and {Meisner}, Aaron and {Myers}, Adam D. and {Moustakas}, John and {Nugent}, Peter and {Patej}, Anna and {Schlafly}, Edward F. and {Walker}, Alistair R. and {Valdes}, Francisco and {Weaver}, Benjamin A. and {Y{\`e}che}, Christophe and {Zou}, Hu and {Zhou}, Xu and {Abareshi}, Behzad and {Abbott}, T.~M.~C. and {Abolfathi}, Bela and {Aguilera}, C. and {Alam}, Shadab and {Allen}, Lori and {Alvarez}, A. and {Annis}, James and {Ansarinejad}, Behzad and {Aubert}, Marie and {Beechert}, Jacqueline and {Bell}, Eric F. and {BenZvi}, Segev Y. and {Beutler}, Florian and {Bielby}, Richard M. and {Bolton}, Adam S. and {Brice{\~n}o}, C{\'e}sar and {Buckley-Geer}, Elizabeth J. and {Butler}, Karen and {Calamida}, Annalisa and {Carlberg}, Raymond G. and {Carter}, Paul and {Casas}, Ricard and {Castander}, Francisco J. and {Choi}, Yumi and {Comparat}, Johan and {Cukanovaite}, Elena and {Delubac}, Timoth{\'e}e and {DeVries}, Kaitlin and {Dey}, Sharmila and {Dhungana}, Govinda and {Dickinson}, Mark and {Ding}, Zhejie and {Donaldson}, John B. and {Duan}, Yutong and {Duckworth}, Christopher J. and {Eftekharzadeh}, Sarah and {Eisenstein}, Daniel J. and {Etourneau}, Thomas and {Fagrelius}, Parker A. and {Farihi}, Jay and {Fitzpatrick}, Mike and {Font-Ribera}, Andreu and {Fulmer}, Leah and {G{\"a}nsicke}, Boris T. and {Gaztanaga}, Enrique and {George}, Koshy and {Gerdes}, David W. and {Gontcho}, Satya Gontcho A. and {Gorgoni}, Claudio and {Green}, Gregory and {Guy}, Julien and {Harmer}, Diane and {Hernandez}, M. and {Honscheid}, Klaus and {Huang}, Lijuan Wendy and {James}, David J. and {Jannuzi}, Buell T. and {Jiang}, Linhua and {Joyce}, Richard and {Karcher}, Armin and {Karkar}, Sonia and {Kehoe}, Robert and {Kneib}, Jean-Paul and {Kueter-Young}, Andrea and {Lan}, Ting-Wen and {Lauer}, Tod R. and {Le Guillou}, Laurent and {Le Van Suu}, Auguste and {Lee}, Jae Hyeon and {Lesser}, Michael and {Perreault Levasseur}, Laurence and {Li}, Ting S. and {Mann}, Justin L. and {Marshall}, Robert and {Mart{\'\i}nez-V{\'a}zquez}, C.~E. and {Martini}, Paul and {du Mas des Bourboux}, H{\'e}lion and {McManus}, Sean and {Meier}, Tobias Gabriel and {M{\'e}nard}, Brice and {Metcalfe}, Nigel and {Mu{\~n}oz-Guti{\'e}rrez}, Andrea and {Najita}, Joan and {Napier}, Kevin and {Narayan}, Gautham and {Newman}, Jeffrey A. and {Nie}, Jundan and {Nord}, Brian and {Norman}, Dara J. and {Olsen}, Knut A.~G. and {Paat}, Anthony and {Palanque-Delabrouille}, Nathalie and {Peng}, Xiyan and {Poppett}, Claire L. and {Poremba}, Megan R. and {Prakash}, Abhishek and {Rabinowitz}, David and {Raichoor}, Anand and {Rezaie}, Mehdi and {Robertson}, A.~N. and {Roe}, Natalie A. and {Ross}, Ashley J. and {Ross}, Nicholas P. and {Rudnick}, Gregory and {Safonova}, Sasha and {Saha}, Abhijit and {S{\'a}nchez}, F. Javier and {Savary}, Elodie and {Schweiker}, Heidi and {Scott}, Adam and {Seo}, Hee-Jong and {Shan}, Huanyuan and {Silva}, David R. and {Slepian}, Zachary and {Soto}, Christian and {Sprayberry}, David and {Staten}, Ryan and {Stillman}, Coley M. and {Stupak}, Robert J. and {Summers}, David L. and {Sien Tie}, Suk and {Tirado}, H. and {Vargas-Maga{\~n}a}, Mariana and {Vivas}, A. Katherina and {Wechsler}, Risa H. and {Williams}, Doug and {Yang}, Jinyi and {Yang}, Qian and {Yapici}, Tolga and {Zaritsky}, Dennis and {Zenteno}, A. and {Zhang}, Kai and {Zhang}, Tianmeng and {Zhou}, Rongpu and {Zhou}, Zhimin},
        title = "{Overview of the DESI Legacy Imaging Surveys}",
      journal = {\aj},
         year = 2019,
        month = may,
       volume = {157},
       number = {5},
          eid = {168},
        pages = {168},
          doi = {10.3847/1538-3881/ab089d},
archivePrefix = {arXiv},
       eprint = {1804.08657},
 primaryClass = {astro-ph.IM},
       adsurl = {https://ui.adsabs.harvard.edu/abs/2019AJ....157..168D}
}

@ARTICLE{Fardal2013,
       author = {{Fardal}, Mark A. and {Weinberg}, Martin D. and {Babul}, Arif and {Irwin}, Mike J. and {Guhathakurta}, Puragra and {Gilbert}, Karoline M. and {Ferguson}, Annette M.~N. and {Ibata}, Rodrigo A. and {Lewis}, Geraint F. and {Tanvir}, Nial R. and {Huxor}, Avon P.},
        title = "{Inferring the Andromeda Galaxy's mass from its giant southern stream with Bayesian simulation sampling}",
      journal = {\mnras},
         year = 2013,
        month = oct,
       volume = {434},
       number = {4},
        pages = {2779-2802},
          doi = {10.1093/mnras/stt1121},
archivePrefix = {arXiv},
       eprint = {1307.3219},
 primaryClass = {astro-ph.CO},
       adsurl = {https://ui.adsabs.harvard.edu/abs/2013MNRAS.434.2779F}
}

@ARTICLE{Varghese2011,
       author = {{Varghese}, A. and {Ibata}, R. and {Lewis}, G.~F.},
        title = "{Stellar streams as probes of dark halo mass and morphology: a Bayesian reconstruction}",
      journal = {\mnras},
         year = 2011,
        month = oct,
       volume = {417},
       number = {1},
        pages = {198-215},
          doi = {10.1111/j.1365-2966.2011.19097.x},
archivePrefix = {arXiv},
       eprint = {1106.1765},
 primaryClass = {astro-ph.GA},
       adsurl = {https://ui.adsabs.harvard.edu/abs/2011MNRAS.417..198V}
}

@article{Choi2007,
   title={The dynamics of tidal tails from massive satellites: The dynamics of tidal tails},
   volume={381},
   ISSN={1365-2966},
   url={http://dx.doi.org/10.1111/j.1365-2966.2007.12313.x},
   DOI={10.1111/j.1365-2966.2007.12313.x},
   number={3},
   journal={Monthly Notices of the Royal Astronomical Society},
   publisher={Oxford University Press (OUP)},
   author={Choi, Jun-Hwan and Weinberg, Martin D. and Katz, Neal},
   year={2007},
   month=oct, pages={987–1000} }

@ARTICLE{Belokurov2014,
       author = {{Belokurov}, V. and {Koposov}, S.~E. and {Evans}, N.~W. and {Pe{\~n}arrubia}, J. and {Irwin}, M.~J. and {Smith}, M.~C. and {Lewis}, G.~F. and {Gieles}, M. and {Wilkinson}, M.~I. and {Gilmore}, G. and {Olszewski}, E.~W. and {Niederste-Ostholt}, M.},
        title = "{Precession of the Sagittarius stream}",
      journal = {\mnras},
         year = 2014,
        month = jan,
       volume = {437},
       number = {1},
        pages = {116-131},
          doi = {10.1093/mnras/stt1862},
archivePrefix = {arXiv},
       eprint = {1301.7069},
 primaryClass = {astro-ph.GA},
       adsurl = {https://ui.adsabs.harvard.edu/abs/2014MNRAS.437..116B}
}

@ARTICLE{Bonaca2018,
       author = {{Bonaca}, Ana and {Hogg}, David W.},
        title = "{The Information Content in Cold Stellar Streams}",
      journal = {\apj},
         year = 2018,
        month = nov,
       volume = {867},
       number = {2},
          eid = {101},
        pages = {101},
          doi = {10.3847/1538-4357/aae4da},
archivePrefix = {arXiv},
       eprint = {1804.06854},
 primaryClass = {astro-ph.GA},
       adsurl = {https://ui.adsabs.harvard.edu/abs/2018ApJ...867..101B}
}

@ARTICLE{Navarro1997,
       author = {{Navarro}, Julio F. and {Frenk}, Carlos S. and {White}, Simon D.~M.},
        title = "{A Universal Density Profile from Hierarchical Clustering}",
      journal = {\apj},
         year = 1997,
        month = dec,
       volume = {490},
       number = {2},
        pages = {493-508},
          doi = {10.1086/304888},
archivePrefix = {arXiv},
       eprint = {astro-ph/9611107},
 primaryClass = {astro-ph},
       adsurl = {https://ui.adsabs.harvard.edu/abs/1997ApJ...490..493N}
}

@ARTICLE{Zhao2003,
       author = {{Zhao}, D.~H. and {Mo}, H.~J. and {Jing}, Y.~P. and {B{\"o}rner}, G.},
        title = "{The growth and structure of dark matter haloes}",
      journal = {\mnras},
         year = 2003,
        month = feb,
       volume = {339},
       number = {1},
        pages = {12-24},
          doi = {10.1046/j.1365-8711.2003.06135.x},
archivePrefix = {arXiv},
       eprint = {astro-ph/0204108},
 primaryClass = {astro-ph},
       adsurl = {https://ui.adsabs.harvard.edu/abs/2003MNRAS.339...12Z}
}

@ARTICLE{Ludlow2014,
       author = {{Ludlow}, Aaron D. and {Navarro}, Julio F. and {Angulo}, Ra{\'u}l E. and {Boylan-Kolchin}, Michael and {Springel}, Volker and {Frenk}, Carlos and {White}, Simon D.~M.},
        title = "{The mass-concentration-redshift relation of cold dark matter haloes}",
      journal = {\mnras},
         year = 2014,
        month = jun,
       volume = {441},
       number = {1},
        pages = {378-388},
          doi = {10.1093/mnras/stu483},
archivePrefix = {arXiv},
       eprint = {1312.0945},
 primaryClass = {astro-ph.CO},
       adsurl = {https://ui.adsabs.harvard.edu/abs/2014MNRAS.441..378L}
}

@ARTICLE{Wechsler2002,
       author = {{Wechsler}, Risa H. and {Bullock}, James S. and {Primack}, Joel R. and {Kravtsov}, Andrey V. and {Dekel}, Avishai},
        title = "{Concentrations of Dark Halos from Their Assembly Histories}",
      journal = {\apj},
         year = 2002,
        month = mar,
       volume = {568},
       number = {1},
        pages = {52-70},
          doi = {10.1086/338765},
archivePrefix = {arXiv},
       eprint = {astro-ph/0108151},
 primaryClass = {astro-ph},
       adsurl = {https://ui.adsabs.harvard.edu/abs/2002ApJ...568...52W}
}

@ARTICLE{Maccio2008,
       author = {{Macci{\`o}}, Andrea V. and {Dutton}, Aaron A. and {van den Bosch}, Frank C.},
        title = "{Concentration, spin and shape of dark matter haloes as a function of the cosmological model: WMAP1, WMAP3 and WMAP5 results}",
      journal = {\mnras},
         year = 2008,
        month = dec,
       volume = {391},
       number = {4},
        pages = {1940-1954},
          doi = {10.1111/j.1365-2966.2008.14029.x},
archivePrefix = {arXiv},
       eprint = {0805.1926},
 primaryClass = {astro-ph},
       adsurl = {https://ui.adsabs.harvard.edu/abs/2008MNRAS.391.1940M}
}




\appendix

\section{Population Relations}\label{app:A}

Here we examine how the accuracy of the inferred flattening varies with simple morphological properties of the stream tracks across our full population. In total, we analyse the 105 individual posteriors obtained from this work. For each posterior we compute the absolute mean error (AME), which measures the average deviation of the posterior samples from the true value. Smaller AME values correspond to more accurate constraints.

For each stream we compute four descriptive quantities: angular length, physical length, mean distance from the origin, and radial distance spread (Figure~\ref{fig:goodness}). Angular lengths are discrete because tracks are binned in angle; the remaining quantities are continuous. Across the population, the scatter in AME at fixed morphological properties is large, and no single parameter produces a clean predictive trend. Nonetheless, we use linear regression to describe the correlation between the AME and our aformentionned quantites. Longer angular tracks exhibit slightly smaller errors, consistent with the idea that a greater number of angular bins provides more information. A similar trend appears with physical length: longer streams tend to have marginally lower AME, though this can be partly a geometric effect since the same angular extent corresponds to larger physical lengths at larger distances. The mean distance shows the opposite behaviour. Streams at smaller mean distance are more likely to cover larger angular extents, and thus modestly lower AME values. Finally, the radial distance spread shows almost no relation with AME. This is expected, as the flattening parameter primarily affects the angular structure of the orbit; probing a wider radial range is not required to constrain $q$ in the projected track setting explored here.

Overall, while some qualitative trends exist, the intrinsic scatter is substantial. A larger and more diverse population of mock streams would be needed to assess the robustness of these relations and to quantify the sources of spread.

\begin{figure*}
    \centering
    \includegraphics[width=1.0\linewidth]{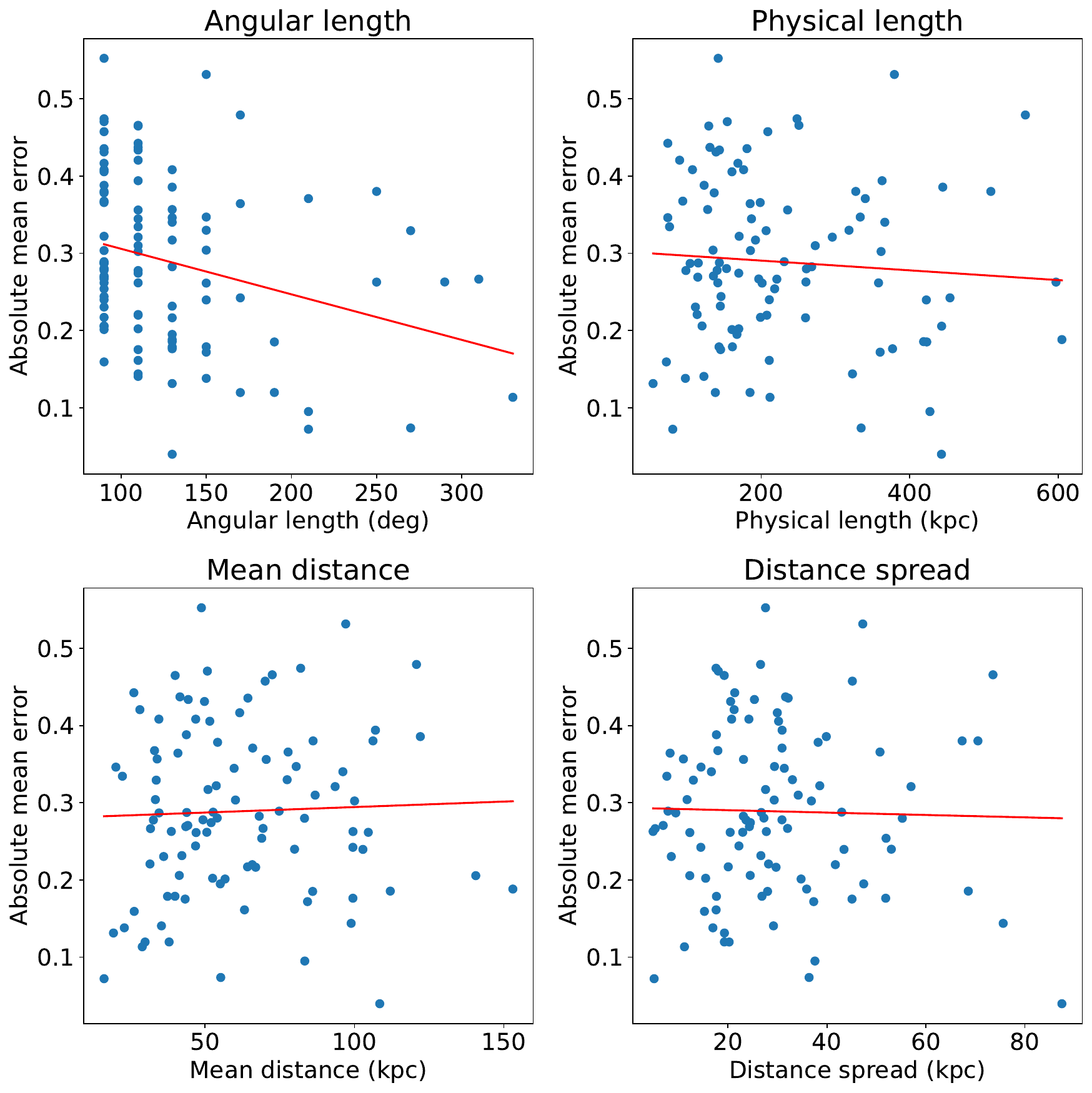}
    \caption[]{Absolute mean error of the inferred flattening for 105 stream fits as a function of four morphological properties of each stream: angular length, linear length, mean distance, and radial distance spread. The red line shows a simple linear fit. Weak trends are present but the overall scatter is large, and no single morphological parameter reliably predicts inference quality. That being said, longer streams (both angular and linear) tend to yield smaller errors, shorter mean distances also lead to smaller errors, and radial distance spread has no effect.}
    \label{fig:goodness}
\end{figure*}


\bsp	
\label{lastpage}
\end{document}